\documentclass{article}

\usepackage{arxiv}

\usepackage[utf8]{inputenc} \usepackage[T1]{fontenc}    \usepackage{hyperref}       \usepackage{url}            \usepackage{booktabs}       \usepackage{amsfonts}       \usepackage{nicefrac}       \usepackage{microtype}      \usepackage{lipsum}		\usepackage{graphicx}
\usepackage[round]{natbib}
\usepackage{doi}
\usepackage[alignedforall]{lpform} \usepackage{bbm}
\def\mydefb#1{\expandafter\def\csname b#1\endcsname{\bm{#1}}}
\def\mydefallb#1{\ifx#1\mydefallb\else\mydefb#1\expandafter\mydefallb\fi}
\mydefallb ABCDEFGHIJKLMNOPQRSTUVWXYZabcdeghijklnopqrstuvwxyz\mydefallb
\usepackage[alignedforall]{lpform} \def\mydefgreek#1{\expandafter\def\csname b#1\endcsname{\text{\boldmath$\mathbf{\csname #1\endcsname}$}}}
\def\mydefallgreek#1{\ifx\mydefallgreek#1\else\mydefgreek{#1}\lowercase{\mydefgreek{#1}}\expandafter\mydefallgreek\fi}
\mydefallgreek {beta}{Gamma}{alpha}{Delta}{epsilon}{Omega}{Theta}{Iota}{Lambda}{kappa}{mu}{nu}{Xi}{Pi}{chi}{rho}{Sigma}{Psi}{tau}\mydefallgreek
\newcommand{\C}{\text{\normalfont C}}
\newcommand{\D}{\text{\normalfont D}}
\usepackage{amsmath,amsfonts,amssymb, bm}

\usepackage{xcolor}
\definecolor{darkgreen}{RGB}{0, 150, 0}
\definecolor{royalblue}{RGB}{65, 105, 225} 
\definecolor{darkorange}{RGB}{255, 140, 0}  

\newcommand{\T}{\text{\normalfont T}}

  \usepackage{placeins}
  \usepackage{makecell}
\usepackage[english]{babel}
\usepackage{fancyhdr}
  \addto\extrasenglish{

}

\usepackage{multirow}
\usepackage{graphicx}
\usepackage{ulem}
\renewcommand{\emph}[1]{{\it #1}}
\usepackage{booktabs}
\usepackage{accents}
\newcommand{\ubar}[1]{\underaccent{\bar}{#1}}
\usepackage{graphicx}
\newcounter{condition}

\newenvironment{condition}
{
  \refstepcounter{condition}
  \par\noindent\textbf{Condition C (Convexity):}}
{
  \par
}

\usepackage{setspace}
\title{An integer programming-based approach to construct exact two-sample binomial tests with maximum power}

\author{Stef Baas\thanks{Corresponding author} \\
    MRC Biostatistics Unit, University~of~Cambridge, East Forvie Building,\\ Forvie Site, Robinson Way, CB2 0SR, Cambridge, United kingdom \\\phantom{.}\\ {\bf Yaron Racah}\\
	PhaseV, 1 Broadway, Cambridge, MA \\\phantom{.}\\
		{\bf Elad Berkman}\\
	PhaseV, 1 Broadway, Cambridge, MA \\\phantom{.}\\
 {\bf Sof\'ia S. Villar}\\
     MRC Biostatistics Unit, University~of~Cambridge, East Forvie Building,\\ Forvie Site, Robinson Way, CB2 0SR, Cambridge, United kingdom 
}

\renewcommand{\shorttitle}{An integer programming approach to construct exact tests with maximum power}

\hypersetup{
pdftitle={An integer programming-based approach to construct exact two-sample binomial tests with maximum power},
pdfauthor={Stef Baas, Yaron Racah, Elad Berkman, Sofia S. Villar},
pdfkeywords={Average power, Berger and Boos' procedure, Fisher's exact test, Knapsack problem, Maximin Power},
}
\usepackage{bm}
\def\mydefb#1{\expandafter\def\csname b#1\endcsname{\bm{#1}}}
\def\mydefallb#1{\ifx#1\mydefallb\else\mydefb#1\expandafter\mydefallb\fi}
\mydefallb ABCDEFGHIJKLMNOPQRSTUVWXYZabcdeghijklnopqrstuyz\mydefallb

\def\mydefgreek#1{\expandafter\def\csname b#1\endcsname{\text{\boldmath$\mathbf{\csname #1\endcsname}$}}}
\def\mydefallgreek#1{\ifx\mydefallgreek#1\else\mydefgreek{#1}\lowercase{\mydefgreek{#1}}\expandafter\mydefallgreek\fi}
\mydefallgreek {beta}{Gamma}{alpha}{Delta}{epsilon}{Omega}{Theta}{Iota}{Lambda}{kappa}{mu}{nu}{Xi}{Pi}{chi}{rho}{Sigma}{Psi}{tau}\mydefallgreek

\usepackage{lipsum}

\usepackage{graphicx}\usepackage{multirow}\usepackage{amsmath,amssymb,amsfonts}\usepackage{amsthm}\usepackage{mathrsfs}\usepackage[title]{appendix}\usepackage{xcolor}\usepackage{textcomp}\usepackage{manyfoot}\usepackage{booktabs}\usepackage{algorithm}\usepackage{algorithmicx}\usepackage{algpseudocode}\usepackage{listings}\algnewcommand{\Inputs}[1]{\State \textbf{Inputs:}
	\Statex \hspace*{\algorithmicindent}\parbox[t]{.8\linewidth}{\raggedright #1}
}
\algnewcommand{\Initialize}[1]{\State \textbf{Initialize:}
	\Statex \hspace*{\algorithmicindent}\parbox[t]{.8\linewidth}{\raggedright #1}
}
\usepackage{placeins}
\usepackage{accents}
\usepackage{float}

\usepackage{orcidlink}
\usepackage{hyperref}                                           

\hypersetup{bookmarksnumbered,
pdfstartview={FitH},
linkcolor=blue,
citecolor=blue,
colorlinks=true,
pdfpagemode={UseNone},
plainpages=false
}

\newtheorem{theorem}{Theorem}

\newtheorem{remark}[theorem]{Remark}

\begin{document}
\maketitle
\begin{abstract} Comparing binomial proportions is a foundational task in clinical trials; however, practitioners often use a liberal likelihood-based test or the conservative Fisher’s exact test. To increase power without sacrificing validity, we use a linear integer program to construct finite-sample exact binomial tests with maximum power. Our proposed Average Power Knapsack (APK) test finds a decision boundary that maximizes average power, guaranteeing it cannot uniformly be improved in terms of power, while enforcing type I error rate control across the null hypothesis parameter space using Lipschitz continuity. Our numerical evaluation reveals consistent pointwise power gains of up to 35\% for the APK test over Fisher’s exact test, as well as systematic power improvements over state-of-the-art exact Berger and Boos implementations of the mid-$p$-value and Z-pooled tests. When compared against non-exact procedures exhibiting moderate type I error rate inflation, the exact APK test maintains comparable power. For sample sizes where APK construction becomes computationally intensive, Fisher’s mid-$p$-value test emerges as the best non-exact alternative. To support practical application, we developed an interactive R Shiny tool that enables seamless implementation of our optimized exact tests for binomial proportions.
\end{abstract}

\section{Introduction}\label{sect:introduction}
Statistical tests for superiority based on binomial proportions 
are among the first concepts taught in statistics courses. They are also ubiquitous in clinical research, as binary outcome measures are very common in confirmatory two-arm superiority clinical trials, with a frequency of use of at least 28\%  reported in~\cite{thompson2025RR}.  

\begin{figure}[h!]
\label{app:occurrances}
    \centering
    \includegraphics[width=.85\linewidth]{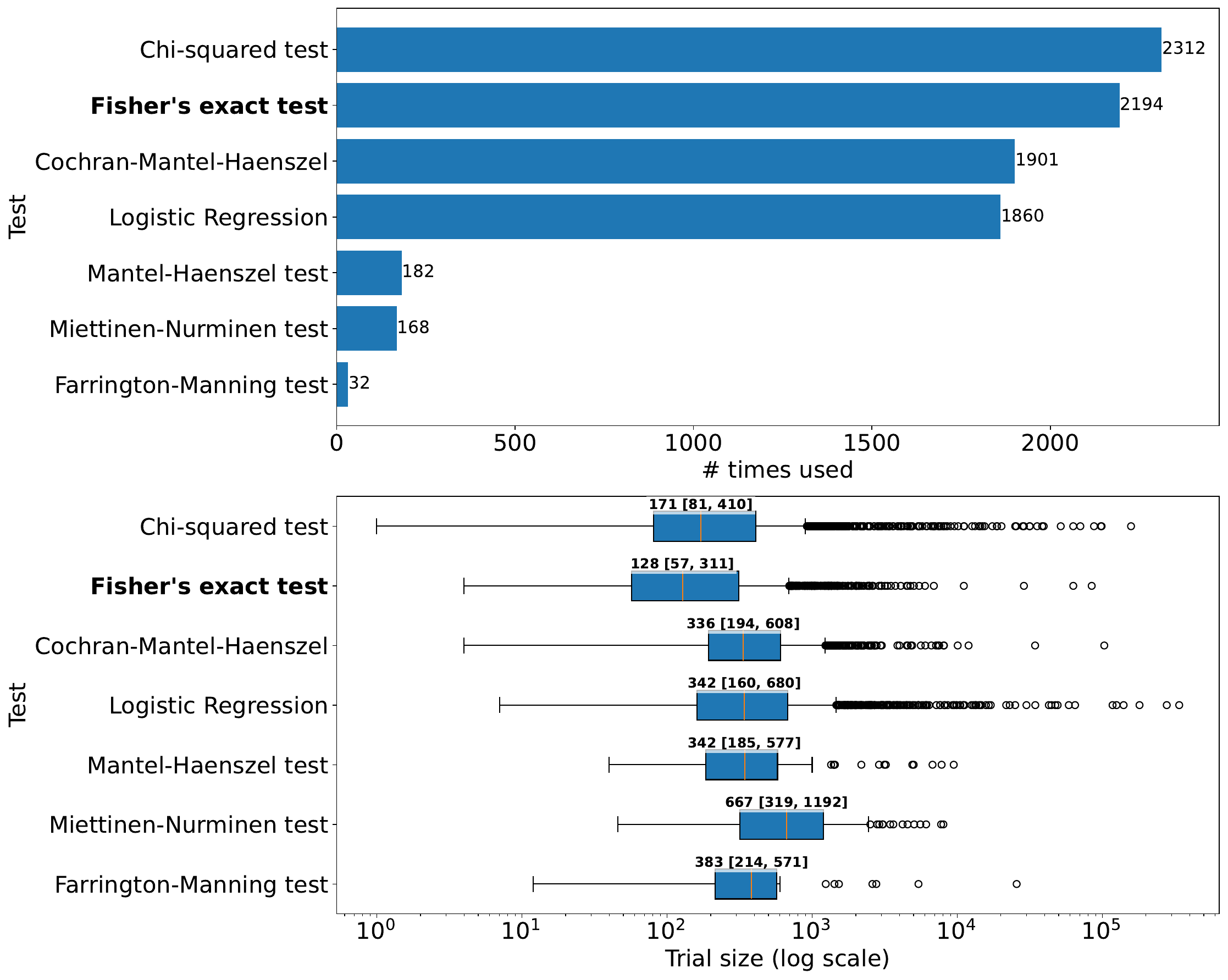}
    \caption{{\bf Occurrences of tests for categorical data and trial sizes}: Top: number of applied two-sample tests for both binary and categorical data for trials reported on~\url{https://clinicaltrials.gov} found using the AACT data~(\url{https://aact.ctti-clinicaltrials.org/downloads}) on 18/7/2026. We grouped entries corresponding to the Chi-squared test (``Chi-square test", ``Chi-squared", ``Chi-squared, Corrected", ``Wald Chi-square''),
                    Fisher's exact test~(``Fisher Exact", ``Fisher's exact test", ``Fisher's Exact Test"), Cochran-Mantel-Haenszel test~
                    (``Cochran-Mantel-Haenszel", ``Cochran-Mantel-Haenszel chi-square test"), 
                    logistic regression~(``Regression, Logistic", ``Logistic Regression", ``Logistic regression model"),
Mantel-Haenszel test~(``Mantel Haenszel"), 
                    Miettinen-Nurminen test~(``Miettinen \& Nurminen method", ``Miettinen and Nurminen method",  ``Miettinen \& Nurminen",  ``Miettinen and Nurminen", ``Miettinen-Nurminen"),
                    Farrington-Manning test~(``Farrington-Manning", ``Farrington-Manning test"). Bottom: boxplots (median and interquartile range written out) of trial sizes (enrollment) in trials that used the respective statistical analysis method.
}
    \label{fig:overview_test_categorical_trials}
\end{figure}

 A search of the~\url{https://clinicaltrials.gov} registry (conducted on July 18 2026) revealed that approximately 25\%--30\% (about 2,300 analyses) of two-sample binary or categorical data analyses in clinical trials used Fisher's exact test~(see the top row of \autoref{fig:overview_test_categorical_trials}). In addition, roughly 2,200 analyses utilized a chi-squared test, indicating that the majority of these analyses did not incorporate covariates or stratification as in the setting our paper focuses on. Methodologically, the use of Fisher’s exact test ~\citep{Fisher1934method} strongly suggests small sample size constraints. 
This statement aligns with the distribution over trial sizes for these respective analyses, shown in the bottom row of~\autoref{fig:overview_test_categorical_trials}, where the median and interquartile range for the trials using Fisher's exact tests were 128 and [57, 311], respectively, while larger values are observed for the other tests.

Fisher's exact test is viewed as~(too) conservative~\citep[see, e.g.,][Section~3.5.6]{agresti_book_cda} which often translates to substantially lower power than other tests. 
Unconditional exact binomial tests have been proposed, which are less restrictive than Fisher's~(conditional) exact test as they treat the total number of successes as a random variable instead of a fixed quantity. As a result, such tests can attain higher unconditional power. Examples of such unconditional exact tests are  
Boschloo's test and Berger and Boos' modification of non-exact tests~\citep[see, e.g.,][]{mehrotra2003cautionary}.
There are several non-exact tests for superiority testing on binomial proportions.  
Commonly used non-exact tests include the (pooled or unpooled) asymptotic Z-test~(coinciding with the Wald or score test, respectively), Fisher's mid-p value test, and estimated p-value tests, which can yield higher power than unconditional exact tests but can also be~too liberal~(i.e., can lead to type I error rate inflation)~\citep{ripamonti_quatto_2017}.

While many tests exist for the two-sample superiority testing problem for binary proportions, the test that achieves the highest power depends entirely on the location within the parameter space (i.e., the true success rates) and the sample sizes; no test is guaranteed to be uniformly most powerful.
\cite{mehrotra2003cautionary} show that a poor choice of test statistic in this setting can actually lead to a test with lower unconditional power than Fisher's exact test.

To address this problem, we examine an integer programming-based approach to construct one-sided two-sample tests for a difference in binomial proportions.  Our proposed method maximizes average power while ensuring type I error rate control across the null hypothesis parameter space. 
The optimization problem can be viewed as a knapsack optimization problem, where items packed correspond to datasets for which the test yields a rejection, the maximized value of the knapsack is average power, and the weight of the knapsack equals the maximum type~I~error rate. This idea (maximizing power under type I error rate constraints)  aligns with the Neyman-Pearson paradigm~\citep{neymanpearson}. 

The integer programming approach to construct binomial tests was first proposed by~\cite{paroush1969integer}, though without evaluating the type~I~error rate or power. More recently, extensions of these tests and corresponding p-value construction methods were considered and evaluated in~\cite{keer2023hypothesis}. Lastly, instead of a testing approach, \cite{peer2024optimal} used integer linear programs to construct optimal exact confidence intervals.

This paper makes five contributions to the literature above:

(i) We introduce linear constraints that guarantee strict type I error control over the entire null parameter space, leading to knapsack-based tests with stronger validity guarantees than integer programming tests relying on average~\citet{paroush1969integer} or pointwise~\citet{keer2023hypothesis} error rate control, making our method directly comparable to other exact tests.
(ii) We leverage exact computations (e.g., an analytic expression for one-sided average power), eliminating the need for simulations or numerical integration.
(iii) Comprehensive evaluations across balanced and unbalanced designs show our test substantially outperforms Fisher's exact test (FET) and almost uniformly beats state-of-the-art Berger and Boos' exact tests. Computation of the knapsack test is computationally tractable~(less than two hours) for trial sizes up to roughly 300 participants—covering around 75\% of clinical trials currently using FET (\autoref{fig:overview_test_categorical_trials}). The appendix includes evaluations of alternative knapsack tests, non-exact tests (up to 10,000 participants), and tests for superiority with a margin.
(iv) We demonstrate real-world utility by computing a knapsack-based-$p$-value for a 280-participant clinical trial which was also considered in~\citet{mehrotra2003cautionary}, where multiple exact tests were compared.
(v) We provide an R Shiny application for practitioners to calculate these proposed exact tests (see \autoref{sect:acknowledgements}).

\autoref{tab:tests} provides an overview of the main differences between our proposed average power knapsack test and test approaches that have been proposed in the literature.

The current paper is organized as follows. \autoref{sect:model} describes the data model, hypotheses, and theoretical properties for the one-sided two-sample binomial testing problem. \autoref{sect:tests} describes the proposed average power knapsack-based test and comparator unconditional exact tests. 
\autoref{sect:numerical_results} provides a detailed type~I~error and power comparison of the considered tests. \autoref{sect:application} considers a case study based on the Merck Research Laboratories Trial, also considered in~\cite{mehrotra2003cautionary}. \autoref{sect:discussion} provides a discussion and conclusion. 

\FloatBarrier
\begin{table}
\caption{Differences between the proposed test and test approaches from the literature.}\label{tab:tests}
\centering
\renewcommand{\arraystretch}{1.5} 
\begin{tabular}{
>{\raggedright\arraybackslash}p{28mm} 
    >{\raggedright\arraybackslash}p{28mm}
    >{\raggedright\arraybackslash}p{28mm}
    >{\raggedright\arraybackslash}p{28mm}
    >{\raggedright\arraybackslash}p{28mm}
}
\hline
Approach (Source) & Example test(s) & Alternative & Type I error rate guarantee & Optimality Guarantee \\
\midrule
Proposed approach (current paper) & Average power knapsack test with exact type I error control & One-sided, e.g, \newline superiority with/without margin & Exact type I error rate control & Maximum average power \\
\citet{paroush1969integer} & Simple vs. simple knapsack test & Point and composite alternative & Pointwise and average type I error rate control & Maximum pointwise or average  power\\
\citet{keer2023hypothesis} & Average power knapsack test with pointwise type I error control & Two-sided & Pointwise type I error rate control (several parameters) &  Maximum average power; maximin and maximum average type I error rate\\
Conditional exact \citep{Fisher1934method}  & Fisher's exact test & One-sided, two-sided & Exact conditional type I error rate control & Uniformly most powerful unbiased conditional test\\
Unconditional exact \citep{mehrotra2003cautionary} & Barnard's test, \newline Berger and Boos test, \newline Boschloo's test & One-/two-sided/ \newline superiority with margin & Exact type I error rate control &  Maximum average  power for two-sided Barnard's test~\citep{ANDRES1994555}\\
Non-exact \citep{ripamonti_quatto_2017,Hernandez02112018} & Fisher's mid-p-value test, \newline  Wald test, score test, estimated p-value test  & One-sided/two-sided/ \newline superiority with margin & Not exact (only asymptotic or for finite set of parameters) & Asymptotic power optimality for likelihood-based tests\\
\hline
\end{tabular}
\end{table}

\section{One-sided two-sample binomial testing problem}\label{sect:model}
In this section, we state the one-sided two-sample binomial testing problem, which formalizes the problem of comparing binomial proportions mentioned in~\autoref{sect:introduction}. Let the unknown success probabilities be denoted by~$\btheta= (\theta_\C,\theta_\D)\in[0,1]^2$, where~$\C$ denotes the control treatment and~$\D$ the developmental~(experimental) treatment. The same convention~(i.e., first C then D) will be used throughout the current paper
to construct tuples from variables for the control and developmental treatment; furthermore, vectors will be indicated in bold, and every vector inequality is meant to be element-wise.
Let~$n_\C,n_\D\in\mathbb{N}$ be the fixed and known treatment group sizes, and~$n=n_\C+n_\D$ be the trial size. The random variables denoting the total sums of treatment successes~$\bS=(S_\C,S_\D)$, lying in sample space~$\mathcal{S}=\{\bs\in\mathbb{N}_0^2:\bs\leq\bn\}$, have a binomial distribution:
\begin{equation}
    \mathbb{P}_{\btheta}(\bS=\bs)=\prod_{a\in\{\C,\D\}}\binom{n_a}{s_a}\theta_a^{s_{a}}(1-\theta_a)^{n_a-s_a}.\label{eqn:datalikelihood}
\end{equation}
\newpage
The one-sided two-sample binomial testing problem involves testing
\begin{equation}H_0:\theta_\D\leq g_0(\theta_\C)\quad\text{vs.}\quad H_1:\theta_\D> g_1(\theta_\C),\label{eqn:hypotheses}\end{equation}
where~$g_h$ is an increasing differentiable function for each~$h$. In this paper, we mainly focus on the standard superiority test for $H_0:\theta_\D\leq \theta_\C\;\text{versus}\;H_1:\theta_\D> \theta_\C$, however, the setting above also covers other testing settings. For instance, superiority testing with a margin on the absolute difference in success rates corresponds \hbox{to~$g_0(\theta_{\C}) = g_1(\theta_{\C}) = \theta_{\C}+\delta_1$} for a margin~$\delta_1\in(0,1]$, and testing superiority with a margin on the relative risk, where~$H_1 = (1-\theta_\C)/(1-\theta_\D)>\delta_2$, corresponds to~$g_0(\theta_{\C})=g_1(\theta_{\C}) = 1-(1-\theta_{\C})/\delta_2$
for~$\delta_2\in(0,1]$. We have~$g_0\equiv g_1$ in the above two examples, but these functions can also be chosen differently such that the union of the parameter space under the null and alternative hypotheses does not equal~$[0,1]^2.$ 

Let
~$d:\mathcal{S}\mapsto\{0,1\}$ be a non-randomized~test decision function, indicating rejection of the null hypothesis~$H_0$ in~$\eqref{eqn:hypotheses}$ for every possible outcome vector~$(S_\C, S_\D)$.
Note there are~$2^{(n_\C+1)(n_\D+1)}$ possible
test decision functions.
We restrict to unconditional exact tests for~$H_0$, i.e., we consider~$d$ in a set~$\mathcal{D}_{\alpha}$ which, for a significance level~$\alpha\in(0,1)$ and the set~$$\Theta_0=\{\btheta\in[0,1]^2:\theta_\D\leq g_0(\theta_\C)\}$$ of parameters satisfying the null hypothesis~$H_0$, is defined as~$$\mathcal{D}_{\alpha} = \{d\in\{0,1\}^\mathcal{S}:\mathbb{P}_{\btheta}(d(\bS)=1)\leq \alpha,\quad \forall\,\btheta\in\Theta_0\}.$$  Letting~$\Theta_1=\{\btheta\in[0,1]^2:\theta_\D>g_1(\theta_\C)\}$ be the parameter range under~$H_1$, the one-sided two-sample binomial exact testing  problem aims to find
\begin{align}
d\in\mathcal{D}_{\alpha}\quad\text{s.t. }\quad\mathbb{P}_{\btheta}(d(\bS)=1)&\geq \mathbb{P}_{\btheta}(d'(\bS)=1),\quad \forall\,\btheta\in\Theta_1,\;d'\in\mathcal{D}_{\alpha}.\label{testprob}\end{align}
Problem~\eqref{testprob} has no solution if the power~$ \mathbb{P}_{\btheta}(d(\bS)=1)$ is maximized at different test decision functions~$d\in\mathcal{D}_{\alpha}$ for different parameters~$\btheta\in\Theta_1$, which is a realistic situation in practice. In this case, the remaining option is to compare several exact tests numerically~\citep[see also, e.g.,][]{lehmann1986testing}

A test is said to satisfy Barnard's convexity condition~(C)~\citep{barnard2x2tab} if, whenever the test rejects for a dataset~$\bs$, the test also rejects for more extreme datasets, i.e., datasets with fewer control successes or more developmental treatment successes~(resulting in a convex rejection region).
\hbox{Let~$\partial\bs_\C = (1,0)$}, $\partial\bs_\D=(0,1)$ be the difference in~$\bs$ when there is one success more for treatment group~$\C,\D$, respectively. Barnard's convexity condition requires:
\begin{condition} \label{condC}$$d(\bs -\partial \bs_\C)\geq d(\bs)\text{ and } d(\bs +\partial \bs_\D)\geq d(\bs)\quad \forall \bs\in\mathcal{S}. 
$$ \end{condition} 
Let~$r_d(\btheta)=\mathbb{P}_{\btheta}(d(\bS)=1)$ be the rejection rate function
and~\begin{align}\partial\Theta_h=\{\btheta\in[0,1]^2:\theta_\D=g_h(\theta_\C)\}\label{defn_boundary}\end{align} be the boundary of the parameter range~$\Theta_h$ under hypothesis~$H_h$ for~$h\in\{0,1\}$.
The following result now follows from the proof of Theorem~1 in~\citet{Rohmel1999unconditional}:
\begin{theorem} \label{thm:monotonicity_power}
    If~$d$ satisfies \autoref{condC} then under~$H_0$~$r_d$  is maximized at~$\partial\Theta_0$ and under~$H_1$~$r_d$ is minimized at~$\partial\Theta_1.$
\end{theorem}

To construct exact integer programming tests, we need the next theorem, which gives an upper bound on the type I error rate across~$\partial\Theta_0$ based on evaluation of parameters in a discretization of this set.
Let~$\bs_1,\bs_2,\dots,\bs_{|\mathcal{S}|}$ be a sequence of   total successes vectors covering~$\mathcal{S}$~(i.e.,~$\cup_{i=1}\{\bs_i\}=\mathcal{S}$) and~$\phi$ map the vector  to the respective index~(i.e.,~$\phi(\bs_i)=i$ for all~$\bs_i$). 
For~$K\in\mathbb{N}$ consider the sequence~$\theta_{1}< \theta_{2}<\dots< \theta_{K}$ of increasing success rates with~$\theta_{1}=0$ and~$\theta_{K}=g_0^{-1}(1)$. The set~$\hat{\partial\Theta_0}=\cup_{j=1}^K\{(\theta_j,g_0(\theta_j))\}$ is the aforementioned discretization of~$\partial\Theta_0.$
 For each~$i\in\mathcal{I}=\{1,\dots,|\mathcal{S}|\}$ and~$j\in\mathcal{J}_0=\{1,\dots,K\}$ let the probability~$p_{i,j,0}=\mathbb{P}_{(\theta_j,g_0(\theta_j))}(\bS=\bs_i)$ be determined by~\eqref{eqn:datalikelihood}  and define~$\bp_{j,0}=[p_{1,j,0},\dots,p_{|\mathcal{S}|,j,0}]$.
Let~$r_{d}(\theta)$ be the type I error rate under~$(\theta_\C,\theta_\D)=(\theta, g_0(\theta))$ for test decision function~$d$, and note that~$r_{d}(\theta_{j})=\bp_{j,0}^\top\bd.$
For each~$i$ let~$\be_i$ be the~$i$-th standard unit vector in~$\mathbb{R}^{|\mathcal{S}|}$. 
Similar as in~\cite{suissashuster1985}, the following result follows from a Lipschitz bound and the mean value theorem~(see~\autoref{sect:proof_thm2} for the proof).
  \begin{theorem}\label{thm:discretization}
    If~\autoref{condC} holds, then we have~$$r_{d}(\theta)\leq \bp_{j,0}^\top\bd + \max(0, ( {\bm m}_{j,\D}^\top\bA_\D- {\bm m}_{j,\C}^\top\bA_\C)\bd)\quad\quad\forall j\in\mathcal{J}_0,\theta\in(\theta_{j},\theta_{j+1}), $$ where for all~$i$ the~$i$-th rows of $\bA_{\C}$ and~$\bA_{\D}$ equal~$\be_i - \be_{\phi(\bs_i+\partial\bs_\C)
}$ and~$\be_i - \be_{\phi(\bs_i-\partial\bs_\D)
}$, and:
\begin{align*}m_{i,j,\D}&= n_\D\cdot(\theta_{j+1}-\theta_j)\cdot\binom{n_\C}{s_{\C,i}}\binom{n_\D-1}{s_{\D,i}-1} \bar{h}(\bs_{i},\theta_j,\theta_{j+1}),\\
m_{i,j,\C}&=n_\C\cdot (\theta_{j+1}-\theta_j)\cdot\binom{n_\C-1}{s_{\C,i}}\binom{n_\D}{s_{\D,i}}  \ubar{h}(\bs_{i},\theta_j,\theta_{j+1}),\\
\bar{h}(\bs_{i},\theta_j,\theta_{j+1}) &= \max_{\theta\in[\theta_j,\theta_{j+1}]}g'_0(\theta)\theta^{s_\C}g_0(\theta)^{s_\D-1}(1-\theta)^{n_\C-s_\C} (1-g_0(\theta))^{n_\D-s_\D},\\
\ubar{h}(\bs_{i},\theta_j,\theta_{j+1})&= \min_{\theta\in[\theta_j,\theta_{j+1}]}\theta^{s_\C}g_0(\theta)^{s_\D}(1-\theta)^{n_\C-s_\C-1} (1-g_0(\theta))^{n_\D-s_\D},
\end{align*}
where we let $\binom{n}{k} = 0$ if $k < 0$ or $k > n$.
\end{theorem}

\section{Proposed exact test approach and comparators}\label{sect:tests}

This section provides our proposed average power knapsack test~(\autoref{sect:APK}), outlines our methodology for deriving valid, knapsack-based $p$-values~(\autoref{section:p_value_knapsack}) and details the non-knapsack benchmark exact tests used for our comparative analysis~(\autoref{sect:non_knapsack_based}). 
\subsection{Proposed approach: The average power knapsack test}\label{sect:APK}
This subsection describes the \emph{integer linear program}~(ILP) underlying the \emph{average power knapsack}~(APK) test.
The APK test maximizes average power under~$H_1$ in~\eqref{eqn:hypotheses} while ensuring type~I~error control under~$H_0$ in~\eqref{eqn:hypotheses}. 
Assume~$g_0\equiv g_1$ is the identity, in which case the APK test tests $H_0:\theta_\D\leq \theta_\C\;\text{versus}\;H_1:\theta_\D> \theta_\C$;  the extension to other choices of~$g_1$ follows in a straightforward manner~(see, e.g., \autoref{sect:superiority_test}).
For a decision vector~$\bd$ coupled to the sequence~$\bs_1,\dots,\bs_{|\mathcal{S}|}$, the average power is given by:
\begin{align}
2\int_{[0,1]^2}\sum_{i\,:\,d_i=1}\mathbb{P}_{\btheta}(\bS=\bs_i)\mathbb{I}(\theta_\D\geq\theta_\C) d\btheta = \bd^\top \bar{\bp}_1,\label{expression_AP}
\end{align}
where the~$i$-th element $\bar{p}_{i,1}$ of~$\bar{\bp}_1$ equals
\begin{align}
   \bar{p}_{i,1} &=2\int_{[0,1]^2}\prod_{a\in\{\C,\D\}}\binom{n_a}{s_{a,i}}\theta_a^{s_{a,i}}(1-\theta)^{(n_a-s_{a,i})}\mathbb{I}(\theta_\D\geq\theta_\C)d\btheta\nonumber\\&=
2\left[\prod_{a\in\{\C,\D\}}\binom{n_a}{s_{a,i}}\right]B(\alpha_{\D}(\bs_i), \beta_{\D}(\bs_i))\cdot\sum_{j=0}^{\alpha_\D(\bs_i)-1}
\frac{{B}(\alpha_\C(\bs_i)+j, \beta_\C(\bs_i)+\beta_\D(\bs_i))}{(\beta_\D(\bs_i)+j)B(j+1,\beta_\D(\bs_i))}\label{avg_power_coefs},
\end{align}
 the last equality follows from~\cite{millerblog} where~$\alpha_a(\bs) = s_a+1$ and~$\beta_a(\bs) = n_a-s_a+1$ for all~$\bs\in\mathcal{S}$ and $B$ denotes the beta function. 
 We multiply by~$2$ in~\eqref{expression_AP} as the area of the triangle~$\{\btheta\in[0,1]^2:\theta_\D\geq\theta_\C\}$ is~$1/2$.

With~$(\bp_{j,0})_j$ and~$\bar{\bp}_1$ in hand, we can now define the \emph{average power knapsack}~(APK) test  by the following ILP, denoted the~\emph{APK program}:
\begin{lpformulation}[\text{(APK program)}]\lplabel{knapsack_test_ILP}
\lpobj[obj]{max}{\bar{\bp}_1^\top\bd}
\lpeq[type_1_error_constr]{\bp^\top_{j,0}\bd\leq \alpha,}{j\in\mathcal{J}_0,}
\lpeq[type_1_error_constr2]{(\bp^\top_{j,0}+{\bm m}_{j,\D}^\top\bA_\D- {\bm m}_{j,\C}^\top\bA_\C)\bd\leq \alpha,}{j\in\mathcal{J}_0,}
\lpeq[mon_cons_1Dsuccmore]{d_{\bs + \partial{\bm f}}\geq d_{\bs}, }{\bs\in\mathcal{S},\partial {\bm f} \in\{-\partial \bs_\C, \partial\bs_\D\},}
\lpeq[binary_vars]{ d_{\bs}\in\{0,1\},}{\bs\in\mathcal{S}.}
\end{lpformulation}

The objective~\eqref{obj} maximizes the average power over parameters under the alternative hypothesis.  
Constraint~\eqref{type_1_error_constr} and \eqref{type_1_error_constr2} bound the  type~I error rate for all parameters in~$\Theta_0$ by the significance level~$\alpha$ through~\autoref{thm:monotonicity_power} and~\autoref{thm:discretization}. Constraint~\eqref{mon_cons_1Dsuccmore} enforces~\autoref{condC}. Constraint~\eqref{binary_vars} enforces a binary domain constraint on the decision function.

The APK test maximizes average power for the one-sided setting. This measure has been justified in the literature as ``long term power'' \citep{ANDRES1994555}, assuming any parameter vector is equally likely a priori. 
In practice, while the minimum clinically relevant effect size is often fixed by domain consensus, the baseline control rate is typically uncertain prior to the trial. Because the precise location of the target alternative depends on this control rate, optimizing for a single point alternative carries a high risk of power loss under misspecification. For scenarios with moderate or high certainty regarding the baseline rate or alternative parameters, \autoref{knapsack_comparators} details a weighted average power knapsack~(WAPK) test and simple hypothesis knapsack test, and provides guidance on when such alternatives are reasonable. The WAPK is  in line with \citet{rice1988new}, who additionally considered a weighted average power measure, and the (weighted) APK test essentially maximizes the assurance~\citep{ohagan2005assurance}, providing a reliable alternative to point-fixed power by incorporating prior uncertainty.
 Lastly, as the (weighted) APK test maximizes average power, it cannot uniformly be improved in terms of power, i.e., is admissible, within the set of unconditionally exact tests satisfying constraints \eqref{type_1_error_constr2} and \eqref{mon_cons_1Dsuccmore}, with the effect of~\eqref{type_1_error_constr2} vanishing as the mesh of $\hat{\partial\Theta_0}$ goes to zero.

\begin{remark}[Discretization of null boundary]\label{rem:disc_choice}
While the APK test ensures type I error rate control across~$\Theta_0$ no matter the choice of $\hat{\partial\Theta_0}$, the decisions for the APK test depend on the discretization~$\hat{\partial\Theta_0}$ of~${\partial\Theta_0}$. In particular, choosing~$\hat{\partial\Theta_0}$  too coarse will lead to a power loss over the APK test with~$K\approx \infty$. To this end, we recommend setting the initial mesh of~$\hat{\partial\Theta_0}$ to at most 0.01 (corresponding to $K \ge 101$ grid points).

Assuming that refined discretizations of~${\partial\Theta_0}$ are nested,  the optimal value of the APK program with constraint~\eqref{type_1_error_constr2} removed provides an upper bound on the maximum average power for~$K\approx\infty$; hence, the difference in the optimal values of the two APK programs provides a bound on the loss in average power. When the resulting power loss bound is too large, we recommend refining the discretization around parameters where~\eqref{type_1_error_constr2} is active.
\end{remark}

\subsection{Knapsack-based p-values}\label{section:p_value_knapsack}

As stated in Section 3.3 of ~\citet{lehmann1986testing}, for a test where rejection regions are nested for smaller significance levels, the p-value is the minimum significance level at which the null hypothesis is rejected. 
We propose taking this minimum over a finite set~$\mathcal{P}\subset [0,1]$ such that~$ \{\alpha, 1\}\subseteq\mathcal{P}$. Similar to the construction of repeated p-values~\citep[Chapter 9.4]{jennison1999group}, this requires solving the APK program for consecutive values in the set of significance levels where, starting from level~$\alpha$, we go down until we reach a non-rejection, or up until we reach a rejection. 
As in Section~4.3.2.2 of~\cite{keer2023hypothesis},  we enforce monotonicity of the rejection region over different significance levels for subsequently solved APK programs, to  ensure that datasets stay in the
 rejection region~(see \autoref{alg:exact_p}).
Letting~$p_{\bs}$ be the resulting p-value for outcome vector~$\bs$, we have~$\mathbb{P}_{\btheta_0}(p_{\bS}\leq \tilde{\alpha})\leq \tilde{\alpha}$ for all~$\tilde{\alpha}\in(0,1)$,~$\theta_0\in\Theta_0$. The test that rejects when~$p_{\bs}\leq\alpha$ equals the original APK test and hence has the same optimality guarantees, while tests rejecting when $p_{\bs}\leq\alpha'$ for~$\alpha'\neq\alpha$ may have lower average power than the APK test based on~$\alpha'$.

\begin{algorithm}[h!]

\setstretch{1.35}
		\caption{Exact p-value calculation via knapsack program.}\label{alg:exact_p}
		\begin{algorithmic}[1]
			\Inputs{Observed successes vector~$\bs$, significance level~$\alpha$, set of p-values~$\mathcal{P}$;}
			
			\State  Solve the integer linear program for significance level~$\alpha$, yielding a decision~$d_{\bs}^\alpha$ for~$\bs$;
            \State Set~$\alpha'=\alpha$ and $d'_{\bs}=d^\alpha_{\bs}$;
            \If{$d'_{\bs}=0$}
\While{$d'_{\bs}=0$}
              \State Set~$\alpha' = \min\{\alpha''\in\mathcal{P}:\alpha''>\alpha'\}$;
              \State  Solve the APK program for significance level~$\alpha'$, restricting~$\bd'\leq\bd^{\alpha'}$;
              \State Set $\bd'=\bd^{\alpha'}$;
            \EndWhile
            \Else
            \While{$d'_{\bs}=1$}
              \State Set~$\alpha' = \max\{\alpha''\in\mathcal{P}:\alpha''<\alpha'\}$;
              \State  Solve the APK program for significance level~$\alpha'$, restricting~$\bd'\geq\bd^{\alpha'}$;
              \State Set $\bd'=\bd^{\alpha'}$;
            \EndWhile
            \State Set~$\alpha' = \min\{\alpha''\in\mathcal{P}:\alpha''>\alpha'\}$;
            \EndIf\\
			{\bf Outputs: }\\
			\;\;\;\;Exact p-value~$\alpha'$.
		\end{algorithmic}
	\end{algorithm}

\subsection{Comparator unconditional exact tests}\label{sect:non_knapsack_based}
In this section, we describe the unconditional exact tests that we compare to our proposed APK test (defined in~\autoref{sect:tests}).
One of the comparators is a one-sided version of \emph{Fisher's exact}~(FE) test, as it remains a widely used~(conditional) exact test for binary data collected in a small two-arm experiment~(as shown in \autoref{fig:overview_test_categorical_trials}).
Hence, comparing the power of our knapsack-based test with that of the ~\emph{FE test}~(FET) provides an indication of the practical benefit of using the optimization framework.  
    The one-sided FET rejects, i.e.,~$d(\bs)=1$, when~$\T_{\text{FE}}(\bs)\leq \alpha$ where, letting~$s=s_\C+s_\D$
    \begin{equation}\T_{\text{FE}}(\bs)=\sum_{\bs':s'_{\D}\geq s_{\D}}\binom{n_\C}{s-s'_\D}\binom{n_\D}{s'_\D}/\binom{n}{s}.\label{Pval_FET}\end{equation}
    The summed terms in the above equation correspond to the (hypergeometric) distribution of~$S_\D\mid (S_\C+S_\D)$, hence~$\T_{\text{FE}}(\bs)$ is the probability that~$S_\D\geq s_\D$ given~$S_\C+S_\D=s$.  Importantly, because FET strictly controls the type I error rate for every discrete success total $s$, it inherently functions as an unconditional exact test when integrating these conditional constraints over the full sample space of total successes. 

In addition to FET, our comparative analysis incorporates advanced, state-of-the-art benchmarks. Specifically,  we include one-sided variants of Berger and Boos'~\citep{bergerboos_paper} version of Fisher's mid-p value test~(denoted~FMP$^*$ test) and Berger and Boos' version of the Z-pooled test~(denoted~$\text{Z}_{\text{P}}^*$ test).  Berger and Boos' procedure and the Z-pooled test statistic are as they appear in~\cite{mehrotra2003cautionary}, while Fisher's mid-p value is found by subtracting~$1/2\cdot \binom{n_\C}{s-s_\D}\binom{n_\D}{s_\D}/\binom{n}{s} $ from~\eqref{Pval_FET}. The Berger and Boos tests use a confidence interval of level~$1-\gamma$ for the common success rate; see, e.g.,~\cite{bergerboos_paper}.

As the p-value as a function of the common success rate is a non-convex~(but smooth and Lipschitz continuous) function, finding the unconditional p-value for the FMP$^*$ and~$\text{Z}_{\text{P}}^*$ tests is not straightforward~(e.g., Quasi-Newton approaches are not guaranteed to find a global optimum) and we propose to use the simplicial homology global optimisation~(SHGO) algorithm as implemented in the \verb|SciPy| \verb|Python| package~\cite[see the][]{shgoref}, we note that this algorithm is also used to compute Boschloo's test in \verb|Python|.  

We have also evaluated the results for the unconditional exact FMP test, as well as Boschloo's test and the unconditional exact Z-pooled test from~\cite{mehrotra2003cautionary}, and a {\it maximin power knapsack test} (MPK test, \autoref{knapsack_comparators}). Results are given in~\autoref{subsect:additional_results_general}.  As the unconditional exact FMP test uniformly outperforms Boschloo's test, while the Berger and Boos' modification greatly balances the performance of the tests for unbalanced treatment group sizes (power gains of up to 30\% without substantial power losses in comparison to the unconditional exact
$\text{Z}_\text{P}$ test), and the MPK test provides a very unstable power performance (sometimes performing worse than FET), we have chosen to only report results for the FMP$^*$ and~$\text{Z}^*_\text{P}$ tests here.

\begin{figure}[h]
    \centering
    \includegraphics[width=.85\linewidth]{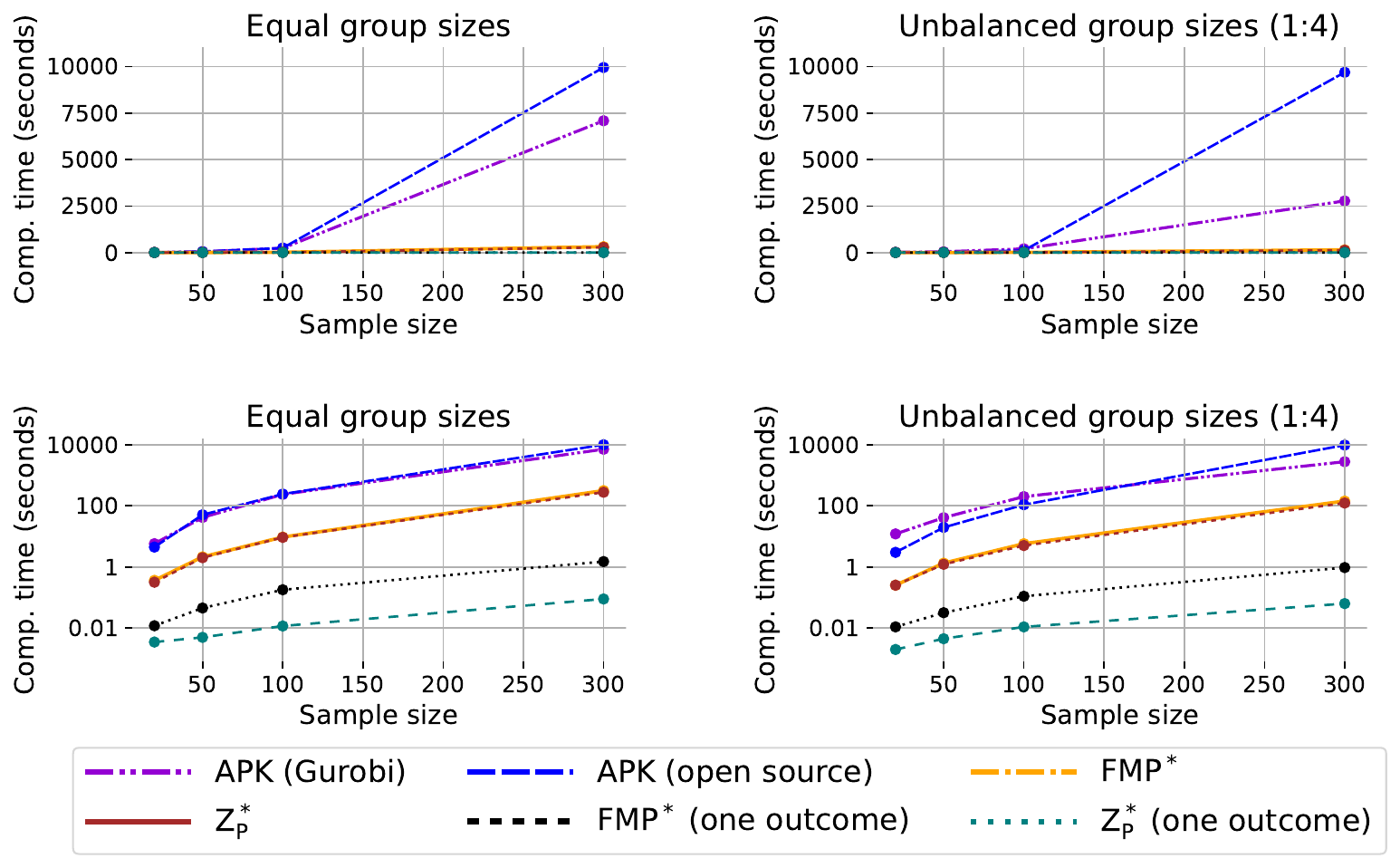}
    \caption{Computation times for the APK test (Gurobi), APK test calculated with the open-source SciPy (HiGHS) ILP solver, Berger and Boos' version of Fisher's mid-p-value test (FMP$^*$), and Berger and Boos' version of the pooled Wald (Z$^*_{\text{P}}$) test for balanced treatment group sizes (left) and a 1:4 allocation ratio (right, more participants allocated to treatment). Computation times are both shown on a linear (first row) and log scale (second row) to indicate absolute and order differences. The Berger and Boos p-values are calculated both for all possible outcome vectors~$\bs\in\mathcal{S}$ and for one possible outcome vector~$\bs_{\bn}=[\lfloor(1/2-\Delta_{\bn}/2)n_{\C}\rfloor, \lceil(1/2+\Delta_{\bn}/2)n_{\D}\rceil]$ with $\Delta_{\bn} = \frac{1}{2}(z_{97.5\%} + z_{80\%})\sqrt{1/n_{\D} + 1/n_{\C}}.$} 
    \label{fig:comptime}
\end{figure}

\begin{remark}[Computation time comparison]

The APK test is solved via branch-and-bound which, for $n_{\text{var}}=(n_{\text{C}}+1)(n_{\text{D}}+1)$ variables and $\Theta(n_{\text{var}})$ constraints,
performs
at most~$2^{n_{\text{var}}}$ branches, each with $O(2^{n_{\text{var}}})$ worst-case complexity~\citep{klee1972good}, yielding a $O(2^{2n_{\text{var}}})$ worst-case computation time.
In contrast, computing a Berger and Boos (BB) p-value requires $O(n_{\text{var}})$ operations for a single dataset $\mathbf{s}$, and $O(n_{\text{var}}^2)$ to evaluate all states $\mathbf{s}\in\mathcal{S}$. Consequently, the APK test is substantially more computationally demanding. Empirically (see \autoref{fig:comptime}), using Gurobi, APK tests take 15–48 times longer than computing all BB p-values over $\mathcal{S}$, and up to 78,700 times longer than computing a single BB p-value. Finally, at a horizon of 300 participants, the open-source \verb|SciPy| solver takes approximately 1–2 hours longer than Gurobi.
Note that for a vector of treatment group sizes and a significance level~$\alpha$ (often 2.5\% or 5\%), the APK test only needs to be computed once, and can be reused for every trial of that size at no additional computational cost.   
\end{remark}

\section{Comparison of tests}\label{sect:numerical_results}
\FloatBarrier

In this section, we compare the~FE, FMP$^*$, Z$_\text{P}^*$,and APK tests for a one-sided significance level~$\alpha$ of~2.5\%. We set the confidence interval parameter $\gamma$ for the Berger and Boos tests FMP$^*$ and Z$_\text{P}^*$ to~$0.0005$~(half the value used in~\cite{mehrotra2003cautionary}, as we are doing one-sided testing). 
We consider trial sizes in ~$n\in\{20,50,100,300\}$. We set~$\theta_1 = 0.000, \theta_2 = 0.001,\dots,\theta_{K}=1.000$ for the APK test. Following~\autoref{rem:disc_choice}, this discretization of~$\partial\Theta_0$ led to a maximum power loss of $9.4\cdot10^{-5}$ with 11 bounds equal to zero.

 For the APK test, we set the relative error tolerance for the optimum to~$2.5\cdot 10^{-4}$, leading to a maximum absolute error on the optimal power of~$2.5\cdot10^{-4}$ as it is bounded by one. 
 All computations were done in \verb|Python| (version 3.12.11), and the Gurobi Optimizer (version 13.0.2) was used to solve the APK programs. The SHGO algorithm~(\verb|Python| package \verb|SciPy|) was used to compute Berger and Boos tests, as well as Boschloo's test.  All results are based on exact binomial computations.

\subsection{Type~I~error rates}
\begin{figure*}[h!]
    \centering
    \includegraphics[width=\linewidth]{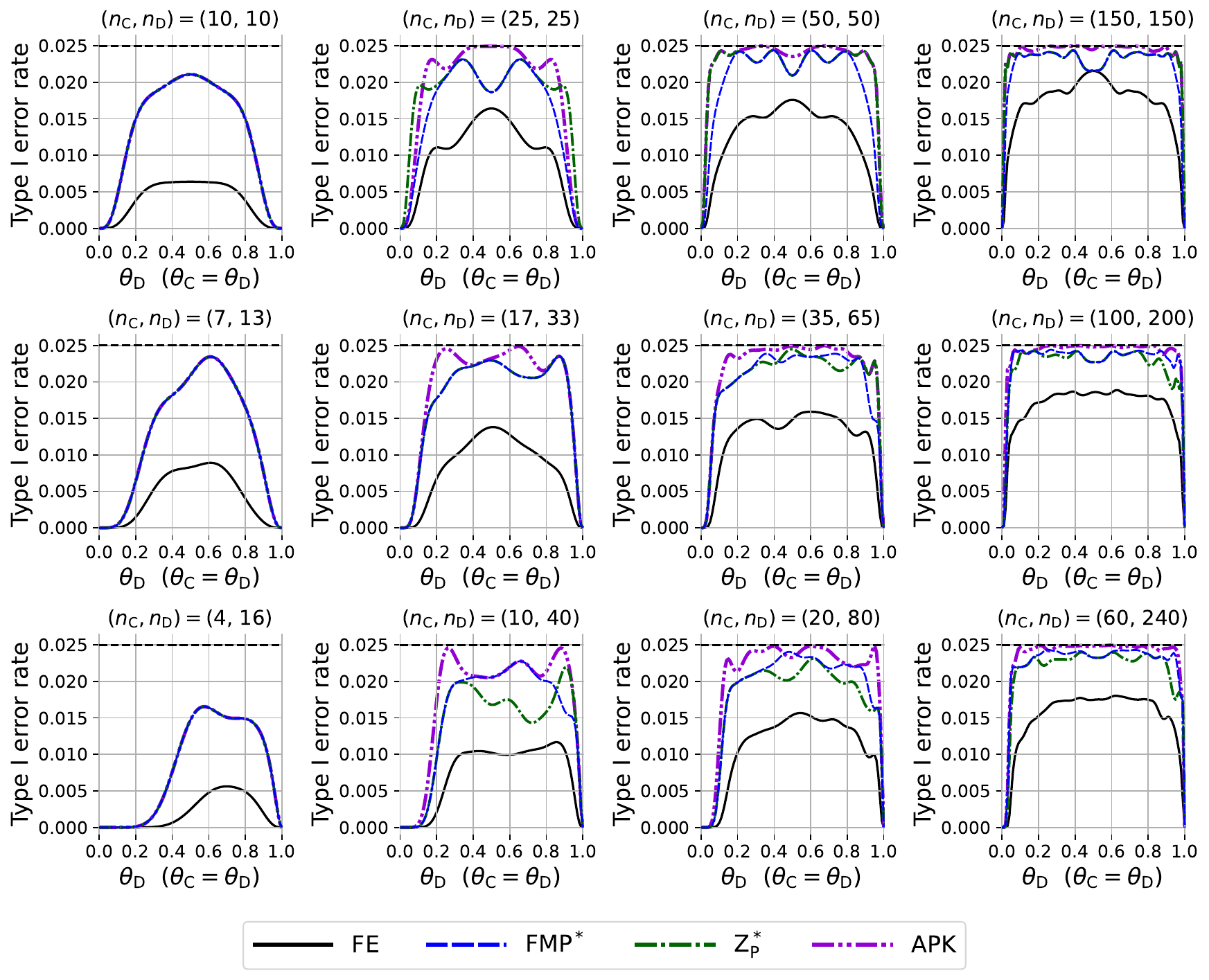}
    \caption{Exact type~I~error rate profiles for different treatment group size configurations under Fisher's exact~(FE) test, 
    Berger and Boos' version of Fisher's mid-p value~(FMP$^*$) test, Berger and Boos' version of the Z-pooled (Z$^*_{\text{P}}$) test, and average power knapsack~(APK) tests. The trial sizes (varying over columns) were chosen as~$20,50,100,300$, and treatment group sizes are indicated in the titles of the subfigures. The one-sided significance level~$\alpha$ is set to~$0.025$ and indicated by a dashed line.}
    \label{fig:T1E}
\end{figure*}

\autoref{fig:T1E} shows the exact type~I~error rate profiles for the considered tests
under different~(small, moderate, and large)  group sizes --chosen in agreement with~\cite{mehrotra2003cautionary}, representing balanced, moderately unbalanced, and severely unbalanced situations--. In the code, it is checked that every considered test satisfies~\autoref{condC}~\cite[where this condition is guaranteed to hold for Fisher's exact test and the knapsack-based tests only; see, e.g.,][]{Rohmel1999unconditional}. 
We reiterate that all considered tests guarantee a type I error bounded by the nominal significance level for all parameters in~$\Theta_0$.

The maximum type~I~error rate for the APK test often lies close to the significance level of~2.5\%, whereas FET often shows the lowest type~I~error rate. The type~I~error rates for the tests~FMP$^*$ and~Z$^*_\text{P}$ are never exactly equal to the significance level~$\alpha$, which is due to the fraction~$\gamma$~(equal to~$0.0005$ in this case) of the significance level~$\alpha$ spent on constructing the confidence interval for the common success rate. 
When the imbalance in treatment group sizes is larger, the type~I~error rate profiles become more asymmetrical around~$0.5$. 
The high degree of conservativeness for the case~$n_\C=4, n_\D=16$ is due to the level of discreteness of the data.

\subsection{Power comparison}
\autoref{tab:power_table} shows the exact power values
for the considered tests under different success rates and treatment group sizes. The parameters are chosen such that the APK test has at least~$80\%$ power. In \autoref{tab:FET80}, we also show power values for the case FET has at least~$80\%$ power, where a similar conclusion can be drawn. The bold power values indicate the highest value in each row, while the underlined power values indicate the lowest row power. 
Power values less than~$75\%$ are italicized.

For~$\bn\in\{(10,10),(16,4),(4,16)\}$, each test except for FET has the same power for all considered parameter configurations, and FET shows lower power. 
The table shows that the APK test often shows highest power; in fact, for 
 all but two considered configurations, the APK test performs best, and the power gains over FET are often substantial, around 2.83\% to 35.11\%. While in the majority of cases the power gains from using the APK test are marginal over non-FE tests~(about 1.2\% on average), the power gains are very consistent, and power gains around 4\%, 5\%, and 7\% over the FMP$^*$ and~Z$^*_\text{P}$ tests are also observed.
These power differences impact required trial sizes and inferior treatment allocations. For example, with $\bn = (10,40)$ and $\btheta = (0.01, 0.35)$, 
reaching 80\% power requires adding 3 or 13 treatment participants for the FMP$^*$ and Z$^*_P$ tests, respectively, or 1 control participant~(corresponding to  $0.99$ more  expected treatment failures). In contrast, this target can be reached for FET by adding 4 control participants or 3 control and 17 treatment participants. Therefore, the APK test can significantly reduce recruitment costs and improve participant benefit over both FET and the state-of-the-art, requiring only additional computation time.

\autoref{tab:comparison_table} summarizes exact pointwise and average power comparisons between the considered tests for different treatment group sizes and the grid of points~$(\theta_\C,\theta_\D)$ such that~$\theta_\D>\theta_\C$ and both~$\theta_\C$ and~$\theta_\D$ are multiples of~$0.01.$
The rows corresponding to the pointwise power comparison show a tuple, where the left entry equals the fraction of parameters in which the column test outperformed the APK test, and vice versa for the right entry. 
The rows corresponding to the average power comparison show the average power for the APK test minus the average power for the column test, and in brackets the average power gain in case the column test does better (left) or in case the APK test does better (right).
\renewcommand{\arraystretch}{0.7} 
\setlength{\tabcolsep}{1.75mm}
\begin{table}[h!]

    \centering
    \caption{Power values~(\%) across treatment group sizes and parameter values (significance level~$2.5\%$). Row maxima (minima) are made bold (underlined). Values below 75\% are italicized.
}
\small
    \begin{tabular}{llllrrrrr}
\toprule
 &  &  &  &   & FE & FMP$^*$ & Z$_{\text{P}}^*$ & APK \\
$n_{\text{C}}$ & $n_{\text{D}}$ & $\theta_{\text{C}}$ & $\theta_{\text{D}}$ &  &  &  &  &  \\
\midrule
\multirow[t]{4}{*}{10} & \multirow[t]{4}{*}{10} & 0.01 & 0.51 &  & ${\it \uline{\textcolor[rgb]{0.734,0.500,0.315}{60.30}}}$ & ${\bf \textcolor[rgb]{0.345,0.428,0.775}{80.08}}$ & ${\bf \textcolor[rgb]{0.345,0.428,0.775}{80.08}}$ & ${\bf \textcolor[rgb]{0.345,0.428,0.775}{80.08}}$ \\

 &  & 0.05 & 0.61 &  & ${\it \uline{\textcolor[rgb]{0.638,0.482,0.429}{65.17}}}$ & ${\bf \textcolor[rgb]{0.328,0.425,0.796}{80.99}}$ & ${\bf \textcolor[rgb]{0.328,0.425,0.796}{80.99}}$ & ${\bf \textcolor[rgb]{0.328,0.425,0.796}{80.99}}$ \\

 &  & 0.20 & 0.80 &  & ${\it \uline{\textcolor[rgb]{0.670,0.488,0.391}{63.53}}}$ & ${\bf \textcolor[rgb]{0.337,0.427,0.785}{80.54}}$ & ${\bf \textcolor[rgb]{0.337,0.427,0.785}{80.54}}$ & ${\bf \textcolor[rgb]{0.337,0.427,0.785}{80.54}}$ \\

 &  & 0.49 & 0.99 &  & ${\it \uline{\textcolor[rgb]{0.734,0.500,0.315}{60.30}}}$ & ${\bf \textcolor[rgb]{0.345,0.428,0.775}{80.08}}$ & ${\bf \textcolor[rgb]{0.345,0.428,0.775}{80.08}}$ & ${\bf \textcolor[rgb]{0.345,0.428,0.775}{80.08}}$ \\
\cline{1-9} \cline{2-9}\noalign{\smallskip} 
\multirow[t]{4}{*}{25} & \multirow[t]{4}{*}{25} & 0.01 & 0.27 &  & ${\it \uline{\textcolor[rgb]{0.629,0.481,0.439}{65.72}}}$ & ${\textcolor[rgb]{0.407,0.440,0.702}{77.03}}$ & ${\bf \textcolor[rgb]{0.267,0.414,0.869}{84.08}}$ & ${\textcolor[rgb]{0.340,0.427,0.782}{80.44}}$ \\

 &  & 0.20 & 0.58 &  & ${\it \uline{\textcolor[rgb]{0.465,0.451,0.633}{74.01}}}$ & ${\textcolor[rgb]{0.354,0.430,0.765}{79.61}}$ & ${\textcolor[rgb]{0.354,0.430,0.765}{79.61}}$ & ${\bf \textcolor[rgb]{0.334,0.426,0.789}{80.71}}$ \\

 &  & 0.40 & 0.79 &  & ${\uline{\textcolor[rgb]{0.419,0.442,0.689}{76.35}}}$ & ${\textcolor[rgb]{0.331,0.426,0.792}{80.83}}$ & ${\textcolor[rgb]{0.331,0.426,0.792}{80.83}}$ & ${\bf \textcolor[rgb]{0.305,0.421,0.824}{82.21}}$ \\

 &  & 0.73 & 0.99 &  & ${\it \uline{\textcolor[rgb]{0.629,0.481,0.439}{65.72}}}$ & ${\textcolor[rgb]{0.407,0.440,0.702}{77.03}}$ & ${\bf \textcolor[rgb]{0.267,0.414,0.869}{84.08}}$ & ${\textcolor[rgb]{0.340,0.427,0.782}{80.44}}$ \\
\cline{1-9} \cline{2-9}\noalign{\smallskip} 
\multirow[t]{4}{*}{50} & \multirow[t]{4}{*}{50} & 0.01 & 0.15 &  & ${\it \uline{\textcolor[rgb]{0.670,0.488,0.391}{63.67}}}$ & ${\textcolor[rgb]{0.424,0.443,0.682}{76.01}}$ & ${\bf \textcolor[rgb]{0.325,0.425,0.799}{81.13}}$ & ${\bf \textcolor[rgb]{0.325,0.425,0.799}{81.13}}$ \\

 &  & 0.30 & 0.58 &  & ${\uline{\textcolor[rgb]{0.407,0.440,0.702}{77.03}}}$ & ${\textcolor[rgb]{0.337,0.427,0.785}{80.59}}$ & ${\textcolor[rgb]{0.337,0.427,0.785}{80.59}}$ & ${\bf \textcolor[rgb]{0.328,0.425,0.796}{80.98}}$ \\

 &  & 0.60 & 0.85 &  & ${\uline{\textcolor[rgb]{0.430,0.444,0.675}{75.77}}}$ & ${\textcolor[rgb]{0.331,0.426,0.792}{80.85}}$ & ${\textcolor[rgb]{0.331,0.426,0.792}{80.85}}$ & ${\bf \textcolor[rgb]{0.316,0.423,0.810}{81.51}}$ \\

 &  & 0.85 & 0.99 &  & ${\it \uline{\textcolor[rgb]{0.670,0.488,0.391}{63.67}}}$ & ${\textcolor[rgb]{0.424,0.443,0.682}{76.01}}$ & ${\bf \textcolor[rgb]{0.325,0.425,0.799}{81.13}}$ & ${\bf \textcolor[rgb]{0.325,0.425,0.799}{81.13}}$ \\
\cline{1-9} \cline{2-9}\noalign{\smallskip} 
\multirow[t]{4}{*}{150} & \multirow[t]{4}{*}{150} & 0.01 & 0.07 &  & ${\it \uline{\textcolor[rgb]{0.538,0.464,0.547}{70.29}}}$ & ${\textcolor[rgb]{0.375,0.434,0.740}{78.59}}$ & ${\bf \textcolor[rgb]{0.337,0.427,0.785}{80.47}}$ & ${\bf \textcolor[rgb]{0.337,0.427,0.785}{80.47}}$ \\

 &  & 0.30 & 0.46 &  & ${\uline{\textcolor[rgb]{0.375,0.434,0.740}{78.55}}}$ & ${\textcolor[rgb]{0.316,0.423,0.810}{81.55}}$ & ${\textcolor[rgb]{0.316,0.423,0.810}{81.55}}$ & ${\bf \textcolor[rgb]{0.310,0.422,0.817}{81.83}}$ \\

 &  & 0.60 & 0.76 &  & ${\uline{\textcolor[rgb]{0.310,0.422,0.817}{81.93}}}$ & ${\textcolor[rgb]{0.264,0.413,0.872}{84.28}}$ & ${\textcolor[rgb]{0.264,0.413,0.872}{84.28}}$ & ${\bf \textcolor[rgb]{0.255,0.412,0.882}{84.76}}$ \\

 &  & 0.93 & 0.99 &  & ${\it \uline{\textcolor[rgb]{0.538,0.464,0.547}{70.29}}}$ & ${\textcolor[rgb]{0.375,0.434,0.740}{78.59}}$ & ${\bf \textcolor[rgb]{0.337,0.427,0.785}{80.47}}$ & ${\bf \textcolor[rgb]{0.337,0.427,0.785}{80.47}}$ \\
\cline{1-9} \cline{2-9}\noalign{\smallskip} 
\multirow[t]{4}{*}{16} & \multirow[t]{4}{*}{4} & 0.01 & 0.63 &  & ${\it \uline{\textcolor[rgb]{0.892,0.529,0.128}{52.36}}}$ & ${\bf \textcolor[rgb]{0.337,0.427,0.785}{80.50}}$ & ${\bf \textcolor[rgb]{0.337,0.427,0.785}{80.50}}$ & ${\bf \textcolor[rgb]{0.337,0.427,0.785}{80.50}}$ \\

 &  & 0.05 & 0.74 &  & ${\it \uline{\textcolor[rgb]{0.658,0.486,0.405}{64.13}}}$ & ${\bf \textcolor[rgb]{0.345,0.428,0.775}{80.10}}$ & ${\bf \textcolor[rgb]{0.345,0.428,0.775}{80.10}}$ & ${\bf \textcolor[rgb]{0.345,0.428,0.775}{80.10}}$ \\

 &  & 0.10 & 0.83 &  & ${\it \uline{\textcolor[rgb]{0.608,0.477,0.464}{66.66}}}$ & ${\bf \textcolor[rgb]{0.340,0.427,0.782}{80.34}}$ & ${\bf \textcolor[rgb]{0.340,0.427,0.782}{80.34}}$ & ${\bf \textcolor[rgb]{0.340,0.427,0.782}{80.34}}$ \\

 &  & 0.29 & 0.99 &  & ${\it \uline{\textcolor[rgb]{1.000,0.549,0.000}{46.74}}}$ & ${\bf \textcolor[rgb]{0.310,0.422,0.817}{81.85}}$ & ${\bf \textcolor[rgb]{0.310,0.422,0.817}{81.85}}$ & ${\bf \textcolor[rgb]{0.310,0.422,0.817}{81.85}}$ \\
\cline{1-9} \cline{2-9}\noalign{\smallskip} 
\multirow[t]{4}{*}{4} & \multirow[t]{4}{*}{16} & 0.01 & 0.71 &  & ${\it \uline{\textcolor[rgb]{1.000,0.549,0.000}{46.74}}}$ & ${\bf \textcolor[rgb]{0.310,0.422,0.817}{81.85}}$ & ${\bf \textcolor[rgb]{0.310,0.422,0.817}{81.85}}$ & ${\bf \textcolor[rgb]{0.310,0.422,0.817}{81.85}}$ \\

 &  & 0.05 & 0.77 &  & ${\it \uline{\textcolor[rgb]{0.769,0.506,0.273}{58.60}}}$ & ${\bf \textcolor[rgb]{0.319,0.424,0.806}{81.42}}$ & ${\bf \textcolor[rgb]{0.319,0.424,0.806}{81.42}}$ & ${\bf \textcolor[rgb]{0.319,0.424,0.806}{81.42}}$ \\

 &  & 0.10 & 0.84 &  & ${\it \uline{\textcolor[rgb]{0.614,0.478,0.457}{66.38}}}$ & ${\bf \textcolor[rgb]{0.340,0.427,0.782}{80.43}}$ & ${\bf \textcolor[rgb]{0.340,0.427,0.782}{80.43}}$ & ${\bf \textcolor[rgb]{0.340,0.427,0.782}{80.43}}$ \\

 &  & 0.37 & 0.99 &  & ${\it \uline{\textcolor[rgb]{0.892,0.529,0.128}{52.36}}}$ & ${\bf \textcolor[rgb]{0.337,0.427,0.785}{80.50}}$ & ${\bf \textcolor[rgb]{0.337,0.427,0.785}{80.50}}$ & ${\bf \textcolor[rgb]{0.337,0.427,0.785}{80.50}}$ \\
\cline{1-9} \cline{2-9}\noalign{\smallskip} 
\multirow[t]{4}{*}{40} & \multirow[t]{4}{*}{10} & 0.01 & 0.32 &  & ${\it \uline{\textcolor[rgb]{0.638,0.482,0.429}{65.25}}}$ & ${\textcolor[rgb]{0.360,0.431,0.758}{79.36}}$ & ${\textcolor[rgb]{0.334,0.426,0.789}{80.75}}$ & ${\bf \textcolor[rgb]{0.331,0.426,0.792}{80.77}}$ \\

 &  & 0.20 & 0.68 &  & ${\it \uline{\textcolor[rgb]{0.500,0.457,0.592}{72.22}}}$ & ${\textcolor[rgb]{0.322,0.424,0.803}{81.20}}$ & ${\textcolor[rgb]{0.433,0.445,0.671}{75.60}}$ & ${\bf \textcolor[rgb]{0.322,0.424,0.803}{81.21}}$ \\

 &  & 0.40 & 0.87 &  & ${\it \uline{\textcolor[rgb]{0.527,0.462,0.561}{70.90}}}$ & ${\bf \textcolor[rgb]{0.334,0.426,0.789}{80.64}}$ & ${\textcolor[rgb]{0.395,0.438,0.716}{77.56}}$ & ${\bf \textcolor[rgb]{0.334,0.426,0.789}{80.64}}$ \\

 &  & 0.65 & 0.99 &  & ${\it \uline{\textcolor[rgb]{0.918,0.534,0.097}{50.91}}}$ & ${\it \textcolor[rgb]{0.486,0.454,0.609}{73.01}}$ & ${\it \textcolor[rgb]{0.486,0.454,0.609}{73.00}}$ & ${\bf \textcolor[rgb]{0.337,0.427,0.785}{80.58}}$ \\
\cline{1-9} \cline{2-9}\noalign{\smallskip} 
\multirow[t]{4}{*}{10} & \multirow[t]{4}{*}{40} & 0.01 & 0.35 &  & ${\it \uline{\textcolor[rgb]{0.918,0.534,0.097}{50.91}}}$ & ${\it \textcolor[rgb]{0.486,0.454,0.609}{73.01}}$ & ${\it \textcolor[rgb]{0.486,0.454,0.609}{73.00}}$ & ${\bf \textcolor[rgb]{0.337,0.427,0.785}{80.58}}$ \\

 &  & 0.20 & 0.68 &  & ${\it \uline{\textcolor[rgb]{0.553,0.467,0.529}{69.60}}}$ & ${\bf \textcolor[rgb]{0.345,0.428,0.775}{80.07}}$ & ${\textcolor[rgb]{0.401,0.439,0.709}{77.28}}$ & ${\bf \textcolor[rgb]{0.345,0.428,0.775}{80.07}}$ \\

 &  & 0.40 & 0.86 &  & ${\it \uline{\textcolor[rgb]{0.489,0.455,0.606}{72.79}}}$ & ${\textcolor[rgb]{0.345,0.428,0.775}{80.07}}$ & ${\textcolor[rgb]{0.416,0.441,0.692}{76.58}}$ & ${\bf \textcolor[rgb]{0.343,0.428,0.779}{80.28}}$ \\

 &  & 0.68 & 0.99 &  & ${\it \uline{\textcolor[rgb]{0.638,0.482,0.429}{65.25}}}$ & ${\textcolor[rgb]{0.360,0.431,0.758}{79.36}}$ & ${\textcolor[rgb]{0.334,0.426,0.789}{80.75}}$ & ${\bf \textcolor[rgb]{0.331,0.426,0.792}{80.77}}$ \\
\cline{1-9} \cline{2-9}\noalign{\smallskip} 
\multirow[t]{4}{*}{80} & \multirow[t]{4}{*}{20} & 0.01 & 0.19 &  & ${\it \uline{\textcolor[rgb]{0.524,0.461,0.564}{70.95}}}$ & ${\textcolor[rgb]{0.372,0.433,0.744}{78.71}}$ & ${\textcolor[rgb]{0.372,0.433,0.744}{78.71}}$ & ${\bf \textcolor[rgb]{0.310,0.422,0.817}{81.85}}$ \\

 &  & 0.30 & 0.65 &  & ${\uline{\textcolor[rgb]{0.427,0.444,0.678}{75.96}}}$ & ${\textcolor[rgb]{0.322,0.424,0.803}{81.31}}$ & ${\textcolor[rgb]{0.322,0.424,0.803}{81.28}}$ & ${\bf \textcolor[rgb]{0.305,0.421,0.824}{82.09}}$ \\

 &  & 0.60 & 0.90 &  & ${\it \uline{\textcolor[rgb]{0.489,0.455,0.606}{72.74}}}$ & ${\textcolor[rgb]{0.351,0.430,0.768}{79.77}}$ & ${\textcolor[rgb]{0.354,0.430,0.765}{79.71}}$ & ${\bf \textcolor[rgb]{0.325,0.425,0.799}{81.14}}$ \\

 &  & 0.79 & 0.99 &  & ${\it \uline{\textcolor[rgb]{0.874,0.526,0.149}{53.23}}}$ & ${\textcolor[rgb]{0.424,0.443,0.682}{76.02}}$ & ${\textcolor[rgb]{0.424,0.443,0.682}{76.02}}$ & ${\bf \textcolor[rgb]{0.348,0.429,0.772}{80.00}}$ \\
\cline{1-9} \cline{2-9}\noalign{\smallskip} 
\multirow[t]{4}{*}{20} & \multirow[t]{4}{*}{80} & 0.01 & 0.21 &  & ${\it \uline{\textcolor[rgb]{0.874,0.526,0.149}{53.23}}}$ & ${\textcolor[rgb]{0.424,0.443,0.682}{76.02}}$ & ${\textcolor[rgb]{0.424,0.443,0.682}{76.02}}$ & ${\bf \textcolor[rgb]{0.348,0.429,0.772}{80.00}}$ \\

 &  & 0.30 & 0.65 &  & ${\uline{\textcolor[rgb]{0.421,0.442,0.685}{76.18}}}$ & ${\textcolor[rgb]{0.325,0.425,0.799}{81.13}}$ & ${\textcolor[rgb]{0.331,0.426,0.792}{80.78}}$ & ${\bf \textcolor[rgb]{0.310,0.422,0.817}{81.86}}$ \\

 &  & 0.60 & 0.90 &  & ${\uline{\textcolor[rgb]{0.398,0.438,0.713}{77.36}}}$ & ${\textcolor[rgb]{0.290,0.418,0.841}{82.84}}$ & ${\textcolor[rgb]{0.313,0.423,0.813}{81.66}}$ & ${\bf \textcolor[rgb]{0.290,0.418,0.841}{82.85}}$ \\

 &  & 0.81 & 0.99 &  & ${\it \uline{\textcolor[rgb]{0.524,0.461,0.564}{70.95}}}$ & ${\textcolor[rgb]{0.372,0.433,0.744}{78.71}}$ & ${\textcolor[rgb]{0.372,0.433,0.744}{78.71}}$ & ${\bf \textcolor[rgb]{0.310,0.422,0.817}{81.85}}$ \\
\cline{1-9} \cline{2-9}\noalign{\smallskip} 
\multirow[t]{4}{*}{240} & \multirow[t]{4}{*}{60} & 0.01 & 0.09 &  & ${\it \uline{\textcolor[rgb]{0.503,0.458,0.588}{72.13}}}$ & ${\textcolor[rgb]{0.351,0.430,0.768}{79.85}}$ & ${\textcolor[rgb]{0.351,0.430,0.768}{79.81}}$ & ${\bf \textcolor[rgb]{0.343,0.428,0.779}{80.24}}$ \\

 &  & 0.30 & 0.50 &  & ${\uline{\textcolor[rgb]{0.389,0.437,0.723}{77.84}}}$ & ${\textcolor[rgb]{0.319,0.424,0.806}{81.38}}$ & ${\textcolor[rgb]{0.331,0.426,0.792}{80.89}}$ & ${\bf \textcolor[rgb]{0.316,0.423,0.810}{81.52}}$ \\

 &  & 0.60 & 0.79 &  & ${\uline{\textcolor[rgb]{0.395,0.438,0.716}{77.58}}}$ & ${\textcolor[rgb]{0.319,0.424,0.806}{81.44}}$ & ${\textcolor[rgb]{0.328,0.425,0.796}{81.04}}$ & ${\bf \textcolor[rgb]{0.307,0.421,0.820}{81.98}}$ \\

 &  & 0.90 & 0.99 &  & ${\it \uline{\textcolor[rgb]{0.532,0.463,0.554}{70.59}}}$ & ${\textcolor[rgb]{0.302,0.420,0.827}{82.30}}$ & ${\textcolor[rgb]{0.307,0.421,0.820}{82.07}}$ & ${\bf \textcolor[rgb]{0.290,0.418,0.841}{82.96}}$ \\
\cline{1-9} \cline{2-9}\noalign{\smallskip} 
\multirow[t]{4}{*}{60} & \multirow[t]{4}{*}{240} & 0.01 & 0.10 &  & ${\it \uline{\textcolor[rgb]{0.532,0.463,0.554}{70.59}}}$ & ${\textcolor[rgb]{0.302,0.420,0.827}{82.30}}$ & ${\textcolor[rgb]{0.307,0.421,0.820}{82.07}}$ & ${\bf \textcolor[rgb]{0.290,0.418,0.841}{82.96}}$ \\

 &  & 0.30 & 0.50 &  & ${\uline{\textcolor[rgb]{0.410,0.440,0.699}{76.88}}}$ & ${\textcolor[rgb]{0.334,0.426,0.789}{80.63}}$ & ${\textcolor[rgb]{0.334,0.426,0.789}{80.63}}$ & ${\bf \textcolor[rgb]{0.325,0.425,0.799}{81.10}}$ \\

 &  & 0.60 & 0.79 &  & ${\uline{\textcolor[rgb]{0.366,0.432,0.751}{79.11}}}$ & ${\textcolor[rgb]{0.296,0.419,0.834}{82.66}}$ & ${\textcolor[rgb]{0.302,0.420,0.827}{82.26}}$ & ${\bf \textcolor[rgb]{0.290,0.418,0.841}{82.90}}$ \\

 &  & 0.91 & 0.99 &  & ${\it \uline{\textcolor[rgb]{0.503,0.458,0.588}{72.13}}}$ & ${\textcolor[rgb]{0.351,0.430,0.768}{79.85}}$ & ${\textcolor[rgb]{0.351,0.430,0.768}{79.81}}$ & ${\bf \textcolor[rgb]{0.343,0.428,0.779}{80.24}}$ \\
\cline{1-9} \cline{2-9}\noalign{\smallskip} 

\end{tabular}
     \\  \vspace{1mm}{\footnotesize  FE: Fisher's exact test, FMP$^*$: Berger and Boos' version of Fisher's mid-p value test, Z$^*_{\text{P}}$: Berger and Boos' version of the Z-pooled test, APK: average power knapsack}
    \label{tab:power_table}
\end{table}

\renewcommand{\arraystretch}{0.1} 
\begin{table*}[h!]
\centering
    \addtolength{\tabcolsep}{-0.3em}
    \caption{ Summary table of power comparisons. The pointwise power comparison rows show the fraction (\%) of parameters for which the test in the column has strictly higher power than the APK test~(left) and the frequency of 
    parameters for which the APK test has strictly higher power~(right), where we consider all parameters~$\btheta$ such that~$\theta_\D>\theta_\C$ and both~$\theta_\C,\theta_\D$ are multiples of~0.01.   The average power comparison shows the difference in average power (\%) for the APK test over the column test, and in brackets the absolute difference in average power (\%) for parameters where the column test has higher power (left) and for parameters where the APK test has higher power (right), where ``$-$'' indicates there are no such parameter values. 
The one-sided significance level equals~$2.5\%$.}
    \centering 
    \renewcommand{\arraystretch}{1.2} 
\setlength{\tabcolsep}{1.4mm}\footnotesize\begin{tabular}{lllccc}
\toprule
 &  &  & FE & FMP$^*$ & Z$_{\text{P}}^*$ \\
$n$ & $(n_{\text{C}}, n_{\text{D}})$ & Power comparison &  &  &  \\
\midrule
\multirow[t]{6}{*}{20} & \multirow[t]{2}{*}{(10, 10)} & \% Pointwise higher (column/APK) & ${ 0.00/ 99.98}$ & ${ 0.00/ 0.00}$ & ${ 0.00/ 0.00}$ \\
 &  & Avg. diff. total (APK worse/better) & \textcolor[RGB]{255,140,0}{${9.84~(-/9.85)}$} & \textcolor[RGB]{65,105,225}{${0.00~(-/-)}$} & \textcolor[RGB]{65,105,225}{${0.00~(-/-)}$} \\
\cline{2-6}
 & \multirow[t]{2}{*}{(7, 13)} & \% Pointwise higher (column/APK) & ${ 0.00/ 99.98}$ & ${ 0.00/ 0.00}$ & ${ 0.00/ 0.00}$ \\
 &  & Avg. diff. total (APK worse/better) & \textcolor[RGB]{235,136,22}{${8.89~(-/8.90)}$} & \textcolor[RGB]{65,105,225}{${0.00~(-/-)}$} & \textcolor[RGB]{65,105,225}{${0.00~(-/-)}$} \\
\cline{2-6}
 & \multirow[t]{2}{*}{(4, 16)} & \% Pointwise higher (column/APK) & ${ 0.00/ 99.98}$ & ${ 0.00/ 0.00}$ & ${ 0.00/ 0.00}$ \\
 &  & Avg. diff. total (APK worse/better) & \textcolor[RGB]{235,136,22}{${9.53~(-/9.53)}$} & \textcolor[RGB]{65,105,225}{${0.00~(-/-)}$} & \textcolor[RGB]{65,105,225}{${0.00~(-/-)}$} \\
\cline{1-6} \cline{2-6}
\multirow[t]{6}{*}{50} & \multirow[t]{2}{*}{(25, 25)} & \% Pointwise higher (column/APK) & ${ 0.00/ 99.43}$ & ${ 0.00/ 95.82}$ & ${ 9.23/ 90.12}$ \\
 &  & Avg. diff. total (APK worse/better) & \textcolor[RGB]{160,122,112}{${4.66~(-/4.69)}$} & \textcolor[RGB]{84,108,202}{${1.14~(-/1.19)}$} & \textcolor[RGB]{84,108,202}{${0.49~(4.04/0.95)}$} \\
\cline{2-6}
 & \multirow[t]{2}{*}{(17, 33)} & \% Pointwise higher (column/APK) & ${ 0.00/ 99.64}$ & ${ 0.00/ 95.90}$ & ${ 0.00/ 95.90}$ \\
 &  & Avg. diff. total (APK worse/better) & \textcolor[RGB]{179,126,90}{${6.14~(-/6.16)}$} & \textcolor[RGB]{84,108,202}{${0.63~(-/0.66)}$} & \textcolor[RGB]{84,108,202}{${0.63~(-/0.66)}$} \\
\cline{2-6}
 & \multirow[t]{2}{*}{(10, 40)} & \% Pointwise higher (column/APK) & ${ 0.00/ 99.78}$ & ${ 0.00/ 97.76}$ & ${ 0.00/ 97.76}$ \\
 &  & Avg. diff. total (APK worse/better) & \textcolor[RGB]{198,129,67}{${6.63~(-/6.64)}$} & \textcolor[RGB]{84,108,202}{${0.88~(-/0.90)}$} & \textcolor[RGB]{103,112,179}{${2.17~(-/2.22)}$} \\
\cline{1-6} \cline{2-6}
\multirow[t]{6}{*}{100} & \multirow[t]{2}{*}{(50, 50)} & \% Pointwise higher (column/APK) & ${ 0.00/ 95.17}$ & ${ 3.07/ 91.21}$ & ${ 3.07/ 87.50}$ \\
 &  & Avg. diff. total (APK worse/better) & \textcolor[RGB]{122,115,157}{${3.13~(-/3.29)}$} & \textcolor[RGB]{84,108,202}{${0.56~(0.00/0.61)}$} & \textcolor[RGB]{65,105,225}{${0.22~(0.00/0.25)}$} \\
\cline{2-6}
 & \multirow[t]{2}{*}{(35, 65)} & \% Pointwise higher (column/APK) & ${ 0.00/ 96.12}$ & ${ 5.56/ 87.11}$ & ${ 0.38/ 92.18}$ \\
 &  & Avg. diff. total (APK worse/better) & \textcolor[RGB]{122,115,157}{${3.23~(-/3.36)}$} & \textcolor[RGB]{65,105,225}{${0.40~(0.00/0.46)}$} & \textcolor[RGB]{65,105,225}{${0.40~(0.00/0.44)}$} \\
\cline{2-6}
 & \multirow[t]{2}{*}{(20, 80)} & \% Pointwise higher (column/APK) & ${ 0.00/ 98.48}$ & ${ 8.83/ 87.39}$ & ${ 0.00/ 96.50}$ \\
 &  & Avg. diff. total (APK worse/better) & \textcolor[RGB]{141,119,135}{${4.09~(-/4.15)}$} & \textcolor[RGB]{84,108,202}{${0.58~(0.19/0.68)}$} & \textcolor[RGB]{84,108,202}{${1.03~(-/1.07)}$} \\
\cline{1-6} \cline{2-6}
\multirow[t]{6}{*}{300} & \multirow[t]{2}{*}{(150, 150)} & \% Pointwise higher (column/APK) & ${ 0.00/ 73.98}$ & ${ 0.00/ 72.93}$ & ${ 0.00/ 71.74}$ \\
 &  & Avg. diff. total (APK worse/better) & \textcolor[RGB]{84,108,202}{${1.08~(-/1.46)}$} & \textcolor[RGB]{65,105,225}{${0.23~(-/0.31)}$} & \textcolor[RGB]{65,105,225}{${0.18~(-/0.25)}$} \\
\cline{2-6}
 & \multirow[t]{2}{*}{(100, 200)} & \% Pointwise higher (column/APK) & ${ 0.00/ 76.85}$ & ${ 2.00/ 72.67}$ & ${ 0.00/ 74.85}$ \\
 &  & Avg. diff. total (APK worse/better) & \textcolor[RGB]{84,108,202}{${1.28~(-/1.67)}$} & \textcolor[RGB]{65,105,225}{${0.16~(0.05/0.22)}$} & \textcolor[RGB]{65,105,225}{${0.24~(-/0.32)}$} \\
\cline{2-6}
 & \multirow[t]{2}{*}{(60, 240)} & \% Pointwise higher (column/APK) & ${ 0.00/ 84.18}$ & ${ 5.13/ 76.08}$ & ${ 0.99/ 81.05}$ \\
 &  & Avg. diff. total (APK worse/better) & \textcolor[RGB]{103,112,179}{${1.78~(-/2.11)}$} & \textcolor[RGB]{65,105,225}{${0.17~(0.02/0.22)}$} & \textcolor[RGB]{65,105,225}{${0.32~(0.00/0.39)}$} \\
\cline{1-6} \cline{2-6}

\end{tabular}
     \label{tab:comparison_table}\\
   {
FE: Fisher's exact test, FMP$^*$: Berger and Boos' version of Fisher's mid-p value test, Z$^*_{\text{P}}$:~Berger and Boos' version of the Z-pooled test }
\end{table*}
\FloatBarrier

\autoref{tab:comparison_table} shows that FET never showed higher power than the APK test, and the increase in average power of the APK test over FET  ranges from~$1.08\%$ to~$9.84\%$.  The APK test outperforms the FMP$^*$ test for smaller sample sizes, while it performs no worse than the FMP$^*$ test in at least~$91.17\%$ of cases for~$n\geq100$.  The Z$^*_{\text{P}}$ is outperformed by the APK test for 
unbalanced designs in at least $99.01\%$ of cases, and in at least~$90.77\%$ of cases for balanced groups. While consistent, the average power gains of the APK test over the FMP$^*$ and Z$_P^*$ tests are observed to lie in the interval~$0.00\% - 2.17\%.$ 
We have split out the average power differences in \autoref{tab:comparison_table}
to investigate whether large power gains offset each other whenever the average power gain from the APK test equals zero. 
\autoref{tab:comparison_table} shows that this only occurs when comparing to the $Z^*_\text{P}$ test in case $n_{\C}=n_{\D}=25$, and in most cases the average gain of the APK test is substantially higher than the value of its comparator in cases where it outperforms, but the average power gains for all tests shrink in the trial size.

In \autoref{sect:comparison_non_exact}, we compare the exact tests in this section with several 
non-exact tests. 
Using a cutoff of about 2-2.5 hours of computation time, we recommend  the APK test for trial sizes up to 300, as this test is exact and yields power values comparable to the ``near-exact'' estimated p-value and FMP tests. 
 For larger trial sizes, the best test depends on the tolerance for type I error rate inflation: the FMP$^*$ and~Z$_{\text{P}}^*$ p-values can be calculated up to relatively high trial sizes for single datasets, but if a slight type I error rate inflation is allowed then we would recommend the FMP test,  which shows lower type I error inflation and shows similar or higher power values than estimated p-value tests.
\autoref{sect:superiority_test} shows that our APK test outperforms the unconditional exact unpooled Wald test for superiority testing with a margin, especially with unbalanced groups, showcasing the wide applicability of the APK test. 

\FloatBarrier
\section{Case study: Merck Research Laboratories trial}\label{sect:application}

In this section, we analyze the trial by Merck Research Laboratories described in Section~3.1 of~\cite{mehrotra2003cautionary}. In order to make the trial setup more consistent with the rest of the current paper, we assume that treatment group~1~(consisting of~132 participants) was the developmental group, while treatment group~2~(consisting of~148 participants) was the control treatment group. 
 Furthermore, the goal of the trial is reformulated to testing whether the proportion of trial participants without a specific type of rash is higher in the developmental treatment group than in the control treatment group.
 The observed proportions of participants that experienced no rash were~$\hat{\theta}_\C=140/148$ and~$\hat{\theta}_\D=131/132$.

\autoref{table_pvalues} shows the one-sided p-values, where the p-value for the APK test is found as outlined in~\autoref{section:p_value_knapsack} tuned to a significance level~$\alpha=2.5\%$ and using $\mathcal{P}=\{0.001,0.002,\dots,0.100, 0.110,\dots, 1.000\}$. As in~\cite{mehrotra2003cautionary}, the choice of the unconditional exact test makes a big difference, as the FE test does not lead to a p-value lower than~$2.5\%$, while the other p-values, when used on their own, would lead to a rejection at the~$2.5\%$ level. The APK test shows a lower p-value than the FMP$^*$ and~Z$^*_\text{P}$ tests.

\begin{table}[h!]
\centering
\setlength{\tabcolsep}{0.8mm}

\caption{{\bf One-sided p-values for the Merck trial}: p-values found under the different tests used in the Merck trial of~\autoref{sect:application}.}\label{table_pvalues}
\small
\renewcommand{\arraystretch}{1.3} 
\begin{tabular}{cccc}
\hline
\multicolumn{1}{l}{\it Conditional} & \multicolumn{3}{c}{ \it Unconditional}        \\ \hline
FE                                                      & FMP$^*$ & Z$^*_\text{P}$ & APK     \\
.0271                                                  & .0144 & .0136          & .0130 \\ \hline
\end{tabular}
\end{table}
\FloatBarrier
\autoref{case_study} contains p-values for other exact tests and other knapsack-based tests, as well as a power sensitivity analysis for the considered tests. 
\FloatBarrier
\section{Discussion}\label{sect:discussion}

We considered an integer programming approach to construct one-sided unconditional exact tests maximizing average power to detect differences in binary proportions. By definition, the power gains realized by such tests do not inflate the type I error rate beyond the nominal significance level, and as it maximizes the average power, it cannot uniformly be improved in terms of power. Numerically, the average power knapsack test has a stable performance, showing substantially higher power than Fisher's exact test~(FET) and often having a higher power than Berger and Boos’ versions of Fisher’s mid-p and Z-pooled tests. 
As both our knapsack-based test and the state-of-the-art Berger and Boos tests uniformly improved power over FET, with power differences reaching 35\%, our recommendation would be to move away from widespread use of FET in small samples~(except if used for randomization-based inference). 
As also investigated in~\citet{keer2023hypothesis}, the integer programming approach readily generalizes to two-sided tests, where we note that~\citet{ANDRES1994555} show that, when also imposing a symmetry condition, Barnard's test optimizes average power in this setting. An equal-tailed two-sided knapsack test at significance level $\alpha$ can furthermore be readily constructed by solving two separate one-sided knapsack problems at level $\alpha/2$ by simply swapping the roles of the control and developmental groups.

  A limitation of the integer programming approach is the computation time:  the Gurobi optimization tool was only able to compute knapsack-based tests for trial sizes of~300 participants within a reasonable time (roughly 2 hours). Open-source solvers are available~(e.g., from the \verb|Python| package \verb|SciPy|), but these often take longer. For computationally prohibitive trial sizes, we recommend to either use Berger and Boos' version of Fisher's-mid-p test or the pooled-Z test, or to use Fisher's mid-p-value test if strict type I error rate control is not required.
 Another limitation is that knapsack tests may be sensitive to solver configurations~(see \autoref{sect:sensitivity_solver_settings}), and before implementation, we recommend performing a sensitivity analysis of a knapsack test against comparator exact tests.

Several areas are left for future research.  
Similar to~\cite{keer2023hypothesis}, future research could consider the application to~$r\times 2$ tables or other discrete distributions. 
Future work could also focus on extending our integer programming approach to general (e.g., continuous) outcomes by, e.g., discretizing the outcome space and using additional Lipschitz bound results.
Future work could also investigate ensuring uniqueness of solutions to knapsack programs.
Despite the fact that the APK test does not have a traditional test statistic and instead optimizes test decisions,  exact p-values for the APK test are readily obtained by the method outlined in \autoref{section:p_value_knapsack}. This method is guaranteed to yield the optimal APK test for at least one significance level, making this approach suitable for traditional trial settings that use a single significance level.
 In more complex (e.g., adaptive) clinical trial settings, p-value combination methods may be used, which may require a homogeneous quality of the p-value over different significance levels.
 As the current APK-based p-values are optimized for only one significance level, they are not uniformly optimal, and hence p-values resulting from the combination of APK-based p-values may be more conservative than combinations of p-values that show a more homogeneous performance.
 Further research is needed to make the integer programming approach better suited to scenarios of this type. 
For instance, \cite{GUTMAN20072380} and~\cite{RISTL20181} use an integer programming approach to construct optimal rejection regions for intersection hypothesis tests based on multiple binary endpoints. It could be interesting to consider an unconditional exact approach that is more in line with ours, as~\cite{GUTMAN20072380} consider the combination of decisions of multiple two-sample integer programming tests, whereas~\cite{RISTL20181} consider a transformation of the multi-outcome setting to the extension of Fisher's exact test to categorical data.
Given the heavy computational burden of the method in \citet{peer2024optimal} (restricted to $n \le 21$), it would be instructive to compare their results with exact confidence intervals inverted from our APK test, which accommodates substantially larger sample sizes.
Future work could also consider settings with large trial sizes, where we start with an already well-performing
 test and iteratively use the APK program to increase average 
 power while ensuring type I error rate control. 

\subsection{Data availability statement}
The Shiny App can be found through this link:~\url{https://mrc-bsu.shinyapps.io/knapsack/}

\subsection{Acknowledgements}\label{sect:acknowledgements}
We thank the anonymous reviewers of this paper for their helpful comments. {We thank Alan Nardo for the creation of the Shiny app accompanying this paper.}

\bibliographystyle{abbrvnat}
\bibliography{main}       

\par\vspace{.5em} 

\appendix

\section{Proof of Theorem 2}\label{sect:proof_thm2}
The proof is based on the one in~\citet{suissashuster1985}. We have
$$    r_d(\theta)=\sum_{\bs:d(\bs)=1}\binom{n_\C}{s_\C}\binom{n_\D}{s_\D} \theta^{s_\C}g_0(\theta)^{s_\D}(1-\theta)^{n_\C-s_\C} (1-g_0(\theta))^{n_\D-s_\D},$$
hence, making use of the fact that for~$x,y\in\mathbb{N}_0$~\begin{align*}
    &y\cdot\binom{x}{y} =y\frac{x!}{y!(x-y)!}=\frac{x!}{(y-1)!(x-y)!}=x\cdot \frac{(x-1)!}{(y-1)!(x-y)!}=x\binom{x-1}{y-1},\\&(x-y)\cdot\binom{x}{y} =(x-y)\frac{x!}{y!(x-y)!}=\frac{x!}{y!(x-1-y)!}=x\cdot \frac{(x-1)!}{y!(x-1-y)!}=x\binom{x-1}{y},
\end{align*}
we have
\begin{align*}
    r'_d(\theta)&= \underbrace{\sum_{{\bs:d(\bs)=1}}n_\C\binom{n_\C-1}{s_\C-1}\binom{n_\D}{s_\D} \theta^{s_\C-1}g_0(\theta)^{s_\D}(1-\theta)^{n_\C-s_\C} (1-g_0(\theta))^{n_\D-s_\D}}_{(a)} \\&+ \underbrace{g_0'(\theta)\sum_{\bs:d(\bs)=1}n_\D\binom{n_\C}{s_\C}\binom{n_\D-1}{s_\D-1} \theta^{s_\C}g_0(\theta)^{s_\D-1}(1-\theta)^{n_\C-s_\C} (1-g_0(\theta))^{n_\D-s_\D}}_{(b)}\\ &- \underbrace{\sum_{\bs:d(\bs)=1}n_\C\binom{n_\C-1}{s_\C}\binom{n_\D}{s_\D} \theta^{s_\C}g_0(\theta)^{s_\D}(1-\theta)^{n_\C-s_\C-1} (1-g_0(\theta))^{n_\D-s_\D}}_{(c)}\\&-\underbrace{g_0'(\theta)\sum_{\bs:d(\bs)=1}n_\D\binom{n_\C}{s_\C}\binom{n_\D-1}{s_\D}\theta^{s_\C}g_0(\theta)^{s_\D}(1-\theta)^{n_\C-s_\C} (1-g_0(\theta))^{n_\D-s_\D-1}}_{(d)}\\
    =g_0'(\theta)&\sum_{\partial\mathcal{R}_{d,\D}}n_\D\binom{n_\C}{s_\C}\binom{n_\D-1}{s_\D-1}\theta^{s_\C}g_0(\theta)^{s_\D-1}(1-\theta)^{n_\C-s_\C} (1-g_0(\theta))^{n_\D-s_\D}\\-&\sum_{\partial\mathcal{R}_{d,\C}}n_\C\binom{n_\C-1}{s_\C}\binom{n_\D}{s_\D} \theta^{s_\C}g_0(\theta)^{s_\D}(1-\theta)^{n_\C-s_\C-1} (1-g_0(\theta))^{n_\D-s_\D}.
\end{align*}
The last equation above follows from combining the sums~$(a)$ and~$(c)$ and~$(b)$ and~$(d)$ (using the fact that Condition (C) holds), where furthermore
\begin{align*}
&\partial\mathcal{R}_{d,\D}=\{\bs\in\mathcal{S}:d(s_\C,s_\D)=1, d(s_\C,s_\D-1)=0\},\\
&\partial\mathcal{R}_{d,\C}=\{\bs\in\mathcal{S}:d(s_\C,s_\D)=1, d(s_\C+1,s_\D)=0\}.
\end{align*}
The result now follows from the mean value theorem and bounding the positive part from above and the negative part from below.

\section{Alternative knapsack formulations}\label{knapsack_comparators}
In addition to the APK test, we have considered other knapsack-based tests, listed below:
\begin{itemize}
    \item {\bf Maximin power knapsack}: 
    The \emph{maximin power knapsack}~(MPK) test maximizes the minimum power over a finite set of parameter configurations under~$H_1$.
    This maximin test is different from the one in~\cite{keer2023hypothesis}, which maximizes the minimum type I error rate while also ensuring type I error rate control.
    The boundary~$\partial\Theta_1$ of parameters under~$H_1$ is discretized to a set~$\partial\tilde{\Theta}_1 = \{\btheta_{j,1}:j\in\mathcal{J}_1\}$  and we define~$\bp_{j,1}$ similar to~$\bp_{j,0}$.
    The objective~(7) in the manuscript is replaced by~$1-\beta$ for a variable~$\beta$ representing the maximum type II error, and we furthermore add the following set of constraints to rewrite the APK program to the MPK program:
$$1-\bp^\top_{j,1}\bd\leq \beta,\quad \forall j\in\mathcal{J}_1.$$
Typically, $\Theta_0\cup\Theta_1\neq[0,1]^2$ for the MPK test, as otherwise the test aims to bring the minimum type I error (occurring at the boundary of the closure of~$\Theta_1$, which is also the boundary of~$\Theta_0$) as close as possible to the nominal significance level~$\alpha$. 

The MPK test optimizes for a worst-case scenario under the alternative hypothesis by maximizing the minimum possible power. It might be preferred over the APK test when there is a large degree of risk-aversion with respect to low power.

If the set~$\partial\tilde{\Theta}_1$ above is chosen too discrete, the MPK test aims to maximize the minimum power over a small set of points and may have low power for parameters in~$\partial\Theta_1\backslash \partial\tilde{\Theta}_1$. This can be checked by taking a finer discretization~$\partial\tilde{\Theta}'_1$  and checking the power values coming from the MPK test for parameters in~$ \partial\tilde{\Theta}'_1\backslash\partial\tilde{\Theta}_1$. As a starting point, we recommend choosing~$\partial\tilde{\Theta}_1$ so that the control success rates include the minimum and maximum possible values and have a maximum distance of~$1\%.$

\item {\bf Weighted average power knapsack}: 
    The \emph{weighted average power knapsack}~(WAPK) test changes the average in~(5) used in the objective~(7) in the manuscript to a weighted average, and~(5) is replaced by the more general expression
    \begin{equation*}\int_{[0,1]^2}\sum_{i\,:\,d_i=1}\mathbb{P}_{\btheta}(\bS=\bs_i)\mathbb{I}(\theta_\D\geq\theta_\C) \mathbb{Q}(d\btheta)=\bd^\top\bar{\bp}_1^{\mathbb{Q}}\end{equation*}
    where~$\mathbb{Q}$ is a probability measure on~$\{\btheta\in[0,1]^2:\theta_\D\geq\theta_\C\}.$

    The WAPK test is appropriate when reliable historical or pilot data are available, as it allows the trialist to place a weight over the parameter space, optimizing the test's rejection boundary specifically to the region of realistic success probabilities.
    
     Although not a strictly Bayesian approach~---type~I~error constraints~(8)+(9) in the manuscript are still enforced--- this approach could be classified as a hybrid Bayesian and frequentist approach, especially when~$\mathbb{Q}$ represents the prior belief of the investigator. If~$\mathbb{Q}$ is a product of two independent beta distributions with parameters~$\alpha^0_a,\beta^0_a\in\mathbb{N}$ for both arms~$a\in\{\C,\D\}$ then, again following~\cite{millerblog}, $\bar{\bp}^\mathbb{Q}_1$ is proportional to~\eqref{avg_power_coefs},
where we now have~$\alpha_a(\bs) = s_a+\alpha_a^0$ and~$\alpha_a(\bs) = n_a-s_a+\beta_a^0$ for all~$\bs\in\mathcal{S}$.
    \item {\bf Simple hypothesis knapsack}: The \emph{simple hypothesis knapsack}~(SHK) test maximizes the power at a single alternative hypothesis parameter while controlling the type I error rate across the null space. The SHK test may be appropriate when there is absolute certainty about the alternative hypothesis parameter.
\end{itemize}

In practice, the weights for the WAPK test may be elicited in a similar manner as in a Bayesian analysis~\citep{hampson2014bayesian},  and may represent prior knowledge. We reiterate that knapsack-based tests provide type I error rate control no matter the weights used for the power; hence the risk lies in potential power losses. We recommend performing a sensitivity analysis to the true alternative parameter to see 
whether there are actual power gains over the APK test, as was done in~\autoref{case_study}, where no power gains for the WAPK and SHK tests over the APK test were found. 

In the following, we will include the MPK test in our comparisons, and will also include the WAPK and SHK tests in the real-life case study (\autoref{case_study}).

\section{Additional computational results}\label{sect:additional_results}
\subsection{Additional results for the comparison of tests}\label{subsect:additional_results_general}

This section presents additional results for \autoref{sect:numerical_results}. The WAPK and SHK tests were not included as they are more suitable for situations where there is relatively high certainty around the true parameter value~(e.g., in the case study of \autoref{sect:application}).

\begin{figure}
    \centering
    \includegraphics[width=\linewidth]{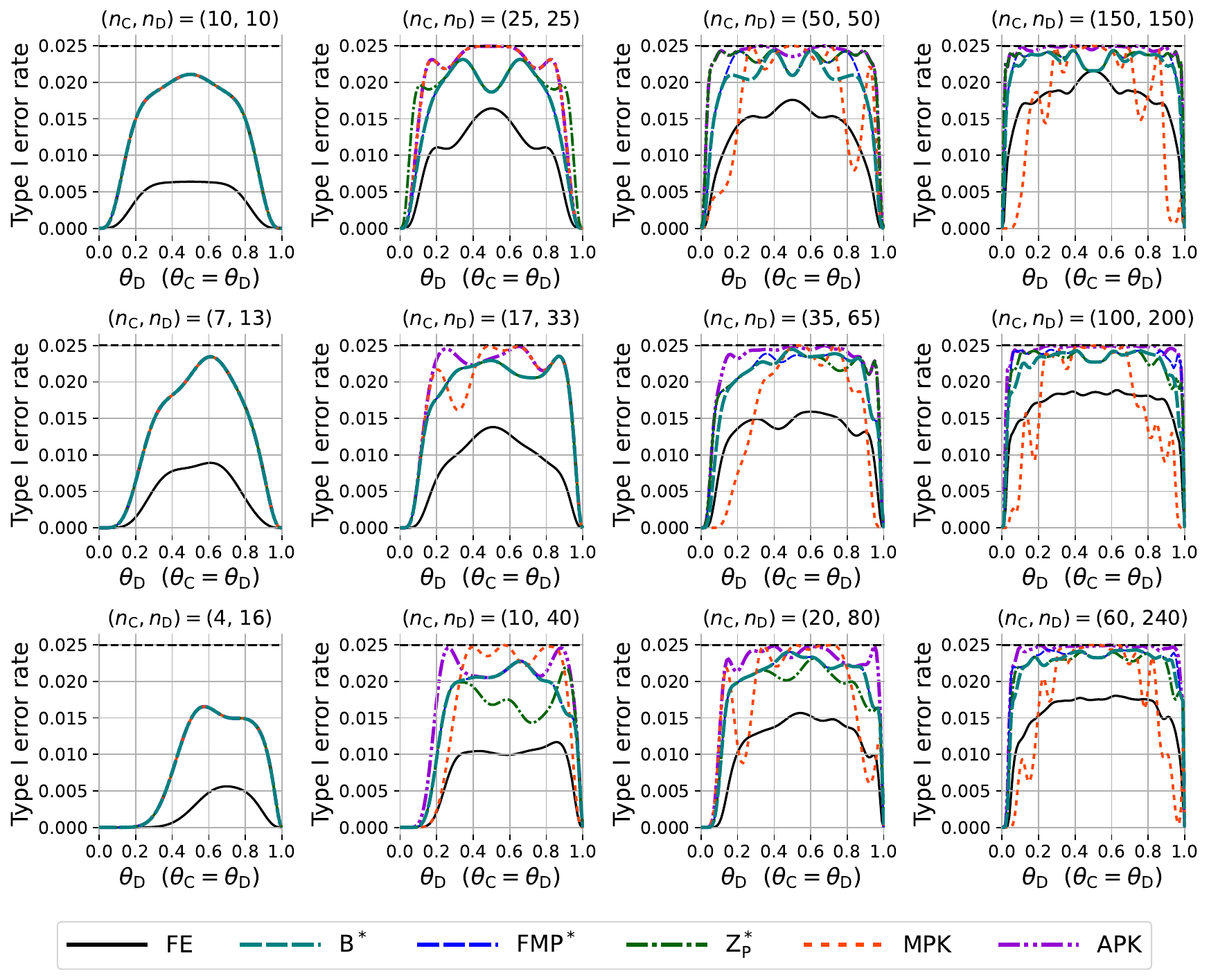}
    \caption{Exact type~I~error rate profiles for different treatment group size configurations under Fisher's exact~(FE) test, Berger and Boos' version of Boschloo's~(B$^*$) test,
    Berger and Boos' version of Fisher's mid-p value~(FMP$^*$) test, Berger and Boos' version of the Z-pooled (Z$^*_{\text{P}}$) test,  maximin power knapsack~(MPK) test, and average power knapsack~(APK) tests. The trial sizes (varying over columns) were chosen as~$20,50,100,300$, and treatment group sizes are indicated in the titles of the subfigures. The one-sided significance level~$\alpha$ is set to~$0.025$ and indicated by a dashed line.}
    \label{fig:T1E_extra}
\end{figure}

\autoref{fig:T1E_extra} shows a type~I~error rate comparison similar to~\autoref{fig:T1E}, but in this case we have included Berger and Boos' version of Boschloo's test~(as described in~\citet{mehrotra2003cautionary}) as well as the MPK test.
The maximum type~I~error rate for the MPK test often lies close to the significance level of~2.5\%, while the B$^*$ test is often more conservative than the FMP$^*$ and Z$^*_\text{P}$ tests.  In comparison to the  APK test, the MPK test shows a substantially more variable type I error rate across null parameters, sometimes even showing lower type I error rates than FET. 

\autoref{tab:power_table_80FET} shows a power comparison similar to~\autoref{tab:comparison_table}, but in this case we evaluate the two additional tests~B$^*$ and MPK, and the parameters are chosen such that FET has at least 80\% power. 
Comparing the FMP$^*,$ Z$^*_\text{P}$, and APK tests, roughly the same conclusion can be drawn as from Table~1 in the paper. For this setting, the gains by using the APK test are smaller as all comparators have power values larger than~$80\%$. As the possible gains for the APK test are smaller in this setting, the highest observed power gains by the APK test over the non-knapsack-based comparators are smaller than in \autoref{tab:comparison_table}, up to around 2\% for the non-FET tests but gains up to around $15\%$ over FET. 
\autoref{tab:power_table_80FET} furthermore shows that the FMP$^*$ often gives the same power as the~B$^*$ test, and where the powers differ, the FMP$^*$ test shows a higher power.  Looking at the power values for the MPK test, similar as for the type I error rates, we see a variable performance. In some settings, the MPK test is the test with the highest power, but in other settings it is outperformed by FET. The most extreme setting occurs when~$\bn=(150,150)$ and $\btheta=(0.01,0.08)$, where MPK shows a power of 12.39\% but the other tests have power above~80\%.

\renewcommand{\arraystretch}{1.2} 
\setlength{\tabcolsep}{1.75mm}
\begin{table}[h!]
    \centering
    \caption{Power values~(\%) across treatment group sizes and parameter values (significance level~$2.5\%$). Row maxima (minima) are made bold (underlined). A color gradient (orange to blue) differentiates power values. Values below 75\% are italicized.}\label{tab:FET80}\small\begin{tabular}{lllllllllll}
\toprule
 &  &  &  &   & FE & B$^*$ & FMP$^*$ & Z$_{\text{P}}^*$ & MPK & APK \\
$n_{\text{C}}$ & $n_{\text{D}}$ & $\theta_{\text{C}}$ & $\theta_{\text{D}}$ &  &  &  &  &  &  &  \\
\midrule
\multirow[t]{4}{*}{10} & \multirow[t]{4}{*}{10} & 0.01 & 0.61 &  & ${\uline{\textcolor[rgb]{0.386,0.436,0.727}{80.67}}}$ & ${\bf \textcolor[rgb]{0.284,0.417,0.848}{92.31}}$ & ${\bf \textcolor[rgb]{0.284,0.417,0.848}{92.31}}$ & ${\bf \textcolor[rgb]{0.284,0.417,0.848}{92.31}}$ & ${\bf \textcolor[rgb]{0.284,0.417,0.848}{92.31}}$ & ${\bf \textcolor[rgb]{0.284,0.417,0.848}{92.31}}$ \\
\cline{3-11}
 &  & 0.05 & 0.70 &  & ${\uline{\textcolor[rgb]{0.389,0.437,0.723}{80.53}}}$ & ${\bf \textcolor[rgb]{0.293,0.419,0.837}{91.28}}$ & ${\bf \textcolor[rgb]{0.293,0.419,0.837}{91.28}}$ & ${\bf \textcolor[rgb]{0.293,0.419,0.837}{91.28}}$ & ${\bf \textcolor[rgb]{0.293,0.419,0.837}{91.28}}$ & ${\bf \textcolor[rgb]{0.293,0.419,0.837}{91.28}}$ \\
\cline{3-11}
 &  & 0.20 & 0.88 &  & ${\uline{\textcolor[rgb]{0.392,0.437,0.720}{80.29}}}$ & ${\bf \textcolor[rgb]{0.290,0.418,0.841}{91.59}}$ & ${\bf \textcolor[rgb]{0.290,0.418,0.841}{91.59}}$ & ${\bf \textcolor[rgb]{0.290,0.418,0.841}{91.59}}$ & ${\bf \textcolor[rgb]{0.290,0.418,0.841}{91.59}}$ & ${\bf \textcolor[rgb]{0.290,0.418,0.841}{91.59}}$ \\
\cline{3-11}
 &  & 0.39 & 0.99 &  & ${\uline{\textcolor[rgb]{0.386,0.436,0.727}{80.67}}}$ & ${\bf \textcolor[rgb]{0.284,0.417,0.848}{92.31}}$ & ${\bf \textcolor[rgb]{0.284,0.417,0.848}{92.31}}$ & ${\bf \textcolor[rgb]{0.284,0.417,0.848}{92.31}}$ & ${\bf \textcolor[rgb]{0.284,0.417,0.848}{92.31}}$ & ${\bf \textcolor[rgb]{0.284,0.417,0.848}{92.31}}$ \\
\cline{1-11} \cline{2-11} \cline{3-11}
\multirow[t]{4}{*}{25} & \multirow[t]{4}{*}{25} & 0.01 & 0.32 &  & ${\uline{\textcolor[rgb]{0.375,0.434,0.740}{82.26}}}$ & ${\textcolor[rgb]{0.313,0.423,0.813}{88.88}}$ & ${\textcolor[rgb]{0.313,0.423,0.813}{88.88}}$ & ${\bf \textcolor[rgb]{0.287,0.418,0.844}{92.01}}$ & ${\textcolor[rgb]{0.293,0.419,0.837}{91.33}}$ & ${\textcolor[rgb]{0.293,0.419,0.837}{91.33}}$ \\
\cline{3-11}
 &  & 0.20 & 0.61 &  & ${\uline{\textcolor[rgb]{0.383,0.435,0.730}{81.03}}}$ & ${\textcolor[rgb]{0.348,0.429,0.772}{84.98}}$ & ${\textcolor[rgb]{0.348,0.429,0.772}{84.98}}$ & ${\textcolor[rgb]{0.348,0.429,0.772}{84.98}}$ & ${\bf \textcolor[rgb]{0.340,0.427,0.782}{86.13}}$ & ${\bf \textcolor[rgb]{0.340,0.427,0.782}{86.13}}$ \\
\cline{3-11}
 &  & 0.40 & 0.81 &  & ${\uline{\textcolor[rgb]{0.383,0.435,0.730}{81.18}}}$ & ${\textcolor[rgb]{0.343,0.428,0.779}{85.60}}$ & ${\textcolor[rgb]{0.343,0.428,0.779}{85.60}}$ & ${\textcolor[rgb]{0.343,0.428,0.779}{85.60}}$ & ${\bf \textcolor[rgb]{0.334,0.426,0.789}{86.53}}$ & ${\bf \textcolor[rgb]{0.334,0.426,0.789}{86.53}}$ \\
\cline{3-11}
 &  & 0.68 & 0.99 &  & ${\uline{\textcolor[rgb]{0.375,0.434,0.740}{82.26}}}$ & ${\textcolor[rgb]{0.313,0.423,0.813}{88.88}}$ & ${\textcolor[rgb]{0.313,0.423,0.813}{88.88}}$ & ${\bf \textcolor[rgb]{0.287,0.418,0.844}{92.01}}$ & ${\textcolor[rgb]{0.293,0.419,0.837}{91.33}}$ & ${\textcolor[rgb]{0.293,0.419,0.837}{91.33}}$ \\
\cline{1-11} \cline{2-11} \cline{3-11}
\multirow[t]{4}{*}{50} & \multirow[t]{4}{*}{50} & 0.01 & 0.19 &  & ${\textcolor[rgb]{0.360,0.431,0.758}{83.72}}$ & ${\textcolor[rgb]{0.299,0.420,0.830}{90.57}}$ & ${\textcolor[rgb]{0.299,0.420,0.830}{90.57}}$ & ${\bf \textcolor[rgb]{0.281,0.417,0.851}{92.68}}$ & ${\uline{\textcolor[rgb]{0.439,0.446,0.664}{75.02}}}$ & ${\bf \textcolor[rgb]{0.281,0.417,0.851}{92.68}}$ \\
\cline{3-11}
 &  & 0.30 & 0.59 &  & ${\uline{\textcolor[rgb]{0.392,0.437,0.720}{80.09}}}$ & ${\textcolor[rgb]{0.366,0.432,0.751}{83.14}}$ & ${\textcolor[rgb]{0.366,0.432,0.751}{83.14}}$ & ${\textcolor[rgb]{0.366,0.432,0.751}{83.14}}$ & ${\bf \textcolor[rgb]{0.360,0.431,0.758}{83.82}}$ & ${\textcolor[rgb]{0.363,0.432,0.754}{83.59}}$ \\
\cline{3-11}
 &  & 0.60 & 0.87 &  & ${\uline{\textcolor[rgb]{0.360,0.431,0.758}{83.79}}}$ & ${\textcolor[rgb]{0.334,0.426,0.789}{86.60}}$ & ${\textcolor[rgb]{0.322,0.424,0.803}{87.90}}$ & ${\textcolor[rgb]{0.322,0.424,0.803}{87.90}}$ & ${\textcolor[rgb]{0.340,0.427,0.782}{85.89}}$ & ${\bf \textcolor[rgb]{0.319,0.424,0.806}{88.24}}$ \\
\cline{3-11}
 &  & 0.81 & 0.99 &  & ${\uline{\textcolor[rgb]{0.360,0.431,0.758}{83.72}}}$ & ${\textcolor[rgb]{0.299,0.420,0.830}{90.57}}$ & ${\textcolor[rgb]{0.299,0.420,0.830}{90.57}}$ & ${\bf \textcolor[rgb]{0.281,0.417,0.851}{92.68}}$ & ${\textcolor[rgb]{0.284,0.417,0.848}{92.20}}$ & ${\bf \textcolor[rgb]{0.281,0.417,0.851}{92.68}}$ \\
\cline{1-11} \cline{2-11} \cline{3-11}
\multirow[t]{4}{*}{150} & \multirow[t]{4}{*}{150} & 0.01 & 0.08 &  & ${\textcolor[rgb]{0.383,0.435,0.730}{81.20}}$ & ${\textcolor[rgb]{0.328,0.425,0.796}{87.22}}$ & ${\textcolor[rgb]{0.328,0.425,0.796}{87.22}}$ & ${\bf \textcolor[rgb]{0.316,0.423,0.810}{88.66}}$ & ${\it \uline{\textcolor[rgb]{1.000,0.549,0.000}{12.39}}}$ & ${\bf \textcolor[rgb]{0.316,0.423,0.810}{88.66}}$ \\
\cline{3-11}
 &  & 0.30 & 0.47 &  & ${\uline{\textcolor[rgb]{0.363,0.432,0.754}{83.36}}}$ & ${\textcolor[rgb]{0.343,0.428,0.779}{85.84}}$ & ${\textcolor[rgb]{0.343,0.428,0.779}{85.85}}$ & ${\textcolor[rgb]{0.343,0.428,0.779}{85.85}}$ & ${\textcolor[rgb]{0.343,0.428,0.779}{85.60}}$ & ${\bf \textcolor[rgb]{0.340,0.427,0.782}{86.06}}$ \\
\cline{3-11}
 &  & 0.60 & 0.76 &  & ${\uline{\textcolor[rgb]{0.378,0.434,0.737}{81.93}}}$ & ${\textcolor[rgb]{0.357,0.431,0.761}{83.98}}$ & ${\textcolor[rgb]{0.354,0.430,0.765}{84.28}}$ & ${\textcolor[rgb]{0.354,0.430,0.765}{84.28}}$ & ${\textcolor[rgb]{0.366,0.432,0.751}{83.14}}$ & ${\bf \textcolor[rgb]{0.351,0.430,0.768}{84.76}}$ \\
\cline{3-11}
 &  & 0.92 & 0.99 &  & ${\textcolor[rgb]{0.383,0.435,0.730}{81.20}}$ & ${\textcolor[rgb]{0.328,0.425,0.796}{87.22}}$ & ${\textcolor[rgb]{0.328,0.425,0.796}{87.22}}$ & ${\bf \textcolor[rgb]{0.316,0.423,0.810}{88.66}}$ & ${\it \uline{\textcolor[rgb]{0.781,0.509,0.260}{36.87}}}$ & ${\bf \textcolor[rgb]{0.316,0.423,0.810}{88.66}}$ \\
\cline{1-11} \cline{2-11} \cline{3-11}
\multirow[t]{4}{*}{16} & \multirow[t]{4}{*}{4} & 0.01 & 0.80 &  & ${\uline{\textcolor[rgb]{0.381,0.435,0.734}{81.47}}}$ & ${\bf \textcolor[rgb]{0.258,0.412,0.879}{94.98}}$ & ${\bf \textcolor[rgb]{0.258,0.412,0.879}{94.98}}$ & ${\bf \textcolor[rgb]{0.258,0.412,0.879}{94.98}}$ & ${\bf \textcolor[rgb]{0.258,0.412,0.879}{94.98}}$ & ${\bf \textcolor[rgb]{0.258,0.412,0.879}{94.98}}$ \\
\cline{3-11}
 &  & 0.05 & 0.84 &  & ${\uline{\textcolor[rgb]{0.389,0.437,0.723}{80.50}}}$ & ${\bf \textcolor[rgb]{0.296,0.419,0.834}{90.86}}$ & ${\bf \textcolor[rgb]{0.296,0.419,0.834}{90.86}}$ & ${\bf \textcolor[rgb]{0.296,0.419,0.834}{90.86}}$ & ${\bf \textcolor[rgb]{0.296,0.419,0.834}{90.86}}$ & ${\bf \textcolor[rgb]{0.296,0.419,0.834}{90.86}}$ \\
\cline{3-11}
 &  & 0.10 & 0.91 &  & ${\uline{\textcolor[rgb]{0.381,0.435,0.734}{81.37}}}$ & ${\bf \textcolor[rgb]{0.299,0.420,0.830}{90.70}}$ & ${\bf \textcolor[rgb]{0.299,0.420,0.830}{90.70}}$ & ${\bf \textcolor[rgb]{0.299,0.420,0.830}{90.70}}$ & ${\bf \textcolor[rgb]{0.299,0.420,0.830}{90.70}}$ & ${\bf \textcolor[rgb]{0.299,0.420,0.830}{90.70}}$ \\
\cline{3-11}
 &  & 0.19 & 0.99 &  & ${\uline{\textcolor[rgb]{0.392,0.437,0.720}{80.10}}}$ & ${\bf \textcolor[rgb]{0.255,0.412,0.882}{95.62}}$ & ${\bf \textcolor[rgb]{0.255,0.412,0.882}{95.62}}$ & ${\bf \textcolor[rgb]{0.255,0.412,0.882}{95.62}}$ & ${\bf \textcolor[rgb]{0.255,0.412,0.882}{95.62}}$ & ${\bf \textcolor[rgb]{0.255,0.412,0.882}{95.62}}$ \\
\cline{1-11} \cline{2-11} \cline{3-11}
\multirow[t]{4}{*}{4} & \multirow[t]{4}{*}{16} & 0.01 & 0.81 &  & ${\uline{\textcolor[rgb]{0.392,0.437,0.720}{80.10}}}$ & ${\bf \textcolor[rgb]{0.255,0.412,0.882}{95.62}}$ & ${\bf \textcolor[rgb]{0.255,0.412,0.882}{95.62}}$ & ${\bf \textcolor[rgb]{0.255,0.412,0.882}{95.62}}$ & ${\bf \textcolor[rgb]{0.255,0.412,0.882}{95.62}}$ & ${\bf \textcolor[rgb]{0.255,0.412,0.882}{95.62}}$ \\
\cline{3-11}
 &  & 0.05 & 0.86 &  & ${\uline{\textcolor[rgb]{0.378,0.434,0.737}{81.95}}}$ & ${\bf \textcolor[rgb]{0.290,0.418,0.841}{91.68}}$ & ${\bf \textcolor[rgb]{0.290,0.418,0.841}{91.68}}$ & ${\bf \textcolor[rgb]{0.290,0.418,0.841}{91.68}}$ & ${\bf \textcolor[rgb]{0.290,0.418,0.841}{91.68}}$ & ${\bf \textcolor[rgb]{0.290,0.418,0.841}{91.68}}$ \\
\cline{3-11}
 &  & 0.10 & 0.91 &  & ${\uline{\textcolor[rgb]{0.381,0.435,0.734}{81.54}}}$ & ${\bf \textcolor[rgb]{0.296,0.419,0.834}{90.89}}$ & ${\bf \textcolor[rgb]{0.296,0.419,0.834}{90.89}}$ & ${\bf \textcolor[rgb]{0.296,0.419,0.834}{90.89}}$ & ${\bf \textcolor[rgb]{0.296,0.419,0.834}{90.89}}$ & ${\bf \textcolor[rgb]{0.296,0.419,0.834}{90.89}}$ \\
\cline{3-11}
 &  & 0.20 & 0.99 &  & ${\uline{\textcolor[rgb]{0.381,0.435,0.734}{81.47}}}$ & ${\bf \textcolor[rgb]{0.258,0.412,0.879}{94.98}}$ & ${\bf \textcolor[rgb]{0.258,0.412,0.879}{94.98}}$ & ${\bf \textcolor[rgb]{0.258,0.412,0.879}{94.98}}$ & ${\bf \textcolor[rgb]{0.258,0.412,0.879}{94.98}}$ & ${\bf \textcolor[rgb]{0.258,0.412,0.879}{94.98}}$ \\
\cline{1-11} \cline{2-11} \cline{3-11}
\multirow[t]{4}{*}{40} & \multirow[t]{4}{*}{10} & 0.01 & 0.39 &  & ${\uline{\textcolor[rgb]{0.392,0.437,0.720}{80.23}}}$ & ${\textcolor[rgb]{0.313,0.423,0.813}{89.02}}$ & ${\textcolor[rgb]{0.313,0.423,0.813}{89.02}}$ & ${\textcolor[rgb]{0.302,0.420,0.827}{90.20}}$ & ${\textcolor[rgb]{0.381,0.435,0.734}{81.44}}$ & ${\bf \textcolor[rgb]{0.302,0.420,0.827}{90.21}}$ \\
\cline{3-11}
 &  & 0.20 & 0.73 &  & ${\uline{\textcolor[rgb]{0.378,0.434,0.737}{81.67}}}$ & ${\bf \textcolor[rgb]{0.316,0.423,0.810}{88.72}}$ & ${\bf \textcolor[rgb]{0.316,0.423,0.810}{88.72}}$ & ${\textcolor[rgb]{0.354,0.430,0.765}{84.44}}$ & ${\textcolor[rgb]{0.319,0.424,0.806}{88.34}}$ & ${\bf \textcolor[rgb]{0.316,0.423,0.810}{88.72}}$ \\
\cline{3-11}
 &  & 0.40 & 0.91 &  & ${\uline{\textcolor[rgb]{0.372,0.433,0.744}{82.45}}}$ & ${\textcolor[rgb]{0.307,0.421,0.820}{89.51}}$ & ${\textcolor[rgb]{0.307,0.421,0.820}{89.51}}$ & ${\textcolor[rgb]{0.328,0.425,0.796}{87.42}}$ & ${\bf \textcolor[rgb]{0.302,0.420,0.827}{90.37}}$ & ${\textcolor[rgb]{0.307,0.421,0.820}{89.51}}$ \\
\cline{3-11}
 &  & 0.57 & 0.99 &  & ${\uline{\textcolor[rgb]{0.378,0.434,0.737}{81.89}}}$ & ${\textcolor[rgb]{0.287,0.418,0.844}{91.73}}$ & ${\textcolor[rgb]{0.287,0.418,0.844}{91.73}}$ & ${\textcolor[rgb]{0.287,0.418,0.844}{91.72}}$ & ${\textcolor[rgb]{0.348,0.429,0.772}{85.15}}$ & ${\bf \textcolor[rgb]{0.272,0.415,0.862}{93.35}}$ \\
\cline{1-11} \cline{2-11} \cline{3-11}
\multirow[t]{4}{*}{10} & \multirow[t]{4}{*}{40} & 0.01 & 0.43 &  & ${\uline{\textcolor[rgb]{0.378,0.434,0.737}{81.89}}}$ & ${\textcolor[rgb]{0.287,0.418,0.844}{91.73}}$ & ${\textcolor[rgb]{0.287,0.418,0.844}{91.73}}$ & ${\textcolor[rgb]{0.287,0.418,0.844}{91.72}}$ & ${\textcolor[rgb]{0.348,0.429,0.772}{85.15}}$ & ${\bf \textcolor[rgb]{0.272,0.415,0.862}{93.35}}$ \\
\cline{3-11}
 &  & 0.20 & 0.73 &  & ${\uline{\textcolor[rgb]{0.389,0.437,0.723}{80.38}}}$ & ${\textcolor[rgb]{0.316,0.423,0.810}{88.49}}$ & ${\textcolor[rgb]{0.316,0.423,0.810}{88.49}}$ & ${\textcolor[rgb]{0.343,0.428,0.779}{85.79}}$ & ${\bf \textcolor[rgb]{0.313,0.423,0.813}{88.86}}$ & ${\textcolor[rgb]{0.316,0.423,0.810}{88.49}}$ \\
\cline{3-11}
 &  & 0.40 & 0.89 &  & ${\uline{\textcolor[rgb]{0.383,0.435,0.730}{81.27}}}$ & ${\textcolor[rgb]{0.334,0.426,0.789}{86.74}}$ & ${\textcolor[rgb]{0.334,0.426,0.789}{86.74}}$ & ${\textcolor[rgb]{0.354,0.430,0.765}{84.37}}$ & ${\bf \textcolor[rgb]{0.316,0.423,0.810}{88.60}}$ & ${\textcolor[rgb]{0.328,0.425,0.796}{87.22}}$ \\
\cline{3-11}
 &  & 0.61 & 0.99 &  & ${\uline{\textcolor[rgb]{0.392,0.437,0.720}{80.23}}}$ & ${\textcolor[rgb]{0.313,0.423,0.813}{89.02}}$ & ${\textcolor[rgb]{0.313,0.423,0.813}{89.02}}$ & ${\textcolor[rgb]{0.302,0.420,0.827}{90.20}}$ & ${\textcolor[rgb]{0.381,0.435,0.734}{81.44}}$ & ${\bf \textcolor[rgb]{0.302,0.420,0.827}{90.21}}$ \\
\cline{1-11} \cline{2-11} \cline{3-11}
\multirow[t]{4}{*}{80} & \multirow[t]{4}{*}{20} & 0.01 & 0.22 &  & ${\uline{\textcolor[rgb]{0.389,0.437,0.723}{80.47}}}$ & ${\textcolor[rgb]{0.340,0.427,0.782}{86.01}}$ & ${\textcolor[rgb]{0.340,0.427,0.782}{86.01}}$ & ${\textcolor[rgb]{0.340,0.427,0.782}{86.00}}$ & ${\textcolor[rgb]{0.375,0.434,0.740}{82.17}}$ & ${\bf \textcolor[rgb]{0.316,0.423,0.810}{88.57}}$ \\
\cline{3-11}
 &  & 0.30 & 0.67 &  & ${\uline{\textcolor[rgb]{0.386,0.436,0.727}{80.98}}}$ & ${\textcolor[rgb]{0.343,0.428,0.779}{85.59}}$ & ${\textcolor[rgb]{0.343,0.428,0.779}{85.59}}$ & ${\textcolor[rgb]{0.343,0.428,0.779}{85.55}}$ & ${\textcolor[rgb]{0.340,0.427,0.782}{85.97}}$ & ${\bf \textcolor[rgb]{0.337,0.427,0.785}{86.22}}$ \\
\cline{3-11}
 &  & 0.60 & 0.92 &  & ${\uline{\textcolor[rgb]{0.381,0.435,0.734}{81.40}}}$ & ${\textcolor[rgb]{0.331,0.426,0.792}{87.01}}$ & ${\textcolor[rgb]{0.331,0.426,0.792}{87.01}}$ & ${\textcolor[rgb]{0.331,0.426,0.792}{86.99}}$ & ${\textcolor[rgb]{0.343,0.428,0.779}{85.75}}$ & ${\bf \textcolor[rgb]{0.322,0.424,0.803}{88.08}}$ \\
\cline{3-11}
 &  & 0.74 & 0.99 &  & ${\textcolor[rgb]{0.378,0.434,0.737}{81.87}}$ & ${\textcolor[rgb]{0.293,0.419,0.837}{91.15}}$ & ${\textcolor[rgb]{0.293,0.419,0.837}{91.15}}$ & ${\textcolor[rgb]{0.293,0.419,0.837}{91.15}}$ & ${\it \uline{\textcolor[rgb]{0.506,0.458,0.585}{67.34}}}$ & ${\bf \textcolor[rgb]{0.287,0.418,0.844}{91.84}}$ \\
\cline{1-11} \cline{2-11} \cline{3-11}
\multirow[t]{4}{*}{20} & \multirow[t]{4}{*}{80} & 0.01 & 0.26 &  & ${\uline{\textcolor[rgb]{0.378,0.434,0.737}{81.87}}}$ & ${\textcolor[rgb]{0.293,0.419,0.837}{91.15}}$ & ${\textcolor[rgb]{0.293,0.419,0.837}{91.15}}$ & ${\textcolor[rgb]{0.293,0.419,0.837}{91.15}}$ & ${\textcolor[rgb]{0.363,0.432,0.754}{83.33}}$ & ${\bf \textcolor[rgb]{0.287,0.418,0.844}{91.84}}$ \\
\cline{3-11}
 &  & 0.30 & 0.67 &  & ${\uline{\textcolor[rgb]{0.383,0.435,0.730}{81.13}}}$ & ${\textcolor[rgb]{0.345,0.428,0.775}{85.53}}$ & ${\textcolor[rgb]{0.345,0.428,0.775}{85.53}}$ & ${\textcolor[rgb]{0.345,0.428,0.775}{85.38}}$ & ${\textcolor[rgb]{0.343,0.428,0.779}{85.78}}$ & ${\bf \textcolor[rgb]{0.340,0.427,0.782}{86.15}}$ \\
\cline{3-11}
 &  & 0.60 & 0.91 &  & ${\uline{\textcolor[rgb]{0.383,0.435,0.730}{81.02}}}$ & ${\textcolor[rgb]{0.340,0.427,0.782}{86.12}}$ & ${\textcolor[rgb]{0.340,0.427,0.782}{86.12}}$ & ${\textcolor[rgb]{0.351,0.430,0.768}{84.89}}$ & ${\textcolor[rgb]{0.383,0.435,0.730}{81.31}}$ & ${\bf \textcolor[rgb]{0.340,0.427,0.782}{86.14}}$ \\
\cline{3-11}
 &  & 0.78 & 0.99 &  & ${\textcolor[rgb]{0.389,0.437,0.723}{80.47}}$ & ${\textcolor[rgb]{0.340,0.427,0.782}{86.01}}$ & ${\textcolor[rgb]{0.340,0.427,0.782}{86.01}}$ & ${\textcolor[rgb]{0.340,0.427,0.782}{86.00}}$ & ${\uline{\textcolor[rgb]{0.413,0.441,0.696}{77.88}}}$ & ${\bf \textcolor[rgb]{0.316,0.423,0.810}{88.57}}$ \\
\cline{1-11} \cline{2-11} \cline{3-11}
\multirow[t]{4}{*}{240} & \multirow[t]{4}{*}{60} & 0.01 & 0.11 &  & ${\textcolor[rgb]{0.345,0.428,0.775}{85.27}}$ & ${\textcolor[rgb]{0.305,0.421,0.824}{89.98}}$ & ${\textcolor[rgb]{0.305,0.421,0.824}{90.02}}$ & ${\textcolor[rgb]{0.305,0.421,0.824}{89.98}}$ & ${\it \uline{\textcolor[rgb]{0.807,0.513,0.228}{33.87}}}$ & ${\bf \textcolor[rgb]{0.302,0.420,0.827}{90.33}}$ \\
\cline{3-11}
 &  & 0.30 & 0.51 &  & ${\uline{\textcolor[rgb]{0.378,0.434,0.737}{81.74}}}$ & ${\textcolor[rgb]{0.351,0.430,0.768}{84.86}}$ & ${\textcolor[rgb]{0.351,0.430,0.768}{84.86}}$ & ${\textcolor[rgb]{0.354,0.430,0.765}{84.44}}$ & ${\textcolor[rgb]{0.351,0.430,0.768}{84.76}}$ & ${\bf \textcolor[rgb]{0.348,0.429,0.772}{84.98}}$ \\
\cline{3-11}
 &  & 0.60 & 0.80 &  & ${\uline{\textcolor[rgb]{0.372,0.433,0.744}{82.44}}}$ & ${\textcolor[rgb]{0.343,0.428,0.779}{85.69}}$ & ${\textcolor[rgb]{0.343,0.428,0.779}{85.75}}$ & ${\textcolor[rgb]{0.345,0.428,0.775}{85.41}}$ & ${\textcolor[rgb]{0.343,0.428,0.779}{85.63}}$ & ${\bf \textcolor[rgb]{0.337,0.427,0.785}{86.20}}$ \\
\cline{3-11}
 &  & 0.88 & 0.99 &  & ${\textcolor[rgb]{0.337,0.427,0.785}{86.49}}$ & ${\textcolor[rgb]{0.287,0.418,0.844}{91.76}}$ & ${\textcolor[rgb]{0.281,0.417,0.851}{92.59}}$ & ${\textcolor[rgb]{0.281,0.417,0.851}{92.57}}$ & ${\uline{\textcolor[rgb]{0.398,0.438,0.713}{79.41}}}$ & ${\bf \textcolor[rgb]{0.275,0.416,0.858}{93.34}}$ \\
\cline{1-11} \cline{2-11} \cline{3-11}
\multirow[t]{4}{*}{60} & \multirow[t]{4}{*}{240} & 0.01 & 0.12 &  & ${\textcolor[rgb]{0.337,0.427,0.785}{86.49}}$ & ${\textcolor[rgb]{0.287,0.418,0.844}{91.76}}$ & ${\textcolor[rgb]{0.281,0.417,0.851}{92.59}}$ & ${\textcolor[rgb]{0.281,0.417,0.851}{92.57}}$ & ${\it \uline{\textcolor[rgb]{0.465,0.451,0.633}{71.93}}}$ & ${\bf \textcolor[rgb]{0.275,0.416,0.858}{93.34}}$ \\
\cline{3-11}
 &  & 0.30 & 0.51 &  & ${\uline{\textcolor[rgb]{0.386,0.436,0.727}{80.92}}}$ & ${\textcolor[rgb]{0.357,0.431,0.761}{84.20}}$ & ${\textcolor[rgb]{0.357,0.431,0.761}{84.20}}$ & ${\textcolor[rgb]{0.357,0.431,0.761}{84.20}}$ & ${\bf \textcolor[rgb]{0.351,0.430,0.768}{84.67}}$ & ${\textcolor[rgb]{0.351,0.430,0.768}{84.66}}$ \\
\cline{3-11}
 &  & 0.60 & 0.80 &  & ${\textcolor[rgb]{0.360,0.431,0.758}{83.71}}$ & ${\textcolor[rgb]{0.334,0.426,0.789}{86.67}}$ & ${\textcolor[rgb]{0.334,0.426,0.789}{86.67}}$ & ${\textcolor[rgb]{0.337,0.427,0.785}{86.44}}$ & ${\uline{\textcolor[rgb]{0.375,0.434,0.740}{82.17}}}$ & ${\bf \textcolor[rgb]{0.331,0.426,0.792}{86.93}}$ \\
\cline{3-11}
 &  & 0.89 & 0.99 &  & ${\textcolor[rgb]{0.345,0.428,0.775}{85.27}}$ & ${\textcolor[rgb]{0.305,0.421,0.824}{89.98}}$ & ${\textcolor[rgb]{0.305,0.421,0.824}{90.02}}$ & ${\textcolor[rgb]{0.305,0.421,0.824}{89.98}}$ & ${\it \uline{\textcolor[rgb]{0.743,0.502,0.304}{41.13}}}$ & ${\bf \textcolor[rgb]{0.302,0.420,0.827}{90.33}}$ \\
\cline{1-11} \cline{2-11} \cline{3-11}

\end{tabular}
  \\  {\footnotesize  FE: Fisher's exact test, B$^*$: Berger and Boos' version of Boschloo's test, FMP$^*$: Berger and Boos' version of Fisher's mid-p value test,  Z$^*_{\text{P}}$: Berger and Boos' version of the Z-pooled test, MPK: maximin power knapsack, APK: average power knapsack}
    \label{tab:power_table_80FET}
\end{table}

\renewcommand{\arraystretch}{1.2} 
\setlength{\tabcolsep}{1.75mm}
\begin{table}[h!]
    \centering
    \caption{Power values~(\%) across treatment group sizes and parameter values (significance level~$2.5\%$). Row maxima (minima) are made bold (underlined). A color gradient (orange to blue) differentiates power values. Values below 75\% are italicized.
}
    \small\begin{tabular}{lllllllllll}
\toprule
 &  &  &  &   & FE & B & UX FMP & UX Z$_{\text{P}}$ & MPK & APK \\
$n_{\text{C}}$ & $n_{\text{D}}$ & $\theta_{\text{C}}$ & $\theta_{\text{D}}$ &  &  &  &  &  &  &  \\
\midrule
\multirow[t]{4}{*}{10} & \multirow[t]{4}{*}{10} & 0.01 & 0.51 &  & ${\it \uline{\textcolor[rgb]{0.483,0.454,0.612}{60.30}}}$ & ${\bf \textcolor[rgb]{0.299,0.420,0.830}{80.08}}$ & ${\bf \textcolor[rgb]{0.299,0.420,0.830}{80.08}}$ & ${\bf \textcolor[rgb]{0.299,0.420,0.830}{80.08}}$ & ${\bf \textcolor[rgb]{0.299,0.420,0.830}{80.08}}$ & ${\bf \textcolor[rgb]{0.299,0.420,0.830}{80.08}}$ \\
\cline{3-11}
 &  & 0.05 & 0.61 &  & ${\it \uline{\textcolor[rgb]{0.439,0.446,0.664}{65.17}}}$ & ${\bf \textcolor[rgb]{0.290,0.418,0.841}{80.99}}$ & ${\bf \textcolor[rgb]{0.290,0.418,0.841}{80.99}}$ & ${\bf \textcolor[rgb]{0.290,0.418,0.841}{80.99}}$ & ${\bf \textcolor[rgb]{0.290,0.418,0.841}{80.99}}$ & ${\bf \textcolor[rgb]{0.290,0.418,0.841}{80.99}}$ \\
\cline{3-11}
 &  & 0.20 & 0.80 &  & ${\it \uline{\textcolor[rgb]{0.454,0.448,0.647}{63.53}}}$ & ${\bf \textcolor[rgb]{0.293,0.419,0.837}{80.54}}$ & ${\bf \textcolor[rgb]{0.293,0.419,0.837}{80.54}}$ & ${\bf \textcolor[rgb]{0.293,0.419,0.837}{80.54}}$ & ${\bf \textcolor[rgb]{0.293,0.419,0.837}{80.54}}$ & ${\bf \textcolor[rgb]{0.293,0.419,0.837}{80.54}}$ \\
\cline{3-11}
 &  & 0.49 & 0.99 &  & ${\it \uline{\textcolor[rgb]{0.483,0.454,0.612}{60.30}}}$ & ${\bf \textcolor[rgb]{0.299,0.420,0.830}{80.08}}$ & ${\bf \textcolor[rgb]{0.299,0.420,0.830}{80.08}}$ & ${\bf \textcolor[rgb]{0.299,0.420,0.830}{80.08}}$ & ${\bf \textcolor[rgb]{0.299,0.420,0.830}{80.08}}$ & ${\bf \textcolor[rgb]{0.299,0.420,0.830}{80.08}}$ \\
\cline{1-11} \cline{2-11} \cline{3-11}
\multirow[t]{4}{*}{25} & \multirow[t]{4}{*}{25} & 0.01 & 0.27 &  & ${\it \uline{\textcolor[rgb]{0.433,0.445,0.671}{65.72}}}$ & ${\textcolor[rgb]{0.325,0.425,0.799}{77.03}}$ & ${\textcolor[rgb]{0.325,0.425,0.799}{77.03}}$ & ${\bf \textcolor[rgb]{0.261,0.413,0.875}{84.08}}$ & ${\textcolor[rgb]{0.293,0.419,0.837}{80.44}}$ & ${\textcolor[rgb]{0.293,0.419,0.837}{80.44}}$ \\
\cline{3-11}
 &  & 0.20 & 0.58 &  & ${\it \uline{\textcolor[rgb]{0.354,0.430,0.765}{74.01}}}$ & ${\textcolor[rgb]{0.302,0.420,0.827}{79.61}}$ & ${\textcolor[rgb]{0.302,0.420,0.827}{79.61}}$ & ${\textcolor[rgb]{0.302,0.420,0.827}{79.61}}$ & ${\bf \textcolor[rgb]{0.293,0.419,0.837}{80.71}}$ & ${\bf \textcolor[rgb]{0.293,0.419,0.837}{80.71}}$ \\
\cline{3-11}
 &  & 0.40 & 0.79 &  & ${\uline{\textcolor[rgb]{0.334,0.426,0.789}{76.35}}}$ & ${\textcolor[rgb]{0.290,0.418,0.841}{80.83}}$ & ${\textcolor[rgb]{0.290,0.418,0.841}{80.83}}$ & ${\textcolor[rgb]{0.290,0.418,0.841}{80.83}}$ & ${\bf \textcolor[rgb]{0.278,0.416,0.855}{82.21}}$ & ${\bf \textcolor[rgb]{0.278,0.416,0.855}{82.21}}$ \\
\cline{3-11}
 &  & 0.73 & 0.99 &  & ${\it \uline{\textcolor[rgb]{0.433,0.445,0.671}{65.72}}}$ & ${\textcolor[rgb]{0.325,0.425,0.799}{77.03}}$ & ${\textcolor[rgb]{0.325,0.425,0.799}{77.03}}$ & ${\bf \textcolor[rgb]{0.261,0.413,0.875}{84.08}}$ & ${\textcolor[rgb]{0.293,0.419,0.837}{80.44}}$ & ${\textcolor[rgb]{0.293,0.419,0.837}{80.44}}$ \\
\cline{1-11} \cline{2-11} \cline{3-11}
\multirow[t]{4}{*}{50} & \multirow[t]{4}{*}{50} & 0.01 & 0.15 &  & ${\it \textcolor[rgb]{0.451,0.448,0.651}{63.67}}$ & ${\textcolor[rgb]{0.337,0.427,0.785}{75.90}}$ & ${\textcolor[rgb]{0.337,0.427,0.785}{76.01}}$ & ${\textcolor[rgb]{0.299,0.420,0.830}{80.01}}$ & ${\it \uline{\textcolor[rgb]{0.532,0.463,0.554}{55.21}}}$ & ${\bf \textcolor[rgb]{0.287,0.418,0.844}{81.13}}$ \\
\cline{3-11}
 &  & 0.30 & 0.58 &  & ${\uline{\textcolor[rgb]{0.325,0.425,0.799}{77.03}}}$ & ${\textcolor[rgb]{0.293,0.419,0.837}{80.59}}$ & ${\textcolor[rgb]{0.293,0.419,0.837}{80.59}}$ & ${\textcolor[rgb]{0.293,0.419,0.837}{80.59}}$ & ${\bf \textcolor[rgb]{0.287,0.418,0.844}{81.14}}$ & ${\textcolor[rgb]{0.290,0.418,0.841}{80.98}}$ \\
\cline{3-11}
 &  & 0.60 & 0.85 &  & ${\uline{\textcolor[rgb]{0.337,0.427,0.785}{75.77}}}$ & ${\textcolor[rgb]{0.305,0.421,0.824}{79.27}}$ & ${\textcolor[rgb]{0.293,0.419,0.837}{80.49}}$ & ${\textcolor[rgb]{0.293,0.419,0.837}{80.49}}$ & ${\textcolor[rgb]{0.302,0.420,0.827}{79.62}}$ & ${\bf \textcolor[rgb]{0.284,0.417,0.848}{81.51}}$ \\
\cline{3-11}
 &  & 0.85 & 0.99 &  & ${\it \uline{\textcolor[rgb]{0.451,0.448,0.651}{63.67}}}$ & ${\textcolor[rgb]{0.337,0.427,0.785}{75.90}}$ & ${\textcolor[rgb]{0.337,0.427,0.785}{76.01}}$ & ${\textcolor[rgb]{0.299,0.420,0.830}{80.01}}$ & ${\textcolor[rgb]{0.299,0.420,0.830}{80.10}}$ & ${\bf \textcolor[rgb]{0.287,0.418,0.844}{81.13}}$ \\
\cline{1-11} \cline{2-11} \cline{3-11}
\multirow[t]{4}{*}{150} & \multirow[t]{4}{*}{150} & 0.01 & 0.07 &  & ${\it \textcolor[rgb]{0.389,0.437,0.723}{70.29}}$ & ${\textcolor[rgb]{0.340,0.427,0.782}{75.55}}$ & ${\textcolor[rgb]{0.310,0.422,0.817}{78.59}}$ & ${\bf \textcolor[rgb]{0.293,0.419,0.837}{80.47}}$ & ${\it \uline{\textcolor[rgb]{1.000,0.549,0.000}{5.29}}}$ & ${\bf \textcolor[rgb]{0.293,0.419,0.837}{80.47}}$ \\
\cline{3-11}
 &  & 0.30 & 0.46 &  & ${\uline{\textcolor[rgb]{0.313,0.423,0.813}{78.55}}}$ & ${\textcolor[rgb]{0.284,0.417,0.848}{81.63}}$ & ${\textcolor[rgb]{0.284,0.417,0.848}{81.63}}$ & ${\textcolor[rgb]{0.284,0.417,0.848}{81.63}}$ & ${\textcolor[rgb]{0.287,0.418,0.844}{81.15}}$ & ${\bf \textcolor[rgb]{0.281,0.417,0.851}{81.83}}$ \\
\cline{3-11}
 &  & 0.60 & 0.76 &  & ${\uline{\textcolor[rgb]{0.281,0.417,0.851}{81.93}}}$ & ${\textcolor[rgb]{0.258,0.412,0.879}{84.26}}$ & ${\textcolor[rgb]{0.258,0.412,0.879}{84.26}}$ & ${\textcolor[rgb]{0.258,0.412,0.879}{84.28}}$ & ${\textcolor[rgb]{0.270,0.414,0.865}{83.14}}$ & ${\bf \textcolor[rgb]{0.255,0.412,0.882}{84.76}}$ \\
\cline{3-11}
 &  & 0.93 & 0.99 &  & ${\it \textcolor[rgb]{0.389,0.437,0.723}{70.29}}$ & ${\textcolor[rgb]{0.340,0.427,0.782}{75.55}}$ & ${\textcolor[rgb]{0.310,0.422,0.817}{78.59}}$ & ${\bf \textcolor[rgb]{0.293,0.419,0.837}{80.47}}$ & ${\it \uline{\textcolor[rgb]{0.787,0.510,0.253}{28.14}}}$ & ${\bf \textcolor[rgb]{0.293,0.419,0.837}{80.47}}$ \\
\cline{1-11} \cline{2-11} \cline{3-11}
\multirow[t]{4}{*}{16} & \multirow[t]{4}{*}{4} & 0.01 & 0.63 &  & ${\it \uline{\textcolor[rgb]{0.559,0.468,0.522}{52.36}}}$ & ${\bf \textcolor[rgb]{0.293,0.419,0.837}{80.50}}$ & ${\bf \textcolor[rgb]{0.293,0.419,0.837}{80.50}}$ & ${\bf \textcolor[rgb]{0.293,0.419,0.837}{80.50}}$ & ${\bf \textcolor[rgb]{0.293,0.419,0.837}{80.50}}$ & ${\bf \textcolor[rgb]{0.293,0.419,0.837}{80.50}}$ \\
\cline{3-11}
 &  & 0.05 & 0.74 &  & ${\it \uline{\textcolor[rgb]{0.448,0.447,0.654}{64.13}}}$ & ${\bf \textcolor[rgb]{0.299,0.420,0.830}{80.10}}$ & ${\bf \textcolor[rgb]{0.299,0.420,0.830}{80.10}}$ & ${\bf \textcolor[rgb]{0.299,0.420,0.830}{80.10}}$ & ${\bf \textcolor[rgb]{0.299,0.420,0.830}{80.10}}$ & ${\bf \textcolor[rgb]{0.299,0.420,0.830}{80.10}}$ \\
\cline{3-11}
 &  & 0.10 & 0.83 &  & ${\it \uline{\textcolor[rgb]{0.424,0.443,0.682}{66.66}}}$ & ${\bf \textcolor[rgb]{0.296,0.419,0.834}{80.34}}$ & ${\bf \textcolor[rgb]{0.296,0.419,0.834}{80.34}}$ & ${\bf \textcolor[rgb]{0.296,0.419,0.834}{80.34}}$ & ${\bf \textcolor[rgb]{0.296,0.419,0.834}{80.34}}$ & ${\bf \textcolor[rgb]{0.296,0.419,0.834}{80.34}}$ \\
\cline{3-11}
 &  & 0.29 & 0.99 &  & ${\it \uline{\textcolor[rgb]{0.611,0.477,0.460}{46.74}}}$ & ${\bf \textcolor[rgb]{0.281,0.417,0.851}{81.85}}$ & ${\bf \textcolor[rgb]{0.281,0.417,0.851}{81.85}}$ & ${\bf \textcolor[rgb]{0.281,0.417,0.851}{81.85}}$ & ${\bf \textcolor[rgb]{0.281,0.417,0.851}{81.85}}$ & ${\bf \textcolor[rgb]{0.281,0.417,0.851}{81.85}}$ \\
\cline{1-11} \cline{2-11} \cline{3-11}
\multirow[t]{4}{*}{4} & \multirow[t]{4}{*}{16} & 0.01 & 0.71 &  & ${\it \uline{\textcolor[rgb]{0.611,0.477,0.460}{46.74}}}$ & ${\bf \textcolor[rgb]{0.281,0.417,0.851}{81.85}}$ & ${\bf \textcolor[rgb]{0.281,0.417,0.851}{81.85}}$ & ${\bf \textcolor[rgb]{0.281,0.417,0.851}{81.85}}$ & ${\bf \textcolor[rgb]{0.281,0.417,0.851}{81.85}}$ & ${\bf \textcolor[rgb]{0.281,0.417,0.851}{81.85}}$ \\
\cline{3-11}
 &  & 0.05 & 0.77 &  & ${\it \uline{\textcolor[rgb]{0.500,0.457,0.592}{58.60}}}$ & ${\bf \textcolor[rgb]{0.284,0.417,0.848}{81.42}}$ & ${\bf \textcolor[rgb]{0.284,0.417,0.848}{81.42}}$ & ${\bf \textcolor[rgb]{0.284,0.417,0.848}{81.42}}$ & ${\bf \textcolor[rgb]{0.284,0.417,0.848}{81.42}}$ & ${\bf \textcolor[rgb]{0.284,0.417,0.848}{81.42}}$ \\
\cline{3-11}
 &  & 0.10 & 0.84 &  & ${\it \uline{\textcolor[rgb]{0.427,0.444,0.678}{66.38}}}$ & ${\bf \textcolor[rgb]{0.293,0.419,0.837}{80.43}}$ & ${\bf \textcolor[rgb]{0.293,0.419,0.837}{80.43}}$ & ${\bf \textcolor[rgb]{0.293,0.419,0.837}{80.43}}$ & ${\bf \textcolor[rgb]{0.293,0.419,0.837}{80.43}}$ & ${\bf \textcolor[rgb]{0.293,0.419,0.837}{80.43}}$ \\
\cline{3-11}
 &  & 0.37 & 0.99 &  & ${\it \uline{\textcolor[rgb]{0.559,0.468,0.522}{52.36}}}$ & ${\bf \textcolor[rgb]{0.293,0.419,0.837}{80.50}}$ & ${\bf \textcolor[rgb]{0.293,0.419,0.837}{80.50}}$ & ${\bf \textcolor[rgb]{0.293,0.419,0.837}{80.50}}$ & ${\bf \textcolor[rgb]{0.293,0.419,0.837}{80.50}}$ & ${\bf \textcolor[rgb]{0.293,0.419,0.837}{80.50}}$ \\
\cline{1-11} \cline{2-11} \cline{3-11}
\multirow[t]{4}{*}{40} & \multirow[t]{4}{*}{10} & 0.01 & 0.32 &  & ${\it \uline{\textcolor[rgb]{0.436,0.445,0.668}{65.25}}}$ & ${\textcolor[rgb]{0.305,0.421,0.824}{79.36}}$ & ${\bf \textcolor[rgb]{0.290,0.418,0.841}{80.77}}$ & ${\bf \textcolor[rgb]{0.290,0.418,0.841}{80.77}}$ & ${\it \textcolor[rgb]{0.424,0.443,0.682}{66.67}}$ & ${\bf \textcolor[rgb]{0.290,0.418,0.841}{80.77}}$ \\
\cline{3-11}
 &  & 0.20 & 0.68 &  & ${\it \uline{\textcolor[rgb]{0.372,0.433,0.744}{72.22}}}$ & ${\textcolor[rgb]{0.287,0.418,0.844}{81.20}}$ & ${\bf \textcolor[rgb]{0.287,0.418,0.844}{81.21}}$ & ${\textcolor[rgb]{0.310,0.422,0.817}{78.85}}$ & ${\textcolor[rgb]{0.290,0.418,0.841}{81.03}}$ & ${\bf \textcolor[rgb]{0.287,0.418,0.844}{81.21}}$ \\
\cline{3-11}
 &  & 0.40 & 0.87 &  & ${\it \uline{\textcolor[rgb]{0.383,0.435,0.730}{70.90}}}$ & ${\textcolor[rgb]{0.293,0.419,0.837}{80.64}}$ & ${\textcolor[rgb]{0.293,0.419,0.837}{80.64}}$ & ${\textcolor[rgb]{0.328,0.425,0.796}{76.82}}$ & ${\bf \textcolor[rgb]{0.278,0.416,0.855}{82.03}}$ & ${\textcolor[rgb]{0.293,0.419,0.837}{80.64}}$ \\
\cline{3-11}
 &  & 0.65 & 0.99 &  & ${\it \uline{\textcolor[rgb]{0.573,0.470,0.505}{50.91}}}$ & ${\it \textcolor[rgb]{0.363,0.432,0.754}{73.01}}$ & ${\it \textcolor[rgb]{0.363,0.432,0.754}{73.01}}$ & ${\it \textcolor[rgb]{0.570,0.470,0.509}{51.22}}$ & ${\it \textcolor[rgb]{0.559,0.468,0.522}{52.42}}$ & ${\bf \textcolor[rgb]{0.293,0.419,0.837}{80.58}}$ \\
\cline{1-11} \cline{2-11} \cline{3-11}
\multirow[t]{4}{*}{10} & \multirow[t]{4}{*}{40} & 0.01 & 0.35 &  & ${\it \uline{\textcolor[rgb]{0.573,0.470,0.505}{50.91}}}$ & ${\it \textcolor[rgb]{0.363,0.432,0.754}{73.01}}$ & ${\it \textcolor[rgb]{0.363,0.432,0.754}{73.01}}$ & ${\it \textcolor[rgb]{0.570,0.470,0.509}{51.22}}$ & ${\it \textcolor[rgb]{0.559,0.468,0.522}{52.42}}$ & ${\bf \textcolor[rgb]{0.293,0.419,0.837}{80.58}}$ \\
\cline{3-11}
 &  & 0.20 & 0.68 &  & ${\it \uline{\textcolor[rgb]{0.395,0.438,0.716}{69.60}}}$ & ${\textcolor[rgb]{0.299,0.420,0.830}{80.07}}$ & ${\textcolor[rgb]{0.299,0.420,0.830}{80.07}}$ & ${\textcolor[rgb]{0.319,0.424,0.806}{77.79}}$ & ${\bf \textcolor[rgb]{0.281,0.417,0.851}{81.78}}$ & ${\textcolor[rgb]{0.299,0.420,0.830}{80.07}}$ \\
\cline{3-11}
 &  & 0.40 & 0.86 &  & ${\it \uline{\textcolor[rgb]{0.366,0.432,0.751}{72.79}}}$ & ${\textcolor[rgb]{0.299,0.420,0.830}{80.07}}$ & ${\textcolor[rgb]{0.296,0.419,0.834}{80.28}}$ & ${\textcolor[rgb]{0.302,0.420,0.827}{79.61}}$ & ${\bf \textcolor[rgb]{0.275,0.416,0.858}{82.34}}$ & ${\textcolor[rgb]{0.296,0.419,0.834}{80.28}}$ \\
\cline{3-11}
 &  & 0.68 & 0.99 &  & ${\it \uline{\textcolor[rgb]{0.436,0.445,0.668}{65.25}}}$ & ${\textcolor[rgb]{0.305,0.421,0.824}{79.36}}$ & ${\bf \textcolor[rgb]{0.290,0.418,0.841}{80.77}}$ & ${\bf \textcolor[rgb]{0.290,0.418,0.841}{80.77}}$ & ${\it \textcolor[rgb]{0.424,0.443,0.682}{66.67}}$ & ${\bf \textcolor[rgb]{0.290,0.418,0.841}{80.77}}$ \\
\cline{1-11} \cline{2-11} \cline{3-11}
\multirow[t]{4}{*}{80} & \multirow[t]{4}{*}{20} & 0.01 & 0.19 &  & ${\it \uline{\textcolor[rgb]{0.383,0.435,0.730}{70.95}}}$ & ${\textcolor[rgb]{0.310,0.422,0.817}{78.71}}$ & ${\textcolor[rgb]{0.310,0.422,0.817}{78.71}}$ & ${\bf \textcolor[rgb]{0.281,0.417,0.851}{81.85}}$ & ${\it \textcolor[rgb]{0.360,0.431,0.758}{73.36}}$ & ${\bf \textcolor[rgb]{0.281,0.417,0.851}{81.85}}$ \\
\cline{3-11}
 &  & 0.30 & 0.65 &  & ${\uline{\textcolor[rgb]{0.337,0.427,0.785}{75.96}}}$ & ${\textcolor[rgb]{0.287,0.418,0.844}{81.31}}$ & ${\textcolor[rgb]{0.287,0.418,0.844}{81.31}}$ & ${\textcolor[rgb]{0.328,0.425,0.796}{76.96}}$ & ${\textcolor[rgb]{0.281,0.417,0.851}{81.84}}$ & ${\bf \textcolor[rgb]{0.278,0.416,0.855}{82.09}}$ \\
\cline{3-11}
 &  & 0.60 & 0.90 &  & ${\it \textcolor[rgb]{0.366,0.432,0.751}{72.74}}$ & ${\textcolor[rgb]{0.302,0.420,0.827}{79.72}}$ & ${\textcolor[rgb]{0.302,0.420,0.827}{79.77}}$ & ${\it \uline{\textcolor[rgb]{0.401,0.439,0.709}{69.13}}}$ & ${\textcolor[rgb]{0.305,0.421,0.824}{79.38}}$ & ${\bf \textcolor[rgb]{0.287,0.418,0.844}{81.14}}$ \\
\cline{3-11}
 &  & 0.79 & 0.99 &  & ${\it \textcolor[rgb]{0.550,0.466,0.533}{53.23}}$ & ${\it \textcolor[rgb]{0.398,0.438,0.713}{69.34}}$ & ${\textcolor[rgb]{0.337,0.427,0.785}{76.02}}$ & ${\it \textcolor[rgb]{0.641,0.483,0.426}{43.75}}$ & ${\it \uline{\textcolor[rgb]{0.731,0.500,0.318}{34.04}}}$ & ${\bf \textcolor[rgb]{0.299,0.420,0.830}{80.00}}$ \\
\cline{1-11} \cline{2-11} \cline{3-11}
\multirow[t]{4}{*}{20} & \multirow[t]{4}{*}{80} & 0.01 & 0.21 &  & ${\it \textcolor[rgb]{0.550,0.466,0.533}{53.23}}$ & ${\it \textcolor[rgb]{0.398,0.438,0.713}{69.34}}$ & ${\textcolor[rgb]{0.337,0.427,0.785}{76.02}}$ & ${\it \uline{\textcolor[rgb]{0.641,0.483,0.426}{43.75}}}$ & ${\textcolor[rgb]{0.331,0.426,0.792}{76.43}}$ & ${\bf \textcolor[rgb]{0.299,0.420,0.830}{80.00}}$ \\
\cline{3-11}
 &  & 0.30 & 0.65 &  & ${\textcolor[rgb]{0.334,0.426,0.789}{76.18}}$ & ${\textcolor[rgb]{0.287,0.418,0.844}{81.13}}$ & ${\textcolor[rgb]{0.287,0.418,0.844}{81.13}}$ & ${\uline{\textcolor[rgb]{0.340,0.427,0.782}{75.72}}}$ & ${\textcolor[rgb]{0.284,0.417,0.848}{81.52}}$ & ${\bf \textcolor[rgb]{0.281,0.417,0.851}{81.86}}$ \\
\cline{3-11}
 &  & 0.60 & 0.90 &  & ${\uline{\textcolor[rgb]{0.322,0.424,0.803}{77.36}}}$ & ${\textcolor[rgb]{0.281,0.417,0.851}{81.92}}$ & ${\textcolor[rgb]{0.272,0.415,0.862}{82.84}}$ & ${\textcolor[rgb]{0.275,0.416,0.858}{82.57}}$ & ${\textcolor[rgb]{0.310,0.422,0.817}{78.77}}$ & ${\bf \textcolor[rgb]{0.272,0.415,0.862}{82.85}}$ \\
\cline{3-11}
 &  & 0.81 & 0.99 &  & ${\it \textcolor[rgb]{0.383,0.435,0.730}{70.95}}$ & ${\textcolor[rgb]{0.310,0.422,0.817}{78.71}}$ & ${\textcolor[rgb]{0.310,0.422,0.817}{78.71}}$ & ${\bf \textcolor[rgb]{0.281,0.417,0.851}{81.85}}$ & ${\it \uline{\textcolor[rgb]{0.398,0.438,0.713}{69.27}}}$ & ${\bf \textcolor[rgb]{0.281,0.417,0.851}{81.85}}$ \\
\cline{1-11} \cline{2-11} \cline{3-11}
\multirow[t]{4}{*}{240} & \multirow[t]{4}{*}{60} & 0.01 & 0.09 &  & ${\it \textcolor[rgb]{0.372,0.433,0.744}{72.13}}$ & ${\textcolor[rgb]{0.305,0.421,0.824}{79.27}}$ & ${\textcolor[rgb]{0.299,0.420,0.830}{79.85}}$ & ${\textcolor[rgb]{0.299,0.420,0.830}{79.81}}$ & ${\it \uline{\textcolor[rgb]{0.886,0.528,0.135}{17.68}}}$ & ${\bf \textcolor[rgb]{0.296,0.419,0.834}{80.24}}$ \\
\cline{3-11}
 &  & 0.30 & 0.50 &  & ${\textcolor[rgb]{0.319,0.424,0.806}{77.84}}$ & ${\textcolor[rgb]{0.284,0.417,0.848}{81.38}}$ & ${\textcolor[rgb]{0.284,0.417,0.848}{81.38}}$ & ${\it \uline{\textcolor[rgb]{0.354,0.430,0.765}{74.02}}}$ & ${\textcolor[rgb]{0.287,0.418,0.844}{81.21}}$ & ${\bf \textcolor[rgb]{0.284,0.417,0.848}{81.52}}$ \\
\cline{3-11}
 &  & 0.60 & 0.79 &  & ${\textcolor[rgb]{0.322,0.424,0.803}{77.58}}$ & ${\textcolor[rgb]{0.284,0.417,0.848}{81.39}}$ & ${\textcolor[rgb]{0.281,0.417,0.851}{81.78}}$ & ${\it \uline{\textcolor[rgb]{0.381,0.435,0.734}{71.33}}}$ & ${\textcolor[rgb]{0.284,0.417,0.848}{81.41}}$ & ${\bf \textcolor[rgb]{0.278,0.416,0.855}{81.98}}$ \\
\cline{3-11}
 &  & 0.90 & 0.99 &  & ${\it \textcolor[rgb]{0.386,0.436,0.727}{70.59}}$ & ${\textcolor[rgb]{0.337,0.427,0.785}{75.99}}$ & ${\textcolor[rgb]{0.278,0.416,0.855}{82.07}}$ & ${\it \textcolor[rgb]{0.530,0.462,0.557}{55.40}}$ & ${\it \uline{\textcolor[rgb]{0.600,0.475,0.474}{47.89}}}$ & ${\bf \textcolor[rgb]{0.270,0.414,0.865}{82.96}}$ \\
\cline{1-11} \cline{2-11} \cline{3-11}
\multirow[t]{4}{*}{60} & \multirow[t]{4}{*}{240} & 0.01 & 0.10 &  & ${\it \textcolor[rgb]{0.386,0.436,0.727}{70.59}}$ & ${\textcolor[rgb]{0.337,0.427,0.785}{75.99}}$ & ${\textcolor[rgb]{0.278,0.416,0.855}{82.07}}$ & ${\it \textcolor[rgb]{0.530,0.462,0.557}{55.40}}$ & ${\it \uline{\textcolor[rgb]{0.699,0.494,0.356}{37.30}}}$ & ${\bf \textcolor[rgb]{0.270,0.414,0.865}{82.96}}$ \\
\cline{3-11}
 &  & 0.30 & 0.50 &  & ${\textcolor[rgb]{0.328,0.425,0.796}{76.88}}$ & ${\textcolor[rgb]{0.290,0.418,0.841}{81.01}}$ & ${\textcolor[rgb]{0.290,0.418,0.841}{81.01}}$ & ${\it \uline{\textcolor[rgb]{0.375,0.434,0.740}{71.76}}}$ & ${\textcolor[rgb]{0.290,0.418,0.841}{81.03}}$ & ${\bf \textcolor[rgb]{0.287,0.418,0.844}{81.10}}$ \\
\cline{3-11}
 &  & 0.60 & 0.79 &  & ${\textcolor[rgb]{0.307,0.421,0.820}{79.11}}$ & ${\textcolor[rgb]{0.272,0.415,0.862}{82.66}}$ & ${\textcolor[rgb]{0.272,0.415,0.862}{82.67}}$ & ${\uline{\textcolor[rgb]{0.328,0.425,0.796}{76.86}}}$ & ${\textcolor[rgb]{0.307,0.421,0.820}{79.08}}$ & ${\bf \textcolor[rgb]{0.270,0.414,0.865}{82.90}}$ \\
\cline{3-11}
 &  & 0.91 & 0.99 &  & ${\it \textcolor[rgb]{0.372,0.433,0.744}{72.13}}$ & ${\textcolor[rgb]{0.305,0.421,0.824}{79.27}}$ & ${\textcolor[rgb]{0.299,0.420,0.830}{79.85}}$ & ${\textcolor[rgb]{0.299,0.420,0.830}{79.81}}$ & ${\it \uline{\textcolor[rgb]{0.798,0.512,0.239}{26.88}}}$ & ${\bf \textcolor[rgb]{0.296,0.419,0.834}{80.24}}$ \\
\cline{1-11} \cline{2-11} \cline{3-11}

\end{tabular}
  \\  {\footnotesize  FE: Fisher's exact test, B:  Boschloo's test, UX FMP: unconditional exact Fisher's mid-P test, UX Z$_{\text{P}}$: unconditional exact Z-pooled test, MPK: maximin power knapsack, APK: average power knapsack}
    \label{tab:power_table_UX}
\end{table}
\FloatBarrier

   \renewcommand{\arraystretch}{1.2} 
\begin{table*}[h!]
\centering
    \addtolength{\tabcolsep}{-0.3em}
      \caption{ Summary table of power comparisons. The pointwise power comparison rows show the fraction (\%) of parameters for which the test in the column has strictly higher power than the APK test~(left) and the frequency of 
    parameters for which the APK test has strictly higher power~(right), where we consider all parameters~$\btheta$ such that~$\theta_\D>\theta_\C$ and both~$\theta_\C,\theta_\D$ are multiples of~0.01.   The average power comparison shows the difference in average power (\%) for the APK test over the column test, and in brackets the absolute difference in average power (\%) for parameters where the column test has higher power (left) and for parameters where the APK test has higher power (right), where ``$-$'' indicates there are no such parameter values. 
The one-sided significance level equals~$2.5\%$.}\label{tab:comp_tests_supp}
    \hspace*{-8mm}\footnotesize\begin{tabular}{llllllll}
\toprule
 &  &  & FE & B$^*$ & FMP$^*$ & Z$_{\text{P}}^*$ & MPK \\
$n$ & $(n_{\text{C}}, n_{\text{D}})$ & Power comparison &  &  &  &  &  \\
\midrule
\multirow[t]{6}{*}{20} & \multirow[t]{2}{*}{(10, 10)} & \% Pointwise higher (column/APK) & ${ 0.00/ 99.98}$ & ${ 0.00/ 0.00}$ & ${ 0.00/ 0.00}$ & ${ 0.00/ 0.00}$ & ${ 0.00/ 0.00}$ \\
 &  & Avg. diff. total (APK worse/better) & \textcolor[RGB]{255,140,0}{${9.84~(-/9.85)}$} & \textcolor[RGB]{65,105,225}{${0.00~(-/-)}$} & \textcolor[RGB]{65,105,225}{${0.00~(-/-)}$} & \textcolor[RGB]{65,105,225}{${0.00~(-/-)}$} & \textcolor[RGB]{65,105,225}{${0.00~(-/-)}$} \\
\cline{2-8}
 & \multirow[t]{2}{*}{(7, 13)} & \% Pointwise higher (column/APK) & ${ 0.00/ 99.98}$ & ${ 0.00/ 0.00}$ & ${ 0.00/ 0.00}$ & ${ 0.00/ 0.00}$ & ${ 0.00/ 0.00}$ \\
 &  & Avg. diff. total (APK worse/better) & \textcolor[RGB]{235,136,22}{${8.89~(-/8.90)}$} & \textcolor[RGB]{65,105,225}{${0.00~(-/-)}$} & \textcolor[RGB]{65,105,225}{${0.00~(-/-)}$} & \textcolor[RGB]{65,105,225}{${0.00~(-/-)}$} & \textcolor[RGB]{65,105,225}{${0.00~(-/-)}$} \\
\cline{2-8}
 & \multirow[t]{2}{*}{(4, 16)} & \% Pointwise higher (column/APK) & ${ 0.00/ 99.98}$ & ${ 0.00/ 0.00}$ & ${ 0.00/ 0.00}$ & ${ 0.00/ 0.00}$ & ${ 0.00/ 0.00}$ \\
 &  & Avg. diff. total (APK worse/better) & \textcolor[RGB]{235,136,22}{${9.53~(-/9.53)}$} & \textcolor[RGB]{65,105,225}{${0.00~(-/-)}$} & \textcolor[RGB]{65,105,225}{${0.00~(-/-)}$} & \textcolor[RGB]{65,105,225}{${0.00~(-/-)}$} & \textcolor[RGB]{65,105,225}{${0.00~(-/-)}$} \\
\cline{1-8} \cline{2-8}
\multirow[t]{6}{*}{50} & \multirow[t]{2}{*}{(25, 25)} & \% Pointwise higher (column/APK) & ${ 0.00/ 99.43}$ & ${ 0.00/ 95.82}$ & ${ 0.00/ 95.82}$ & ${ 9.23/ 90.12}$ & ${ 0.00/ 0.00}$ \\
 &  & Avg. diff. total (APK worse/better) & \textcolor[RGB]{160,122,112}{${4.66~(-/4.69)}$} & \textcolor[RGB]{84,108,202}{${1.14~(-/1.19)}$} & \textcolor[RGB]{84,108,202}{${1.14~(-/1.19)}$} & \textcolor[RGB]{84,108,202}{${0.49~(4.04/0.95)}$} & \textcolor[RGB]{65,105,225}{${0.00~(-/-)}$} \\
\cline{2-8}
 & \multirow[t]{2}{*}{(17, 33)} & \% Pointwise higher (column/APK) & ${ 0.00/ 99.64}$ & ${ 0.00/ 95.90}$ & ${ 0.00/ 95.90}$ & ${ 0.00/ 95.90}$ & ${ 55.35/ 33.33}$ \\
 &  & Avg. diff. total (APK worse/better) & \textcolor[RGB]{179,126,90}{${6.14~(-/6.16)}$} & \textcolor[RGB]{84,108,202}{${0.63~(-/0.66)}$} & \textcolor[RGB]{84,108,202}{${0.63~(-/0.66)}$} & \textcolor[RGB]{84,108,202}{${0.63~(-/0.66)}$} & \textcolor[RGB]{84,108,202}{${0.64~(0.09/2.06)}$} \\
\cline{2-8}
 & \multirow[t]{2}{*}{(10, 40)} & \% Pointwise higher (column/APK) & ${ 0.00/ 99.78}$ & ${ 0.00/ 97.76}$ & ${ 0.00/ 97.76}$ & ${ 0.00/ 97.76}$ & ${ 58.97/ 40.81}$ \\
 &  & Avg. diff. total (APK worse/better) & \textcolor[RGB]{198,129,67}{${6.63~(-/6.64)}$} & \textcolor[RGB]{84,108,202}{${0.88~(-/0.90)}$} & \textcolor[RGB]{84,108,202}{${0.88~(-/0.90)}$} & \textcolor[RGB]{103,112,179}{${2.17~(-/2.22)}$} & \textcolor[RGB]{84,108,202}{${1.03~(1.07/4.07)}$} \\
\cline{1-8} \cline{2-8}
\multirow[t]{6}{*}{100} & \multirow[t]{2}{*}{(50, 50)} & \% Pointwise higher (column/APK) & ${ 0.00/ 95.17}$ & ${ 2.40/ 92.06}$ & ${ 3.07/ 91.21}$ & ${ 3.07/ 87.50}$ & ${ 18.48/ 76.57}$ \\
 &  & Avg. diff. total (APK worse/better) & \textcolor[RGB]{122,115,157}{${3.13~(-/3.29)}$} & \textcolor[RGB]{84,108,202}{${0.85~(0.00/0.93)}$} & \textcolor[RGB]{84,108,202}{${0.56~(0.00/0.61)}$} & \textcolor[RGB]{65,105,225}{${0.22~(0.00/0.25)}$} & \textcolor[RGB]{103,112,179}{${2.31~(0.27/3.08)}$} \\
\cline{2-8}
 & \multirow[t]{2}{*}{(35, 65)} & \% Pointwise higher (column/APK) & ${ 0.00/ 96.12}$ & ${ 5.56/ 88.32}$ & ${ 5.56/ 87.11}$ & ${ 0.38/ 92.18}$ & ${ 12.77/ 83.27}$ \\
 &  & Avg. diff. total (APK worse/better) & \textcolor[RGB]{122,115,157}{${3.23~(-/3.36)}$} & \textcolor[RGB]{65,105,225}{${0.50~(0.00/0.56)}$} & \textcolor[RGB]{65,105,225}{${0.40~(0.00/0.46)}$} & \textcolor[RGB]{65,105,225}{${0.40~(0.00/0.44)}$} & \textcolor[RGB]{122,115,157}{${3.16~(0.07/3.80)}$} \\
\cline{2-8}
 & \multirow[t]{2}{*}{(20, 80)} & \% Pointwise higher (column/APK) & ${ 0.00/ 98.48}$ & ${ 8.83/ 87.39}$ & ${ 8.83/ 87.39}$ & ${ 0.00/ 96.50}$ & ${ 27.64/ 67.64}$ \\
 &  & Avg. diff. total (APK worse/better) & \textcolor[RGB]{141,119,135}{${4.09~(-/4.15)}$} & \textcolor[RGB]{84,108,202}{${0.58~(0.19/0.68)}$} & \textcolor[RGB]{84,108,202}{${0.58~(0.19/0.68)}$} & \textcolor[RGB]{84,108,202}{${1.03~(-/1.07)}$} & \textcolor[RGB]{103,112,179}{${1.71~(0.28/2.65)}$} \\
\cline{1-8} \cline{2-8}
\multirow[t]{6}{*}{300} & \multirow[t]{2}{*}{(150, 150)} & \% Pointwise higher (column/APK) & ${ 0.00/ 73.98}$ & ${ 0.00/ 73.07}$ & ${ 0.00/ 72.93}$ & ${ 0.00/ 71.74}$ & ${ 7.03/ 67.64}$ \\
 &  & Avg. diff. total (APK worse/better) & \textcolor[RGB]{84,108,202}{${1.08~(-/1.46)}$} & \textcolor[RGB]{65,105,225}{${0.26~(-/0.36)}$} & \textcolor[RGB]{65,105,225}{${0.23~(-/0.31)}$} & \textcolor[RGB]{65,105,225}{${0.18~(-/0.25)}$} & \textcolor[RGB]{103,112,179}{${2.01~(0.06/2.98)}$} \\
\cline{2-8}
 & \multirow[t]{2}{*}{(100, 200)} & \% Pointwise higher (column/APK) & ${ 0.00/ 76.85}$ & ${ 0.00/ 74.87}$ & ${ 2.00/ 72.67}$ & ${ 0.00/ 74.85}$ & ${ 7.23/ 69.96}$ \\
 &  & Avg. diff. total (APK worse/better) & \textcolor[RGB]{84,108,202}{${1.28~(-/1.67)}$} & \textcolor[RGB]{65,105,225}{${0.26~(-/0.34)}$} & \textcolor[RGB]{65,105,225}{${0.16~(0.05/0.22)}$} & \textcolor[RGB]{65,105,225}{${0.24~(-/0.32)}$} & \textcolor[RGB]{103,112,179}{${1.84~(0.03/2.63)}$} \\
\cline{2-8}
 & \multirow[t]{2}{*}{(60, 240)} & \% Pointwise higher (column/APK) & ${ 0.00/ 84.18}$ & ${ 1.01/ 80.69}$ & ${ 5.13/ 76.08}$ & ${ 0.99/ 81.05}$ & ${ 8.51/ 74.97}$ \\
 &  & Avg. diff. total (APK worse/better) & \textcolor[RGB]{103,112,179}{${1.78~(-/2.11)}$} & \textcolor[RGB]{65,105,225}{${0.25~(0.00/0.31)}$} & \textcolor[RGB]{65,105,225}{${0.17~(0.02/0.22)}$} & \textcolor[RGB]{65,105,225}{${0.32~(0.00/0.39)}$} & \textcolor[RGB]{103,112,179}{${1.89~(0.02/2.52)}$} \\
\cline{1-8} \cline{2-8}

\end{tabular}
 \end{table*}{\footnotesize  FE: Fisher's exact test, B$^*$: Berger and Boos' version of Boschloo's test, FMP$^*$: Berger and Boos' version of Fisher's mid-p value test,  Z$^*_{\text{P}}$: Berger and Boos' version of the Z-pooled test, MPK: maximin power knapsack, APK: average power knapsack}

\renewcommand{\arraystretch}{0.5} 
\setlength{\tabcolsep}{0.5mm}

\autoref{tab:power_table_UX}  shows a power comparison similar to \autoref{tab:comparison_table} (again, the APK test reaches a power of~$80\%$), where in this case we evaluate the unconditional exact~(UX) versions of the non-knapsack-based tests instead of Berger and Boos' version of each test. 
Mainly the same conclusion can be drawn as from Table~1 in the paper. It can be argued that the UX FMP test does slightly better than the FMP$^*$ test in terms of power, but nothing substantial changes when comparing it to the APK test. The UX Z$_\text{P}$ test has substantially worse power than the Z$_\text{P}^*$ test, where for larger trial sizes it often shows the lowest power.  

\autoref{tab:comp_tests_supp} shows a similar comparison as in Table~2 of the paper, but now includes the B$^*$ and MPK tests. After a sample size of 100, the B$^*$ test is more often outperformed by the APK test than FMP$^*$ test in terms of power. For $n=50$ and unbalanced group sizes, the MPK test more often outperforms the APK test in terms of power than comparators, but on average the absolute power differences are smaller when the MPK test outperforms the APK test than vice versa, and for~$n>50$ the APK test often outperforms the MPK test.

The MPK test's variable power can be explained by this test raising the power at the least favorable parameter configuration, which occurs typically at a control success rate around 30-60\%. From~\autoref{fig:T1E}, we can see that around these control success rates, the type I error rate for MPK is closer to~$\alpha=0.025$ than for the APK test. By focusing on the setting where the outcome variance is roughly highest, the MPK test exhausts the critical region to slightly lift the power for the worst-case scenario and omits rejection decisions that yield broader power gains across the alternative parameter space, which are the test decisions included by the APK test.
\FloatBarrier
\subsection{Additional results for the  Merck Research Laboratories trial Case study}\label{case_study}

This section presents additional results for the   Merck Research Laboratories trial Case study described in \autoref{sect:application}. Next to the FE, FMP$^*$, Z$^*_\text{P}$, APK and MPK tests, we consider the WAPK test ($\balpha=(140,131), \bm\beta = (8,1)$) a second MPK (MPK2) test with~$\theta_\C\in[0.896,1-(131/132-140/148)]$ (grid size 100, based on a 95\% Clopper-Pearson interval and the true effect size) and $\theta_\D= \theta_\C + (131/132-140/148)$. Lastly, we also considered the 
simple (alternative) hypothesis knapsack~(SHK) with ~$\btheta_1=(140/148,131/132)$. The extension of \autoref{table_pvalues} is given in \autoref{table_pvalues_extra}. The  MPK and MPK2 tests lead to a p-value higher than 2.5\% and hence would not have resulted in a rejection. The SHK test gives the lowest p-value, but also assumes the most about the true alternative hypothesis.
\FloatBarrier
\begin{table}[h!]
\centering
\setlength{\tabcolsep}{0.8mm}

\caption{{\bf One-sided p-values for the Merck trial}: p-values found under the different tests used in the Merck trial.}\label{table_pvalues_extra}
\normalsize
\renewcommand{\arraystretch}{1.3} 
\begin{tabular}{cccccccc}
\hline
\multicolumn{1}{l}{\it Conditional} & \multicolumn{7}{c}{ \it Unconditional}        \\ \hline
FE                                                      & FMP$^*$ & Z$^*_\text{P}$ & APK    & MPK   & WAPK  & MPK2  & SHK   \\
.0271                                                  & .0144 & .0136          & .0130 & .1300 & .0160 & .0260 & .0080 \\ \hline
\end{tabular}
\end{table}
\autoref{tab:reallife_table} shows a sensitivity analysis by investigating the type~I~error rate and power for all considered tests, where we consider parameter configurations close to~$\btheta = (140/148,131/132)$.
The APK test most often shows the highest power, while the MPK test often shows the lowest power and only shows a superior power value for~$\theta_\C=0.5$. 
The WAPK test doesn't lead to a substantially better performance than the APK test: the WAPK test shows the same power   values as the APK test for~$\theta_\C> 0.914$, while for~$\theta_\C\leq 0.88$ the WAPK test has lower power than the APK test, reaching power values equal to zero. 
The weighted average power values under~$\mathbb{Q}$ are 0.622580 and 0.622579  for the WAPK and APK test, respectively~(hence the APK test attains a slightly lower weighted power), while the average powers~(found by~(6)) are 0.581917 and 0.816704 for the WAPK and APK test, respectively.
The MPK2 test only provides a good power performance for one specific alternative hypothesis, namely~(0.896, 0.944).
The SHK test shows a slightly worse performance than the WAPK test and provides the highest power for~$\btheta=(0.946,0.992)$ by definition, but so do the APK and WAPK tests.
 In conclusion, based on the observed p-values and evaluation of~\autoref{tab:reallife_table} depending on the degree of certainty about the parameter values, the APK, WAPK, or SHK test might be preferred. 

\subsection{Sensitivity of knapsack tests to solver settings}\label{sect:sensitivity_solver_settings}
Knapsack tests, e.g., the APK, MPK, WAPK, and SHK tests, are found as solutions to knapsack programs, which are variants of the APK program introduced in the main text. Uniqueness of these solutions is not guaranteed;  multiple knapsack tests may yield the same average, pointwise, or maximin power within the required suboptimality gap tolerance.
For instance, the MPK test, maximizing the minimum power and focusing on a worst-case alternative, will often have many optimal solutions that yield the same maximin power.

While these knapsack tests are exact and yield the desired optimality guarantees specified in their respective knapsack programs, different solutions may yield different pointwise type I error rate and power values. As Gurobi uses random initialization and multi-threading, the obtained knapsack test may depend on the random seed and number of cores used to solve the knapsack program, and reproducibility of the knapsack solution, if desired, hence requires fixing the random seed and limiting the number of cores to one. 

In the numerical investigation (\autoref{sect:numerical_results} and \autoref{sect:additional_results}), the MPK test was most sensitive to the solver configuration, while the rejection rate values for the  APK test were least sensitive, and the WAPK and SHK tests showed moderate sensitivity. 
This computational stability offers another compelling reason to prefer the APK test over targeted or maximin alternatives. Before implementation, we recommend performing a sensitivity analysis to verify a knapsack test's performance by comparing its power against standard benchmarks across relevant regions of the parametric space.
    
\FloatBarrier
\renewcommand{\arraystretch}{1.17} 
\setlength{\tabcolsep}{1.75mm}
\begin{table}[h!]

    \centering
\caption{ Type~I~error rates and power values~(in percentages)  for the Merck Research Laboratories Trial. True observed proportions were $(0.946, 0.992)$. The one-sided significance level was set to~$2.5\%$. Row maxima (minima) are made bold (underlined). A color gradient (orange to blue) differentiates power values.
}   \label{tab:reallife_table} \small\begin{tabular}{lllllllllll}
\toprule
 &  &   & FE & FMP$^*$ & Z$_{\text{P}}^*$ & APK & MPK & WAPK & MPK2 & SHK \\
$\theta_{\text{C}}$ & $\theta_{\text{D}}$ &  &  &  &  &  &  &  &  &  \\
\midrule
\multirow[t]{4}{*}{0.000} & 0.000 &  & ${\uline{\bf \textcolor[rgb]{1.000,0.549,0.000}{0.00}}}$ & ${\uline{\bf \textcolor[rgb]{1.000,0.549,0.000}{0.00}}}$ & ${\uline{\bf \textcolor[rgb]{1.000,0.549,0.000}{0.00}}}$ & ${\uline{\bf \textcolor[rgb]{1.000,0.549,0.000}{0.00}}}$ & ${\uline{\bf \textcolor[rgb]{1.000,0.549,0.000}{0.00}}}$ & ${\uline{\bf \textcolor[rgb]{1.000,0.549,0.000}{0.00}}}$ & ${\uline{\bf \textcolor[rgb]{1.000,0.549,0.000}{0.00}}}$ & ${\uline{\bf \textcolor[rgb]{1.000,0.549,0.000}{0.00}}}$ \\
 & 0.025 &  & ${\textcolor[rgb]{0.255,0.412,0.882}{23.58}}$ & ${\bf \textcolor[rgb]{0.255,0.412,0.882}{42.05}}$ & ${\bf \textcolor[rgb]{0.255,0.412,0.882}{42.05}}$ & ${\bf \textcolor[rgb]{0.255,0.412,0.882}{42.05}}$ & ${\textcolor[rgb]{0.255,0.412,0.882}{1.83}}$ & ${\uline{\textcolor[rgb]{1.000,0.549,0.000}{0.00}}}$ & ${\uline{\textcolor[rgb]{1.000,0.549,0.000}{0.00}}}$ & ${\uline{\textcolor[rgb]{1.000,0.549,0.000}{0.00}}}$ \\
 & 0.046 &  & ${\textcolor[rgb]{0.255,0.412,0.882}{73.91}}$ & ${\bf \textcolor[rgb]{0.255,0.412,0.882}{86.66}}$ & ${\bf \textcolor[rgb]{0.255,0.412,0.882}{86.66}}$ & ${\bf \textcolor[rgb]{0.255,0.412,0.882}{86.66}}$ & ${\textcolor[rgb]{0.255,0.412,0.882}{27.19}}$ & ${\uline{\textcolor[rgb]{1.000,0.549,0.000}{0.00}}}$ & ${\uline{\textcolor[rgb]{1.000,0.549,0.000}{0.00}}}$ & ${\uline{\textcolor[rgb]{1.000,0.549,0.000}{0.00}}}$ \\
 & 0.075 &  & ${\textcolor[rgb]{0.255,0.412,0.882}{97.32}}$ & ${\bf \textcolor[rgb]{0.255,0.412,0.882}{99.09}}$ & ${\bf \textcolor[rgb]{0.255,0.412,0.882}{99.09}}$ & ${\bf \textcolor[rgb]{0.255,0.412,0.882}{99.09}}$ & ${\textcolor[rgb]{0.255,0.412,0.882}{78.11}}$ & ${\uline{\textcolor[rgb]{1.000,0.549,0.000}{0.00}}}$ & ${\uline{\textcolor[rgb]{1.000,0.549,0.000}{0.00}}}$ & ${\uline{\textcolor[rgb]{1.000,0.549,0.000}{0.00}}}$ \\
\cline{1-11}
\multirow[t]{4}{*}{0.250} & 0.250 &  & ${\textcolor[rgb]{0.255,0.412,0.882}{1.66}}$ & ${\textcolor[rgb]{0.255,0.412,0.882}{2.25}}$ & ${\textcolor[rgb]{0.255,0.412,0.882}{2.25}}$ & ${\bf \textcolor[rgb]{0.255,0.412,0.882}{2.49}}$ & ${\textcolor[rgb]{0.255,0.412,0.882}{2.40}}$ & ${\uline{\textcolor[rgb]{1.000,0.549,0.000}{0.00}}}$ & ${\uline{\textcolor[rgb]{1.000,0.549,0.000}{0.00}}}$ & ${\uline{\textcolor[rgb]{1.000,0.549,0.000}{0.00}}}$ \\
 & 0.296 &  & ${\textcolor[rgb]{0.255,0.412,0.882}{10.41}}$ & ${\textcolor[rgb]{0.255,0.412,0.882}{13.01}}$ & ${\textcolor[rgb]{0.255,0.412,0.882}{13.01}}$ & ${\bf \textcolor[rgb]{0.255,0.412,0.882}{13.72}}$ & ${\textcolor[rgb]{0.255,0.412,0.882}{13.08}}$ & ${\uline{\textcolor[rgb]{1.000,0.549,0.000}{0.00}}}$ & ${\uline{\textcolor[rgb]{1.000,0.549,0.000}{0.00}}}$ & ${\uline{\textcolor[rgb]{1.000,0.549,0.000}{0.00}}}$ \\
 & 0.350 &  & ${\textcolor[rgb]{0.255,0.412,0.882}{38.31}}$ & ${\textcolor[rgb]{0.255,0.412,0.882}{43.28}}$ & ${\textcolor[rgb]{0.255,0.412,0.882}{43.28}}$ & ${\bf \textcolor[rgb]{0.255,0.412,0.882}{44.51}}$ & ${\textcolor[rgb]{0.255,0.412,0.882}{41.55}}$ & ${\textcolor[rgb]{0.874,0.526,0.149}{0.17}}$ & ${\textcolor[rgb]{0.874,0.526,0.149}{0.17}}$ & ${\uline{\textcolor[rgb]{0.889,0.529,0.131}{0.15}}}$ \\
 & 0.400 &  & ${\textcolor[rgb]{0.255,0.412,0.882}{71.78}}$ & ${\textcolor[rgb]{0.255,0.412,0.882}{75.51}}$ & ${\textcolor[rgb]{0.255,0.412,0.882}{75.51}}$ & ${\bf \textcolor[rgb]{0.255,0.412,0.882}{76.64}}$ & ${\textcolor[rgb]{0.255,0.412,0.882}{74.74}}$ & ${\textcolor[rgb]{0.255,0.412,0.882}{2.01}}$ & ${\textcolor[rgb]{0.255,0.412,0.882}{2.01}}$ & ${\uline{\textcolor[rgb]{0.255,0.412,0.882}{1.85}}}$ \\
\cline{1-11}
\multirow[t]{4}{*}{0.500} & 0.500 &  & ${\textcolor[rgb]{0.255,0.412,0.882}{1.90}}$ & ${\textcolor[rgb]{0.255,0.412,0.882}{2.39}}$ & ${\textcolor[rgb]{0.255,0.412,0.882}{2.39}}$ & ${\textcolor[rgb]{0.255,0.412,0.882}{2.49}}$ & ${\bf \textcolor[rgb]{0.255,0.412,0.882}{2.50}}$ & ${\uline{\textcolor[rgb]{1.000,0.549,0.000}{0.00}}}$ & ${\uline{\textcolor[rgb]{1.000,0.549,0.000}{0.00}}}$ & ${\uline{\textcolor[rgb]{1.000,0.549,0.000}{0.00}}}$ \\
 & 0.546 &  & ${\textcolor[rgb]{0.255,0.412,0.882}{9.68}}$ & ${\textcolor[rgb]{0.255,0.412,0.882}{11.47}}$ & ${\textcolor[rgb]{0.255,0.412,0.882}{11.47}}$ & ${\textcolor[rgb]{0.255,0.412,0.882}{11.81}}$ & ${\bf \textcolor[rgb]{0.255,0.412,0.882}{11.82}}$ & ${\uline{\textcolor[rgb]{1.000,0.549,0.000}{0.00}}}$ & ${\uline{\textcolor[rgb]{1.000,0.549,0.000}{0.00}}}$ & ${\uline{\textcolor[rgb]{1.000,0.549,0.000}{0.00}}}$ \\
 & 0.700 &  & ${\textcolor[rgb]{0.255,0.412,0.882}{91.36}}$ & ${\textcolor[rgb]{0.255,0.412,0.882}{92.87}}$ & ${\textcolor[rgb]{0.255,0.412,0.882}{92.87}}$ & ${\bf \textcolor[rgb]{0.255,0.412,0.882}{93.00}}$ & ${\textcolor[rgb]{0.255,0.412,0.882}{92.86}}$ & ${\uline{\textcolor[rgb]{0.538,0.464,0.547}{0.62}}}$ & ${\textcolor[rgb]{0.255,0.412,0.882}{1.25}}$ & ${\textcolor[rgb]{0.255,0.412,0.882}{6.17}}$ \\
 & 0.800 &  & ${\textcolor[rgb]{0.255,0.412,0.882}{99.96}}$ & ${\bf \textcolor[rgb]{0.255,0.412,0.882}{99.97}}$ & ${\bf \textcolor[rgb]{0.255,0.412,0.882}{99.97}}$ & ${\bf \textcolor[rgb]{0.255,0.412,0.882}{99.97}}$ & ${\textcolor[rgb]{0.255,0.412,0.882}{99.96}}$ & ${\uline{\textcolor[rgb]{0.255,0.412,0.882}{31.85}}}$ & ${\textcolor[rgb]{0.255,0.412,0.882}{33.59}}$ & ${\textcolor[rgb]{0.255,0.412,0.882}{63.33}}$ \\
\cline{1-11}
\multirow[t]{4}{*}{0.750} & 0.750 &  & ${\textcolor[rgb]{0.255,0.412,0.882}{1.83}}$ & ${\textcolor[rgb]{0.255,0.412,0.882}{2.37}}$ & ${\textcolor[rgb]{0.255,0.412,0.882}{2.37}}$ & ${\bf \textcolor[rgb]{0.255,0.412,0.882}{2.49}}$ & ${\textcolor[rgb]{0.255,0.412,0.882}{1.99}}$ & ${\textcolor[rgb]{0.971,0.544,0.035}{0.04}}$ & ${\textcolor[rgb]{0.971,0.544,0.035}{0.04}}$ & ${\uline{\textcolor[rgb]{1.000,0.549,0.000}{0.00}}}$ \\
 & 0.796 &  & ${\textcolor[rgb]{0.255,0.412,0.882}{12.00}}$ & ${\textcolor[rgb]{0.255,0.412,0.882}{14.41}}$ & ${\textcolor[rgb]{0.255,0.412,0.882}{14.41}}$ & ${\bf \textcolor[rgb]{0.255,0.412,0.882}{14.95}}$ & ${\textcolor[rgb]{0.255,0.412,0.882}{12.60}}$ & ${\textcolor[rgb]{0.255,0.412,0.882}{1.61}}$ & ${\textcolor[rgb]{0.255,0.412,0.882}{1.52}}$ & ${\uline{\textcolor[rgb]{0.994,0.548,0.007}{0.01}}}$ \\
 & 0.850 &  & ${\textcolor[rgb]{0.255,0.412,0.882}{49.18}}$ & ${\textcolor[rgb]{0.255,0.412,0.882}{54.29}}$ & ${\textcolor[rgb]{0.255,0.412,0.882}{54.29}}$ & ${\bf \textcolor[rgb]{0.255,0.412,0.882}{55.20}}$ & ${\textcolor[rgb]{0.255,0.412,0.882}{48.67}}$ & ${\textcolor[rgb]{0.255,0.412,0.882}{27.16}}$ & ${\textcolor[rgb]{0.255,0.412,0.882}{26.42}}$ & ${\uline{\textcolor[rgb]{0.255,0.412,0.882}{1.53}}}$ \\
 & 0.900 &  & ${\textcolor[rgb]{0.255,0.412,0.882}{89.24}}$ & ${\textcolor[rgb]{0.255,0.412,0.882}{91.58}}$ & ${\textcolor[rgb]{0.255,0.412,0.882}{91.58}}$ & ${\bf \textcolor[rgb]{0.255,0.412,0.882}{91.95}}$ & ${\textcolor[rgb]{0.255,0.412,0.882}{88.16}}$ & ${\textcolor[rgb]{0.255,0.412,0.882}{85.14}}$ & ${\textcolor[rgb]{0.255,0.412,0.882}{85.36}}$ & ${\uline{\textcolor[rgb]{0.255,0.412,0.882}{30.82}}}$ \\
\cline{1-11}
\multirow[t]{4}{*}{0.870} & 0.870 &  & ${\textcolor[rgb]{0.255,0.412,0.882}{1.70}}$ & ${\textcolor[rgb]{0.255,0.412,0.882}{2.35}}$ & ${\textcolor[rgb]{0.255,0.412,0.882}{2.35}}$ & ${\bf \textcolor[rgb]{0.255,0.412,0.882}{2.47}}$ & ${\textcolor[rgb]{0.255,0.412,0.882}{1.46}}$ & ${\textcolor[rgb]{0.255,0.412,0.882}{1.68}}$ & ${\textcolor[rgb]{0.255,0.412,0.882}{2.43}}$ & ${\uline{\textcolor[rgb]{0.439,0.446,0.664}{0.75}}}$ \\
 & 0.931 &  & ${\textcolor[rgb]{0.255,0.412,0.882}{33.38}}$ & ${\textcolor[rgb]{0.255,0.412,0.882}{38.50}}$ & ${\textcolor[rgb]{0.255,0.412,0.882}{38.50}}$ & ${\bf \textcolor[rgb]{0.255,0.412,0.882}{39.85}}$ & ${\uline{\textcolor[rgb]{0.255,0.412,0.882}{23.75}}}$ & ${\textcolor[rgb]{0.255,0.412,0.882}{38.16}}$ & ${\textcolor[rgb]{0.255,0.412,0.882}{38.83}}$ & ${\textcolor[rgb]{0.255,0.412,0.882}{34.28}}$ \\
 & 0.992 &  & ${\textcolor[rgb]{0.255,0.412,0.882}{99.26}}$ & ${\textcolor[rgb]{0.255,0.412,0.882}{99.62}}$ & ${\textcolor[rgb]{0.255,0.412,0.882}{99.64}}$ & ${\bf \textcolor[rgb]{0.255,0.412,0.882}{99.71}}$ & ${\textcolor[rgb]{0.255,0.412,0.882}{99.26}}$ & ${\bf \textcolor[rgb]{0.255,0.412,0.882}{99.71}}$ & ${\uline{\textcolor[rgb]{0.255,0.412,0.882}{99.12}}}$ & ${\bf \textcolor[rgb]{0.255,0.412,0.882}{99.71}}$ \\
 & 1.000 &  & ${\textcolor[rgb]{0.255,0.412,0.882}{99.99}}$ & ${\bf \textcolor[rgb]{0.255,0.412,0.882}{100.00}}$ & ${\bf \textcolor[rgb]{0.255,0.412,0.882}{100.00}}$ & ${\bf \textcolor[rgb]{0.255,0.412,0.882}{100.00}}$ & ${\uline{\textcolor[rgb]{0.255,0.412,0.882}{99.93}}}$ & ${\bf \textcolor[rgb]{0.255,0.412,0.882}{100.00}}$ & ${\textcolor[rgb]{0.255,0.412,0.882}{99.98}}$ & ${\bf \textcolor[rgb]{0.255,0.412,0.882}{100.00}}$ \\
\cline{1-11}
\multirow[t]{4}{*}{0.880} & 0.880 &  & ${\textcolor[rgb]{0.255,0.412,0.882}{1.69}}$ & ${\textcolor[rgb]{0.255,0.412,0.882}{2.32}}$ & ${\textcolor[rgb]{0.255,0.412,0.882}{2.32}}$ & ${\bf \textcolor[rgb]{0.255,0.412,0.882}{2.45}}$ & ${\textcolor[rgb]{0.255,0.412,0.882}{1.29}}$ & ${\textcolor[rgb]{0.255,0.412,0.882}{1.82}}$ & ${\textcolor[rgb]{0.255,0.412,0.882}{2.34}}$ & ${\uline{\textcolor[rgb]{0.255,0.412,0.882}{1.14}}}$ \\
 & 0.936 &  & ${\textcolor[rgb]{0.255,0.412,0.882}{29.95}}$ & ${\textcolor[rgb]{0.255,0.412,0.882}{35.28}}$ & ${\textcolor[rgb]{0.255,0.412,0.882}{35.28}}$ & ${\bf \textcolor[rgb]{0.255,0.412,0.882}{36.66}}$ & ${\uline{\textcolor[rgb]{0.255,0.412,0.882}{19.29}}}$ & ${\textcolor[rgb]{0.255,0.412,0.882}{36.14}}$ & ${\textcolor[rgb]{0.255,0.412,0.882}{36.11}}$ & ${\textcolor[rgb]{0.255,0.412,0.882}{33.84}}$ \\
 & 0.992 &  & ${\textcolor[rgb]{0.255,0.412,0.882}{98.44}}$ & ${\textcolor[rgb]{0.255,0.412,0.882}{99.16}}$ & ${\textcolor[rgb]{0.255,0.412,0.882}{99.22}}$ & ${\bf \textcolor[rgb]{0.255,0.412,0.882}{99.36}}$ & ${\textcolor[rgb]{0.255,0.412,0.882}{98.61}}$ & ${\bf \textcolor[rgb]{0.255,0.412,0.882}{99.36}}$ & ${\uline{\textcolor[rgb]{0.255,0.412,0.882}{98.06}}}$ & ${\bf \textcolor[rgb]{0.255,0.412,0.882}{99.36}}$ \\
 & 1.000 &  & ${\textcolor[rgb]{0.255,0.412,0.882}{99.98}}$ & ${\bf \textcolor[rgb]{0.255,0.412,0.882}{100.00}}$ & ${\bf \textcolor[rgb]{0.255,0.412,0.882}{100.00}}$ & ${\bf \textcolor[rgb]{0.255,0.412,0.882}{100.00}}$ & ${\uline{\textcolor[rgb]{0.255,0.412,0.882}{99.79}}}$ & ${\bf \textcolor[rgb]{0.255,0.412,0.882}{100.00}}$ & ${\textcolor[rgb]{0.255,0.412,0.882}{99.93}}$ & ${\bf \textcolor[rgb]{0.255,0.412,0.882}{100.00}}$ \\
\cline{1-11}
\multirow[t]{4}{*}{0.896} & 0.896 &  & ${\textcolor[rgb]{0.255,0.412,0.882}{1.64}}$ & ${\textcolor[rgb]{0.255,0.412,0.882}{2.27}}$ & ${\textcolor[rgb]{0.255,0.412,0.882}{2.27}}$ & ${\bf \textcolor[rgb]{0.255,0.412,0.882}{2.47}}$ & ${\uline{\textcolor[rgb]{0.305,0.421,0.824}{0.93}}}$ & ${\textcolor[rgb]{0.255,0.412,0.882}{2.27}}$ & ${\textcolor[rgb]{0.255,0.412,0.882}{2.32}}$ & ${\textcolor[rgb]{0.255,0.412,0.882}{1.87}}$ \\
 & 0.944 &  & ${\textcolor[rgb]{0.255,0.412,0.882}{24.45}}$ & ${\textcolor[rgb]{0.255,0.412,0.882}{30.11}}$ & ${\textcolor[rgb]{0.255,0.412,0.882}{30.11}}$ & ${\textcolor[rgb]{0.255,0.412,0.882}{31.21}}$ & ${\uline{\textcolor[rgb]{0.255,0.412,0.882}{15.12}}}$ & ${\textcolor[rgb]{0.255,0.412,0.882}{31.30}}$ & ${\bf \textcolor[rgb]{0.255,0.412,0.882}{31.38}}$ & ${\textcolor[rgb]{0.255,0.412,0.882}{30.59}}$ \\
 & 0.992 &  & ${\textcolor[rgb]{0.255,0.412,0.882}{95.31}}$ & ${\textcolor[rgb]{0.255,0.412,0.882}{97.25}}$ & ${\textcolor[rgb]{0.255,0.412,0.882}{97.50}}$ & ${\bf \textcolor[rgb]{0.255,0.412,0.882}{97.85}}$ & ${\textcolor[rgb]{0.255,0.412,0.882}{96.03}}$ & ${\bf \textcolor[rgb]{0.255,0.412,0.882}{97.85}}$ & ${\uline{\textcolor[rgb]{0.255,0.412,0.882}{93.86}}}$ & ${\bf \textcolor[rgb]{0.255,0.412,0.882}{97.85}}$ \\
 & 1.000 &  & ${\textcolor[rgb]{0.255,0.412,0.882}{99.85}}$ & ${\bf \textcolor[rgb]{0.255,0.412,0.882}{99.96}}$ & ${\bf \textcolor[rgb]{0.255,0.412,0.882}{99.96}}$ & ${\bf \textcolor[rgb]{0.255,0.412,0.882}{99.96}}$ & ${\uline{\textcolor[rgb]{0.255,0.412,0.882}{98.89}}}$ & ${\bf \textcolor[rgb]{0.255,0.412,0.882}{99.96}}$ & ${\textcolor[rgb]{0.255,0.412,0.882}{99.57}}$ & ${\bf \textcolor[rgb]{0.255,0.412,0.882}{99.96}}$ \\
\cline{1-11}
\multirow[t]{4}{*}{0.914} & 0.914 &  & ${\textcolor[rgb]{0.255,0.412,0.882}{1.54}}$ & ${\textcolor[rgb]{0.255,0.412,0.882}{2.28}}$ & ${\textcolor[rgb]{0.255,0.412,0.882}{2.28}}$ & ${\textcolor[rgb]{0.255,0.412,0.882}{2.48}}$ & ${\uline{\textcolor[rgb]{0.457,0.449,0.644}{0.73}}}$ & ${\bf \textcolor[rgb]{0.255,0.412,0.882}{2.49}}$ & ${\textcolor[rgb]{0.255,0.412,0.882}{2.47}}$ & ${\textcolor[rgb]{0.255,0.412,0.882}{2.35}}$ \\
 & 0.960 &  & ${\textcolor[rgb]{0.255,0.412,0.882}{26.92}}$ & ${\textcolor[rgb]{0.255,0.412,0.882}{33.93}}$ & ${\textcolor[rgb]{0.255,0.412,0.882}{34.00}}$ & ${\textcolor[rgb]{0.255,0.412,0.882}{35.52}}$ & ${\uline{\textcolor[rgb]{0.255,0.412,0.882}{22.99}}}$ & ${\bf \textcolor[rgb]{0.255,0.412,0.882}{35.54}}$ & ${\textcolor[rgb]{0.255,0.412,0.882}{33.32}}$ & ${\textcolor[rgb]{0.255,0.412,0.882}{35.49}}$ \\
 & 0.992 &  & ${\textcolor[rgb]{0.255,0.412,0.882}{87.00}}$ & ${\textcolor[rgb]{0.255,0.412,0.882}{91.57}}$ & ${\textcolor[rgb]{0.255,0.412,0.882}{92.62}}$ & ${\bf \textcolor[rgb]{0.255,0.412,0.882}{93.28}}$ & ${\textcolor[rgb]{0.255,0.412,0.882}{88.32}}$ & ${\bf \textcolor[rgb]{0.255,0.412,0.882}{93.28}}$ & ${\uline{\textcolor[rgb]{0.255,0.412,0.882}{82.83}}}$ & ${\bf \textcolor[rgb]{0.255,0.412,0.882}{93.28}}$ \\
 & 1.000 &  & ${\textcolor[rgb]{0.255,0.412,0.882}{99.02}}$ & ${\bf \textcolor[rgb]{0.255,0.412,0.882}{99.67}}$ & ${\bf \textcolor[rgb]{0.255,0.412,0.882}{99.67}}$ & ${\bf \textcolor[rgb]{0.255,0.412,0.882}{99.67}}$ & ${\uline{\textcolor[rgb]{0.255,0.412,0.882}{94.71}}}$ & ${\bf \textcolor[rgb]{0.255,0.412,0.882}{99.67}}$ & ${\textcolor[rgb]{0.255,0.412,0.882}{97.54}}$ & ${\bf \textcolor[rgb]{0.255,0.412,0.882}{99.67}}$ \\
\cline{1-11}
\multirow[t]{4}{*}{0.946} & 0.946 &  & ${\uline{\textcolor[rgb]{0.255,0.412,0.882}{1.21}}}$ & ${\textcolor[rgb]{0.255,0.412,0.882}{2.06}}$ & ${\textcolor[rgb]{0.255,0.412,0.882}{2.13}}$ & ${\bf \textcolor[rgb]{0.255,0.412,0.882}{2.46}}$ & ${\textcolor[rgb]{0.255,0.412,0.882}{1.51}}$ & ${\bf \textcolor[rgb]{0.255,0.412,0.882}{2.46}}$ & ${\textcolor[rgb]{0.255,0.412,0.882}{1.65}}$ & ${\bf \textcolor[rgb]{0.255,0.412,0.882}{2.46}}$ \\
 & 0.969 &  & ${\textcolor[rgb]{0.255,0.412,0.882}{8.41}}$ & ${\textcolor[rgb]{0.255,0.412,0.882}{12.58}}$ & ${\textcolor[rgb]{0.255,0.412,0.882}{13.54}}$ & ${\bf \textcolor[rgb]{0.255,0.412,0.882}{14.94}}$ & ${\textcolor[rgb]{0.255,0.412,0.882}{11.08}}$ & ${\bf \textcolor[rgb]{0.255,0.412,0.882}{14.94}}$ & ${\uline{\textcolor[rgb]{0.255,0.412,0.882}{8.01}}}$ & ${\bf \textcolor[rgb]{0.255,0.412,0.882}{14.94}}$ \\
 & 0.992 &  & ${\textcolor[rgb]{0.255,0.412,0.882}{48.66}}$ & ${\textcolor[rgb]{0.255,0.412,0.882}{59.48}}$ & ${\textcolor[rgb]{0.255,0.412,0.882}{64.74}}$ & ${\bf \textcolor[rgb]{0.255,0.412,0.882}{65.18}}$ & ${\textcolor[rgb]{0.255,0.412,0.882}{43.16}}$ & ${\bf \textcolor[rgb]{0.255,0.412,0.882}{65.18}}$ & ${\uline{\textcolor[rgb]{0.255,0.412,0.882}{38.44}}}$ & ${\bf \textcolor[rgb]{0.255,0.412,0.882}{65.18}}$ \\
 & 1.000 &  & ${\textcolor[rgb]{0.255,0.412,0.882}{81.64}}$ & ${\bf \textcolor[rgb]{0.255,0.412,0.882}{90.66}}$ & ${\bf \textcolor[rgb]{0.255,0.412,0.882}{90.66}}$ & ${\bf \textcolor[rgb]{0.255,0.412,0.882}{90.66}}$ & ${\uline{\textcolor[rgb]{0.255,0.412,0.882}{55.09}}}$ & ${\bf \textcolor[rgb]{0.255,0.412,0.882}{90.66}}$ & ${\textcolor[rgb]{0.255,0.412,0.882}{69.34}}$ & ${\bf \textcolor[rgb]{0.255,0.412,0.882}{90.66}}$ \\
\cline{1-11}
\multirow[t]{4}{*}{0.968} & 0.968 &  & ${\textcolor[rgb]{0.343,0.428,0.779}{0.88}}$ & ${\textcolor[rgb]{0.255,0.412,0.882}{1.67}}$ & ${\textcolor[rgb]{0.255,0.412,0.882}{2.23}}$ & ${\bf \textcolor[rgb]{0.255,0.412,0.882}{2.33}}$ & ${\textcolor[rgb]{0.255,0.412,0.882}{1.07}}$ & ${\bf \textcolor[rgb]{0.255,0.412,0.882}{2.33}}$ & ${\uline{\textcolor[rgb]{0.620,0.479,0.450}{0.51}}}$ & ${\bf \textcolor[rgb]{0.255,0.412,0.882}{2.33}}$ \\
 & 0.980 &  & ${\textcolor[rgb]{0.255,0.412,0.882}{3.53}}$ & ${\textcolor[rgb]{0.255,0.412,0.882}{6.20}}$ & ${\textcolor[rgb]{0.255,0.412,0.882}{7.97}}$ & ${\bf \textcolor[rgb]{0.255,0.412,0.882}{8.08}}$ & ${\textcolor[rgb]{0.255,0.412,0.882}{2.85}}$ & ${\bf \textcolor[rgb]{0.255,0.412,0.882}{8.08}}$ & ${\uline{\textcolor[rgb]{0.255,0.412,0.882}{1.93}}}$ & ${\bf \textcolor[rgb]{0.255,0.412,0.882}{8.08}}$ \\
 & 0.992 &  & ${\textcolor[rgb]{0.255,0.412,0.882}{13.97}}$ & ${\textcolor[rgb]{0.255,0.412,0.882}{22.64}}$ & ${\textcolor[rgb]{0.255,0.412,0.882}{26.02}}$ & ${\bf \textcolor[rgb]{0.255,0.412,0.882}{26.05}}$ & ${\uline{\textcolor[rgb]{0.255,0.412,0.882}{6.28}}}$ & ${\bf \textcolor[rgb]{0.255,0.412,0.882}{26.05}}$ & ${\textcolor[rgb]{0.255,0.412,0.882}{7.84}}$ & ${\bf \textcolor[rgb]{0.255,0.412,0.882}{26.05}}$ \\
 & 1.000 &  & ${\textcolor[rgb]{0.255,0.412,0.882}{32.96}}$ & ${\bf \textcolor[rgb]{0.255,0.412,0.882}{50.63}}$ & ${\bf \textcolor[rgb]{0.255,0.412,0.882}{50.63}}$ & ${\bf \textcolor[rgb]{0.255,0.412,0.882}{50.63}}$ & ${\uline{\textcolor[rgb]{0.255,0.412,0.882}{10.00}}}$ & ${\bf \textcolor[rgb]{0.255,0.412,0.882}{50.63}}$ & ${\textcolor[rgb]{0.255,0.412,0.882}{19.17}}$ & ${\bf \textcolor[rgb]{0.255,0.412,0.882}{50.63}}$ \\
\cline{1-11}
\multirow[t]{4}{*}{0.976} & 0.976 &  & ${\textcolor[rgb]{0.448,0.447,0.654}{0.74}}$ & ${\textcolor[rgb]{0.255,0.412,0.882}{1.57}}$ & ${\textcolor[rgb]{0.255,0.412,0.882}{2.08}}$ & ${\bf \textcolor[rgb]{0.255,0.412,0.882}{2.10}}$ & ${\textcolor[rgb]{0.687,0.491,0.370}{0.42}}$ & ${\bf \textcolor[rgb]{0.255,0.412,0.882}{2.10}}$ & ${\uline{\textcolor[rgb]{0.763,0.505,0.280}{0.32}}}$ & ${\bf \textcolor[rgb]{0.255,0.412,0.882}{2.10}}$ \\
 & 0.984 &  & ${\textcolor[rgb]{0.255,0.412,0.882}{2.01}}$ & ${\textcolor[rgb]{0.255,0.412,0.882}{4.16}}$ & ${\textcolor[rgb]{0.255,0.412,0.882}{5.15}}$ & ${\bf \textcolor[rgb]{0.255,0.412,0.882}{5.16}}$ & ${\uline{\textcolor[rgb]{0.410,0.440,0.699}{0.79}}}$ & ${\bf \textcolor[rgb]{0.255,0.412,0.882}{5.16}}$ & ${\textcolor[rgb]{0.351,0.430,0.768}{0.87}}$ & ${\bf \textcolor[rgb]{0.255,0.412,0.882}{5.16}}$ \\
 & 0.992 &  & ${\textcolor[rgb]{0.255,0.412,0.882}{5.46}}$ & ${\textcolor[rgb]{0.255,0.412,0.882}{10.97}}$ & ${\textcolor[rgb]{0.255,0.412,0.882}{12.36}}$ & ${\bf \textcolor[rgb]{0.255,0.412,0.882}{12.37}}$ & ${\uline{\textcolor[rgb]{0.255,0.412,0.882}{1.41}}}$ & ${\bf \textcolor[rgb]{0.255,0.412,0.882}{12.37}}$ & ${\textcolor[rgb]{0.255,0.412,0.882}{2.41}}$ & ${\bf \textcolor[rgb]{0.255,0.412,0.882}{12.37}}$ \\
 & 1.000 &  & ${\textcolor[rgb]{0.255,0.412,0.882}{13.95}}$ & ${\bf \textcolor[rgb]{0.255,0.412,0.882}{27.27}}$ & ${\bf \textcolor[rgb]{0.255,0.412,0.882}{27.27}}$ & ${\bf \textcolor[rgb]{0.255,0.412,0.882}{27.27}}$ & ${\uline{\textcolor[rgb]{0.255,0.412,0.882}{2.50}}}$ & ${\bf \textcolor[rgb]{0.255,0.412,0.882}{27.27}}$ & ${\textcolor[rgb]{0.255,0.412,0.882}{6.27}}$ & ${\bf \textcolor[rgb]{0.255,0.412,0.882}{27.27}}$ \\
\cline{1-11}
\multirow[t]{4}{*}{0.985} & 0.985 &  & ${\textcolor[rgb]{0.740,0.501,0.308}{0.35}}$ & ${\textcolor[rgb]{0.255,0.412,0.882}{1.05}}$ & ${\bf \textcolor[rgb]{0.255,0.412,0.882}{1.20}}$ & ${\bf \textcolor[rgb]{0.255,0.412,0.882}{1.20}}$ & ${\uline{\textcolor[rgb]{0.965,0.543,0.042}{0.05}}}$ & ${\bf \textcolor[rgb]{0.255,0.412,0.882}{1.20}}$ & ${\textcolor[rgb]{0.927,0.536,0.087}{0.10}}$ & ${\bf \textcolor[rgb]{0.255,0.412,0.882}{1.20}}$ \\
 & 0.990 &  & ${\textcolor[rgb]{0.492,0.455,0.602}{0.68}}$ & ${\textcolor[rgb]{0.255,0.412,0.882}{2.01}}$ & ${\bf \textcolor[rgb]{0.255,0.412,0.882}{2.20}}$ & ${\bf \textcolor[rgb]{0.255,0.412,0.882}{2.20}}$ & ${\uline{\textcolor[rgb]{0.942,0.538,0.069}{0.08}}}$ & ${\bf \textcolor[rgb]{0.255,0.412,0.882}{2.20}}$ & ${\textcolor[rgb]{0.851,0.522,0.176}{0.20}}$ & ${\bf \textcolor[rgb]{0.255,0.412,0.882}{2.20}}$ \\
 & 0.995 &  & ${\textcolor[rgb]{0.255,0.412,0.882}{1.30}}$ & ${\textcolor[rgb]{0.255,0.412,0.882}{3.84}}$ & ${\bf \textcolor[rgb]{0.255,0.412,0.882}{4.03}}$ & ${\bf \textcolor[rgb]{0.255,0.412,0.882}{4.03}}$ & ${\uline{\textcolor[rgb]{0.912,0.533,0.104}{0.12}}}$ & ${\bf \textcolor[rgb]{0.255,0.412,0.882}{4.03}}$ & ${\textcolor[rgb]{0.717,0.497,0.336}{0.38}}$ & ${\bf \textcolor[rgb]{0.255,0.412,0.882}{4.03}}$ \\
 & 1.000 &  & ${\textcolor[rgb]{0.255,0.412,0.882}{2.49}}$ & ${\bf \textcolor[rgb]{0.255,0.412,0.882}{7.32}}$ & ${\bf \textcolor[rgb]{0.255,0.412,0.882}{7.32}}$ & ${\bf \textcolor[rgb]{0.255,0.412,0.882}{7.32}}$ & ${\uline{\textcolor[rgb]{0.860,0.523,0.166}{0.19}}}$ & ${\bf \textcolor[rgb]{0.255,0.412,0.882}{7.32}}$ & ${\textcolor[rgb]{0.457,0.449,0.644}{0.73}}$ & ${\bf \textcolor[rgb]{0.255,0.412,0.882}{7.32}}$ \\
\cline{1-11}

\end{tabular}
 \\
{ \small  FE: Fisher's exact test, FMP$^*$: Berger and Boos' version of Fisher's-mid-p test, Z$^*_\text{P}$: Berger and Boos' version of the Pooled-Z~ test,  APK: average power knapsack, WAPK: weighted average power knapsack, MPK(2): first (second) maximin power knapsack, SHK: simple hypothesis knapsack}
\end{table}
\FloatBarrier

\FloatBarrier

\section{Superiority testing with a margin}\label{sect:superiority_test}
Let~$g_0(\theta) = g_1(\theta) =\theta+\delta_n$ for~$\delta_n>0$, hence  we are testing $H_0:\theta_\D\leq \theta_\C+\delta_n\;\text{vs.}\;H_1:\theta_\D> \theta_\C+\delta_n$.

 We first derive results to compute the matrices in Theorem~2 to ensure exact type I error control. We have~$g_0'(\theta)=1$ and the derivative of the  function maximized in~$\bar{h}$ equals
\begin{align*}&(s_\C(\theta+\delta_n)(1-\theta) (1-(\theta+\delta_n)) + (s_\D-1)\theta(1-\theta) (1-(\theta+\delta_n)) \\&-(n_\C-s_\C)\theta(\theta+\delta_n) (1-(\theta+\delta_n)) -(n_\D-s_\D)\theta(\theta+\delta_n) (1-\theta) )\\&\cdot \theta^{s_\C-1}(\theta+\delta_n)^{s_\D-2}(1-\theta)^{n_\C-s_\C-1} (1-(\theta+\delta_n))^{n_\D-s_\D-1}\\&=
(a\theta^3 + b\theta^2+c\theta+d)\theta^{s_\C-1}(\theta+\delta_n)^{s_\D-2}(1-\theta)^{n_\C-s_\C-1} (1-(\theta+\delta_n))^{n_\D-s_\D-1}
\end{align*}
where
\begin{align*}
    &a = n - 1, \quad &&b= -2s_\C(1-\delta_n) + (s_\D-1)(\delta_n-2) - (n_\C-s_\C)(1-2\delta_n) - (n_\D-s_\D)(1-\delta_n),\\
    &d = s_\C\delta_n(1-\delta_n),\quad  &&c= s_\C(1-3\delta_n+\delta_n^2) + (s_\D-1)(1-\delta_n) -(n_\C-s_\C)(\delta_n-\delta_n^2)-(n_\D-s_\D)\delta_n.
\end{align*}
By Cardano's formula, the roots of the above derivative can now be written as
\begin{align*}
     \theta^*_{k}(\bs)=-(b + \xi^kC + D_0/(\xi^kC))/(3a),\quad k\in\{0,1,2\}
\end{align*}
where
\begin{align*}
   &C = \sqrt[3]{\left(D_1 + \sqrt{D_1^2-4D_0^3}\right)/2}, &&D_0= b^2-3ac,\\
   &D_1=2b^3-9abc + 27a^2d,&&\xi = (-1 + \sqrt{-3})/2.
\end{align*}
To find~$\bar{h}(\bs_{i},\theta_j,\theta_{j+1})$ we hence only have to check the function to be maximized at the values~$(\max(\theta_j,\min(\theta_{j+1},Re(\theta^*_{k}))))_k$, and at $\theta_j$, and~$\theta_{j+1}$ (as the other roots of the derivative above occur at~$0$ and~$1-\delta_n$ which equal~$\theta_1,\theta_K$ respectively). The above focused on computing~$\bar{h}(\bs_{i},\theta_j,\theta_{j+1})$ but computation of~$\ubar{h}(\bs_{i},\theta_j,\theta_{j+1})$ follows similarly.

We now consider computing the average power.  Consider the average behavior of the test under~$H_1.$
For a decision vector~$\bd$ coupled to the sequence~$\bs_1,\dots,\bs_{|\mathcal{S}|}$, the average power is given by:
\begin{align}
\frac{2}{(1-\delta_n)^2}\int_{[0,1]^2}\sum_{i\,:\,d_i=1}\mathbb{P}_{\btheta}(\bS=\bs_i)\mathbb{I}(\theta_\D\geq\theta_\C+\delta_n) d\btheta = \bd^\top \bar{\bp}_1,\label{expression_AP}
\end{align}
where we multiply by~$2/(1-\delta_n)^2$ as the area of the triangle~$\{\btheta\in[0,1]^2:\theta_\D\geq\theta_\C+\delta_n\}$ is~$(1-\delta_n)^2/2$. 
For~$I_{x}(y,z)$ the regularized incomplete Beta function evaluated at~$x$ for parameters~$y,z\in\mathbb{R}_+,$ we have~\citep[following the formulae in][]{millerblog}
\begin{align*}
\bar{p}_{\bs,1}=&\frac{2}{(1-\delta_n)^2}\int_{\btheta:\theta_\D> \theta_\C+\delta_n}\prod_{a\in\{\C,\D\}}\theta_a^{s_{a}}(1-\theta_a)^{n_a-s_{a}} d\btheta\\
    &=\frac{2B(\alpha_{\D}(\bs_i), \beta_{\D}(\bs_i))}{(1-\delta_n)^2}\int_{0}^{1-\delta_n} \int_{\theta_\C+\delta_n}^{1}\theta_\C^{\alpha_\C(\bs_i)-1}(1-\theta_\C)^{\beta_\C(\bs_i)-1}\frac{\theta_\D^{\alpha_\D(\bs_i)-1}(1-\theta_\D)^{\beta_\D(\bs_i)-1}}{B(\alpha_\D(\bs_i),\beta_\D(\bs_i))} d\theta_\D d\theta_\C\\&=
   \frac{2B(\alpha_{\D}(\bs_i), \beta_{\D}(\bs_i))}{(1-\delta_n)^2}\int_{0}^{1-\delta_n} \theta_\C^{\alpha_\C(\bs_i)-1}(1-\theta_\C)^{\beta_\C(\bs_i)-1}\left[I_{1}(\alpha_\D(\bs_i), \beta_\D(\bs_i))-I_{\theta_\C+\delta_n}(\alpha_\D(\bs_i), \beta_\D(\bs_i))\right]d\theta_\C\\&=
\frac{2B(\alpha_{\D}(\bs_i), \beta_{\D}(\bs_i))}{(1-\delta_n)^2}\int_{0}^{1-\delta_n} \theta_\C^{\alpha_\C(\bs_i)-1}(1-\theta_\C)^{\beta_\C(\bs_i)-1}\left[\sum_{i=0}^{\alpha_\D(\bs_i)-1}\frac{(\theta_\C+\delta_n)^i(1-\theta_\C-\delta_n)^{\beta_\D(\bs_i)}}{(\beta_\D(\bs_i)+i)B(i+1,\beta_\D(\bs_i))}\right]d\theta_\C\\
&=
\frac{2B(\alpha_{\D}(\bs_i), \beta_{\D}(\bs_i))}{(1-\delta_n)^2}\sum_{i=0}^{\alpha_\D(\bs_i)-1}
\frac{\int_0^{1-\delta_n}(\theta_\C+\delta_n)^i(1-\theta_\C-\delta_n)^{\beta_\D(\bs_i)} \theta_\C^{\alpha_\C(\bs_i)-1}(1-\theta_\C)^{\beta_\C(\bs_i)-1}d\theta_{\C}}{(\beta_\D(\bs_i)+i)B(i+1,\beta_\D(\bs_i))} 
\end{align*}
where~$\alpha_a(\bs) = s_a+1$ and~$\beta_a(\bs) = n_a-s_a+1$ for all~$\bs\in\mathcal{S}$ and~$B$ denotes the beta function. 
The coefficients above can be determined by numerical integration.

\subsection{Results for superiority testing with a margin}
In this section we present the results for the knapsack-based tests based on superiority testing with a margin. We chose to evaluate~$\delta_{20}=0.2$, $\delta_{50}=0.1$, $\delta_{100}=0.05$, $\delta_{300}=0.01$. The points~$\theta_1,\dots,\theta_K$ used for the type I error rate constraints were chosen equidistant with~$K=1000$. The integrals to determine~$\bar{p}_{\bs,1}$ are computed up to a relative error tolerance of $1\cdot 10^{-5}$. Due to linearity, the relative numerical error will propagate to the error on the average power, and as the power is upper bounded by 1, the absolute error on the power will in most cases be accurate up to four digits (absolute tolerance~$10^{-5}$). 

For the MPK test, we choose the same margins for the alternative hypothesis as in \autoref{sect:additional_results}. As comparator~(benchmark) we chose the unconditional exact~(UX) Z-unpooled test with statistic~\begin{equation}\begin{cases}\frac{\hat{\theta}_\C+ \delta_n - \hat{\theta}_\D }{ \sqrt{\hat{\theta}_\C(1-\hat{\theta}_\C)/n_\C + \hat{\theta}_\D(1-\hat{\theta}_\D)/n_\D}},\quad&\text{if $\hat{\theta}_\C(1-\hat{\theta}_\C)/n_\C + \hat{\theta}_\D(1-\hat{\theta}_\D)/n_\D>0,$}\\
(\hat{\theta}_\C+ \delta_n - \hat{\theta}_\D )\cdot\infty,&\text{else.}\label{stat_unpooled}
\end{cases}\end{equation}
In the above, $\theta_a=s_a/n_a$ for~$a\in\{\C,\D\}$ and the convention~$0\cdot\infty=0$ is used. 

\begin{figure*}[h!]

    \centering
    \includegraphics[width=\linewidth]{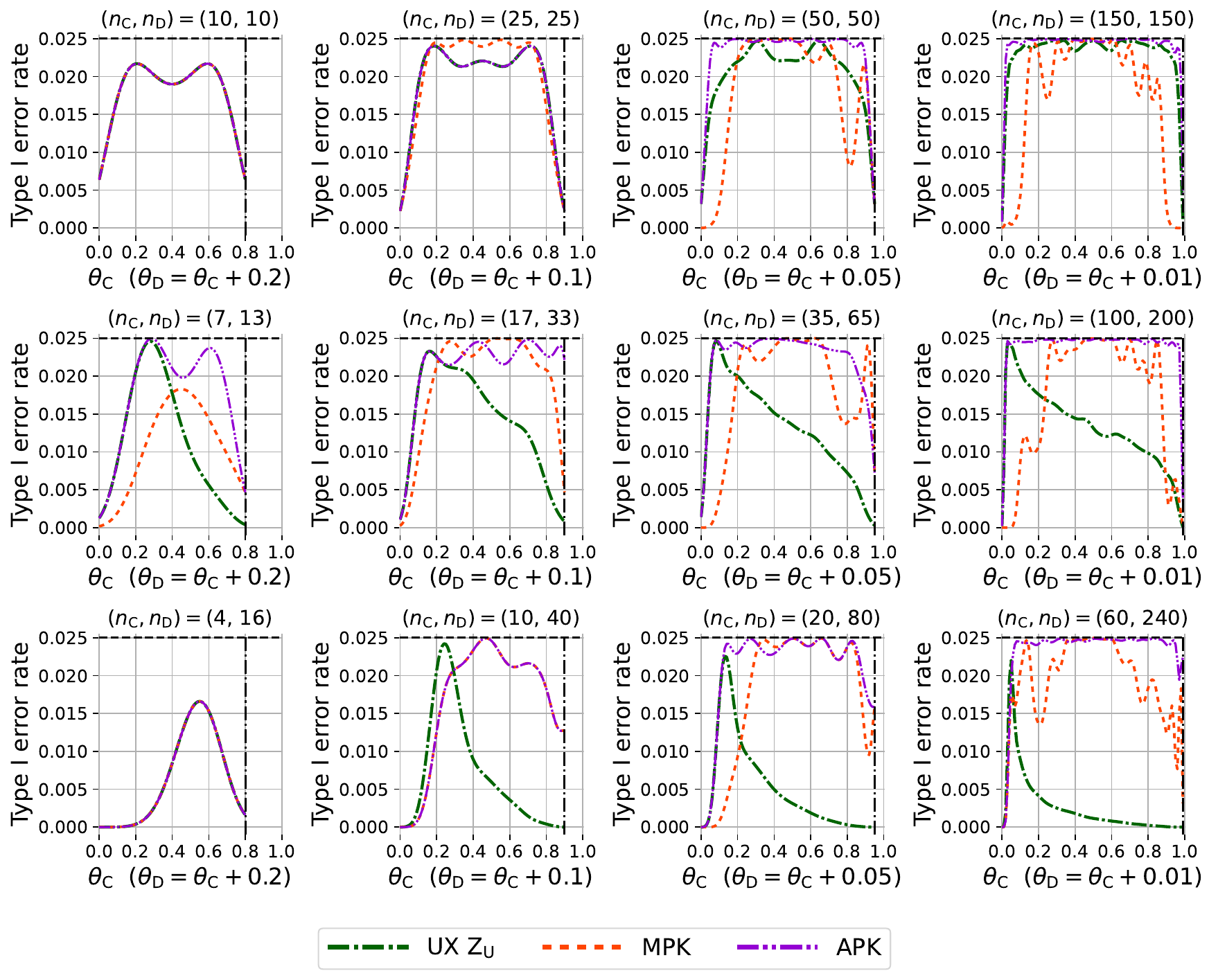}
    \caption{ Type~I~error rate profiles for the one-sided superiority test with a margin (minimum relevant treatment effect~$\delta_n$ indicated below each figure) for different treatment group size configurations under the unconditional exact Unpooled-Z~(UX Z$_\text{U}$) test, maximin power knapsack~(MPK) test, and average power knapsack~(APK) tests. The trial sizes~(varying over columns) were chosen as~$20,50,100,300$, where configurations of the treatment group sizes are indicated in the titles of the subfigures. The one-sided significance level~$\alpha$ is set to~$0.025$ and indicated by a dashed line. The dash-dotted lines indicate the end of the possible range for~$\theta_\C$.}
    \label{fig:T1E_marginsup}
\end{figure*}

\autoref{fig:T1E_marginsup} shows the type~I~error rate profiles for the same treatment group size configurations as in the main paper. 
Again, it was checked that every considered test satisfies Condition~(C); hence, we only need to verify that type~I~error control holds for~$\btheta\in\partial\Theta_0$. 

One thing of note about~\autoref{fig:T1E_marginsup} is that, in contrast with the results in the main paper, already for~$n_\C=7$,~$n_\D=13$, the type I error rate curves look very different across tests. A second thing of note is that the type I error rate profile for the UX Z-unpooled test becomes very skewed when the imbalance in the treatment group sizes becomes severe. 
The type I error rate profile for the MPK test shows the same behavior as in the main paper, where the type I error rate profile fluctuates a lot between being very conservative and showing a type I error close to the nominal level. Again, the type I error rate for the APK test is often close to the nominal significance level and shows stable behavior.

\autoref{tab:power_table_marginsup} shows power values
for similar scenarios as Table~1 in the main manuscript.  Again, the bold power values indicate the highest value in each row, while the underlined power values indicate the lowest power.
For~$\bn\in\{(10,10),(16,4),(4,16)\}$, each test has the same power for all considered parameter configurations. 
The UX Z-unpooled test often shows the lowest power, while the APK test often shows highest power. 
The power gains from using the APK test over the UX Z-unpooled test are often substantial in the case of unbalanced treatment groups, where power gains of around 60\% and 40\%  are observed. 

\autoref{tab:comparison_table_marginsup} summarizes pairwise power comparisons between the considered tests similar to Table~2 in the paper.
The~UX Z$_\text{U}$ test has roughly equal performance to the APK test for balanced small treatment group sizes, whereas for balanced larger treatment group sizes, the APK test often outperforms the~UX Z$_\text{U}$ test. For unbalanced treatment group sizes the~UX Z$_\text{U}$ test is often outperformed in terms of power by the APK test, except in case~$n_\C=4$, $n_\D=16$, where it was observed that the type I error profile was the same for the~UX Z$_\text{U}$ and the APK test. A similar conclusion as in~\autoref{tab:comp_tests_supp} holds for the MPK test: For large trial sizes~($n>50$), the APK test often outperforms the MPK test, and before that the MPK test shows marginal average power gains.

\FloatBarrier
\renewcommand{\arraystretch}{1.11} 
\setlength{\tabcolsep}{1.75mm}
\begin{table}[h!]

    \centering
    \caption{ Power values~(\%) for different superiority tests with a margin for different treatment group size configurations and parameter values~(significance level~$2.5\%$). Row maxima (minima) are made bold (underlined). A color gradient (orange to blue) differentiates power values.
}
\begin{tabular}{llllrrrr}
\toprule
 &  &  &  &   & Z$_{\text{U}}$ & MPK & APK \\
$n_{\text{C}}$ & $n_{\text{D}}$ & $\theta_{\text{C}}$ & $\theta_{\text{D}}$ &  &  &  &  \\
\midrule
\multirow[t]{4}{*}{10} & \multirow[t]{4}{*}{10} & 0.01 & 0.51 &  & ${\uline{\bf \textcolor[rgb]{0.664,0.487,0.398}{36.91}}}$ & ${\uline{\bf \textcolor[rgb]{0.664,0.487,0.398}{36.91}}}$ & ${\uline{\bf \textcolor[rgb]{0.664,0.487,0.398}{36.91}}}$ \\

 &  & 0.05 & 0.61 &  & ${\uline{\bf \textcolor[rgb]{0.582,0.472,0.495}{45.59}}}$ & ${\uline{\bf \textcolor[rgb]{0.582,0.472,0.495}{45.59}}}$ & ${\uline{\bf \textcolor[rgb]{0.582,0.472,0.495}{45.59}}}$ \\

 &  & 0.20 & 0.80 &  & ${\uline{\bf \textcolor[rgb]{0.606,0.476,0.467}{43.04}}}$ & ${\uline{\bf \textcolor[rgb]{0.606,0.476,0.467}{43.04}}}$ & ${\uline{\bf \textcolor[rgb]{0.606,0.476,0.467}{43.04}}}$ \\

 &  & 0.49 & 0.99 &  & ${\uline{\bf \textcolor[rgb]{0.664,0.487,0.398}{36.91}}}$ & ${\uline{\bf \textcolor[rgb]{0.664,0.487,0.398}{36.91}}}$ & ${\uline{\bf \textcolor[rgb]{0.664,0.487,0.398}{36.91}}}$ \\
\cline{1-8} \cline{2-8} 
\multirow[t]{4}{*}{25} & \multirow[t]{4}{*}{25} & 0.01 & 0.27 &  & ${\bf \textcolor[rgb]{0.711,0.496,0.343}{32.01}}$ & ${\uline{\textcolor[rgb]{0.728,0.499,0.322}{30.16}}}$ & ${\bf \textcolor[rgb]{0.711,0.496,0.343}{32.01}}$ \\

 &  & 0.20 & 0.58 &  & ${\uline{\textcolor[rgb]{0.515,0.460,0.574}{52.54}}}$ & ${\bf \textcolor[rgb]{0.497,0.456,0.595}{54.49}}$ & ${\uline{\textcolor[rgb]{0.515,0.460,0.574}{52.54}}}$ \\

 &  & 0.40 & 0.79 &  & ${\uline{\textcolor[rgb]{0.492,0.455,0.602}{55.08}}}$ & ${\bf \textcolor[rgb]{0.471,0.452,0.626}{57.15}}$ & ${\uline{\textcolor[rgb]{0.492,0.455,0.602}{55.08}}}$ \\

 &  & 0.73 & 0.99 &  & ${\bf \textcolor[rgb]{0.711,0.496,0.343}{32.01}}$ & ${\uline{\textcolor[rgb]{0.728,0.499,0.322}{30.16}}}$ & ${\bf \textcolor[rgb]{0.711,0.496,0.343}{32.01}}$ \\
\cline{1-8} \cline{2-8} 
\multirow[t]{4}{*}{50} & \multirow[t]{4}{*}{50} & 0.01 & 0.15 &  & ${\textcolor[rgb]{0.673,0.489,0.388}{36.01}}$ & ${\uline{\textcolor[rgb]{0.974,0.544,0.031}{4.25}}}$ & ${\bf \textcolor[rgb]{0.638,0.482,0.429}{39.77}}$ \\

 &  & 0.30 & 0.58 &  & ${\uline{\textcolor[rgb]{0.413,0.441,0.696}{63.26}}}$ & ${\bf \textcolor[rgb]{0.398,0.438,0.713}{64.88}}$ & ${\textcolor[rgb]{0.401,0.439,0.709}{64.68}}$ \\

 &  & 0.60 & 0.85 &  & ${\textcolor[rgb]{0.430,0.444,0.675}{61.52}}$ & ${\uline{\textcolor[rgb]{0.430,0.444,0.675}{61.33}}}$ & ${\bf \textcolor[rgb]{0.421,0.442,0.685}{62.52}}$ \\

 &  & 0.85 & 0.99 &  & ${\uline{\textcolor[rgb]{0.673,0.489,0.388}{36.01}}}$ & ${\textcolor[rgb]{0.641,0.483,0.426}{39.27}}$ & ${\bf \textcolor[rgb]{0.638,0.482,0.429}{39.77}}$ \\
\cline{1-8} \cline{2-8} 
\multirow[t]{4}{*}{150} & \multirow[t]{4}{*}{150} & 0.01 & 0.07 &  & ${\textcolor[rgb]{0.424,0.443,0.682}{62.11}}$ & ${\uline{\textcolor[rgb]{0.848,0.521,0.180}{17.51}}}$ & ${\bf \textcolor[rgb]{0.392,0.437,0.720}{65.58}}$ \\

 &  & 0.30 & 0.46 &  & ${\textcolor[rgb]{0.290,0.418,0.841}{76.12}}$ & ${\uline{\textcolor[rgb]{0.296,0.419,0.834}{75.60}}}$ & ${\bf \textcolor[rgb]{0.287,0.418,0.844}{76.63}}$ \\

 &  & 0.60 & 0.76 &  & ${\textcolor[rgb]{0.255,0.412,0.882}{79.88}}$ & ${\uline{\textcolor[rgb]{0.278,0.416,0.855}{77.49}}}$ & ${\bf \textcolor[rgb]{0.255,0.412,0.882}{80.03}}$ \\

 &  & 0.93 & 0.99 &  & ${\textcolor[rgb]{0.424,0.443,0.682}{62.11}}$ & ${\uline{\textcolor[rgb]{0.994,0.548,0.007}{2.22}}}$ & ${\bf \textcolor[rgb]{0.395,0.438,0.716}{65.14}}$ \\
\cline{1-8} \cline{2-8} 
\multirow[t]{4}{*}{16} & \multirow[t]{4}{*}{4} & 0.01 & 0.63 &  & ${\uline{\bf \textcolor[rgb]{0.866,0.524,0.159}{15.75}}}$ & ${\uline{\bf \textcolor[rgb]{0.866,0.524,0.159}{15.75}}}$ & ${\uline{\bf \textcolor[rgb]{0.866,0.524,0.159}{15.75}}}$ \\

 &  & 0.05 & 0.74 &  & ${\uline{\bf \textcolor[rgb]{0.731,0.500,0.318}{29.78}}}$ & ${\uline{\bf \textcolor[rgb]{0.731,0.500,0.318}{29.78}}}$ & ${\uline{\bf \textcolor[rgb]{0.731,0.500,0.318}{29.78}}}$ \\

 &  & 0.10 & 0.83 &  & ${\uline{\bf \textcolor[rgb]{0.594,0.474,0.481}{44.21}}}$ & ${\uline{\bf \textcolor[rgb]{0.594,0.474,0.481}{44.21}}}$ & ${\uline{\bf \textcolor[rgb]{0.594,0.474,0.481}{44.21}}}$ \\

 &  & 0.29 & 0.99 &  & ${\uline{\bf \textcolor[rgb]{0.763,0.505,0.280}{26.32}}}$ & ${\uline{\bf \textcolor[rgb]{0.763,0.505,0.280}{26.32}}}$ & ${\uline{\bf \textcolor[rgb]{0.763,0.505,0.280}{26.32}}}$ \\
\cline{1-8} \cline{2-8} 
\multirow[t]{4}{*}{4} & \multirow[t]{4}{*}{16} & 0.01 & 0.71 &  & ${\uline{\bf \textcolor[rgb]{0.763,0.505,0.280}{26.32}}}$ & ${\uline{\bf \textcolor[rgb]{0.763,0.505,0.280}{26.32}}}$ & ${\uline{\bf \textcolor[rgb]{0.763,0.505,0.280}{26.32}}}$ \\

 &  & 0.05 & 0.77 &  & ${\uline{\bf \textcolor[rgb]{0.644,0.483,0.422}{39.07}}}$ & ${\uline{\bf \textcolor[rgb]{0.644,0.483,0.422}{39.07}}}$ & ${\uline{\bf \textcolor[rgb]{0.644,0.483,0.422}{39.07}}}$ \\

 &  & 0.10 & 0.84 &  & ${\uline{\bf \textcolor[rgb]{0.544,0.465,0.540}{49.47}}}$ & ${\uline{\bf \textcolor[rgb]{0.544,0.465,0.540}{49.47}}}$ & ${\uline{\bf \textcolor[rgb]{0.544,0.465,0.540}{49.47}}}$ \\

 &  & 0.37 & 0.99 &  & ${\uline{\bf \textcolor[rgb]{0.866,0.524,0.159}{15.75}}}$ & ${\uline{\bf \textcolor[rgb]{0.866,0.524,0.159}{15.75}}}$ & ${\uline{\bf \textcolor[rgb]{0.866,0.524,0.159}{15.75}}}$ \\
\cline{1-8} \cline{2-8} 
\multirow[t]{4}{*}{40} & \multirow[t]{4}{*}{10} & 0.01 & 0.32 &  & ${\uline{\textcolor[rgb]{1.000,0.549,0.000}{1.55}}}$ & ${\bf \textcolor[rgb]{0.699,0.494,0.356}{33.22}}$ & ${\bf \textcolor[rgb]{0.699,0.494,0.356}{33.22}}$ \\

 &  & 0.20 & 0.68 &  & ${\uline{\textcolor[rgb]{0.746,0.502,0.301}{28.20}}}$ & ${\bf \textcolor[rgb]{0.454,0.448,0.647}{58.88}}$ & ${\bf \textcolor[rgb]{0.454,0.448,0.647}{58.88}}$ \\

 &  & 0.40 & 0.87 &  & ${\uline{\textcolor[rgb]{0.620,0.479,0.450}{41.54}}}$ & ${\bf \textcolor[rgb]{0.424,0.443,0.682}{62.20}}$ & ${\bf \textcolor[rgb]{0.424,0.443,0.682}{62.20}}$ \\

 &  & 0.65 & 0.99 &  & ${\bf \textcolor[rgb]{0.646,0.484,0.419}{38.70}}$ & ${\uline{\textcolor[rgb]{0.752,0.503,0.294}{27.68}}}$ & ${\uline{\textcolor[rgb]{0.752,0.503,0.294}{27.68}}}$ \\
\cline{1-8} \cline{2-8} 
\multirow[t]{4}{*}{10} & \multirow[t]{4}{*}{40} & 0.01 & 0.35 &  & ${\bf \textcolor[rgb]{0.646,0.484,0.419}{38.70}}$ & ${\uline{\textcolor[rgb]{0.752,0.503,0.294}{27.68}}}$ & ${\uline{\textcolor[rgb]{0.752,0.503,0.294}{27.68}}}$ \\

 &  & 0.20 & 0.68 &  & ${\uline{\textcolor[rgb]{0.667,0.488,0.394}{36.71}}}$ & ${\bf \textcolor[rgb]{0.430,0.444,0.675}{61.33}}$ & ${\bf \textcolor[rgb]{0.430,0.444,0.675}{61.33}}$ \\

 &  & 0.40 & 0.86 &  & ${\uline{\textcolor[rgb]{0.828,0.517,0.204}{19.83}}}$ & ${\bf \textcolor[rgb]{0.465,0.451,0.633}{57.63}}$ & ${\bf \textcolor[rgb]{0.465,0.451,0.633}{57.63}}$ \\

 &  & 0.68 & 0.99 &  & ${\uline{\textcolor[rgb]{1.000,0.549,0.000}{1.55}}}$ & ${\bf \textcolor[rgb]{0.699,0.494,0.356}{33.22}}$ & ${\bf \textcolor[rgb]{0.699,0.494,0.356}{33.22}}$ \\
\cline{1-8} \cline{2-8} 
\multirow[t]{4}{*}{80} & \multirow[t]{4}{*}{20} & 0.01 & 0.19 &  & ${\uline{\textcolor[rgb]{1.000,0.549,0.000}{1.47}}}$ & ${\textcolor[rgb]{0.641,0.483,0.426}{39.30}}$ & ${\bf \textcolor[rgb]{0.620,0.479,0.450}{41.65}}$ \\

 &  & 0.30 & 0.65 &  & ${\uline{\textcolor[rgb]{0.708,0.495,0.346}{32.29}}}$ & ${\bf \textcolor[rgb]{0.366,0.432,0.751}{68.14}}$ & ${\bf \textcolor[rgb]{0.366,0.432,0.751}{68.14}}$ \\

 &  & 0.60 & 0.90 &  & ${\uline{\textcolor[rgb]{0.565,0.469,0.516}{47.26}}}$ & ${\textcolor[rgb]{0.407,0.440,0.702}{64.06}}$ & ${\bf \textcolor[rgb]{0.381,0.435,0.734}{66.74}}$ \\

 &  & 0.79 & 0.99 &  & ${\uline{\textcolor[rgb]{0.524,0.461,0.564}{51.65}}}$ & ${\uline{\textcolor[rgb]{0.524,0.461,0.564}{51.65}}}$ & ${\bf \textcolor[rgb]{0.518,0.460,0.571}{52.17}}$ \\
\cline{1-8} \cline{2-8} 
\multirow[t]{4}{*}{20} & \multirow[t]{4}{*}{80} & 0.01 & 0.21 &  & ${\textcolor[rgb]{0.524,0.461,0.564}{51.65}}$ & ${\uline{\textcolor[rgb]{0.895,0.530,0.125}{12.76}}}$ & ${\bf \textcolor[rgb]{0.518,0.460,0.571}{52.17}}$ \\

 &  & 0.30 & 0.65 &  & ${\uline{\textcolor[rgb]{0.679,0.490,0.381}{35.25}}}$ & ${\bf \textcolor[rgb]{0.363,0.432,0.754}{68.67}}$ & ${\bf \textcolor[rgb]{0.363,0.432,0.754}{68.67}}$ \\

 &  & 0.60 & 0.90 &  & ${\uline{\textcolor[rgb]{0.869,0.525,0.156}{15.56}}}$ & ${\textcolor[rgb]{0.378,0.434,0.737}{66.89}}$ & ${\bf \textcolor[rgb]{0.378,0.434,0.737}{66.97}}$ \\

 &  & 0.81 & 0.99 &  & ${\uline{\textcolor[rgb]{1.000,0.549,0.000}{1.47}}}$ & ${\textcolor[rgb]{0.696,0.493,0.360}{33.42}}$ & ${\bf \textcolor[rgb]{0.620,0.479,0.450}{41.65}}$ \\
\cline{1-8} \cline{2-8} 
\multirow[t]{4}{*}{240} & \multirow[t]{4}{*}{60} & 0.01 & 0.09 &  & ${\uline{\textcolor[rgb]{1.000,0.549,0.000}{1.45}}}$ & ${\textcolor[rgb]{0.451,0.448,0.651}{59.29}}$ & ${\bf \textcolor[rgb]{0.372,0.433,0.744}{67.47}}$ \\

 &  & 0.30 & 0.50 &  & ${\uline{\textcolor[rgb]{0.719,0.497,0.332}{30.97}}}$ & ${\textcolor[rgb]{0.296,0.419,0.834}{75.55}}$ & ${\bf \textcolor[rgb]{0.278,0.416,0.855}{77.40}}$ \\

 &  & 0.60 & 0.79 &  & ${\uline{\textcolor[rgb]{0.606,0.476,0.467}{43.00}}}$ & ${\textcolor[rgb]{0.290,0.418,0.841}{76.23}}$ & ${\bf \textcolor[rgb]{0.275,0.416,0.858}{77.85}}$ \\

 &  & 0.90 & 0.99 &  & ${\textcolor[rgb]{0.410,0.440,0.699}{63.53}}$ & ${\uline{\textcolor[rgb]{0.430,0.444,0.675}{61.38}}}$ & ${\bf \textcolor[rgb]{0.302,0.420,0.827}{74.84}}$ \\
\cline{1-8} \cline{2-8} 
\multirow[t]{4}{*}{60} & \multirow[t]{4}{*}{240} & 0.01 & 0.10 &  & ${\uline{\textcolor[rgb]{0.410,0.440,0.699}{63.53}}}$ & ${\textcolor[rgb]{0.383,0.435,0.730}{66.44}}$ & ${\bf \textcolor[rgb]{0.302,0.420,0.827}{74.84}}$ \\

 &  & 0.30 & 0.50 &  & ${\uline{\textcolor[rgb]{0.658,0.486,0.405}{37.67}}}$ & ${\textcolor[rgb]{0.284,0.417,0.848}{76.79}}$ & ${\bf \textcolor[rgb]{0.284,0.417,0.848}{76.92}}$ \\

 &  & 0.60 & 0.79 &  & ${\uline{\textcolor[rgb]{0.752,0.503,0.294}{27.78}}}$ & ${\textcolor[rgb]{0.313,0.423,0.813}{73.68}}$ & ${\bf \textcolor[rgb]{0.270,0.414,0.865}{78.45}}$ \\

 &  & 0.91 & 0.99 &  & ${\uline{\textcolor[rgb]{1.000,0.549,0.000}{1.45}}}$ & ${\textcolor[rgb]{0.512,0.459,0.578}{52.82}}$ & ${\bf \textcolor[rgb]{0.392,0.437,0.720}{65.59}}$ \\
\cline{1-8} \cline{2-8} 

\end{tabular}
  \\  {\footnotesize  UX Z$_\text{U}$: unconditional exact Z-unpooled test, MPK: maximin power knapsack, APK: average power knapsack}
    \label{tab:power_table_marginsup}
\end{table}
\renewcommand{\arraystretch}{1.2} 
\begin{table*}[h!]
\centering
    \addtolength{\tabcolsep}{-0.3em}
    \caption{Summary table of power comparisons for superiority testing with a margin. The pointwise power comparison rows show the fraction (\%) of parameters for which the test in the column has strictly higher power than the APK test~(left) and the frequency of 
    parameters for which the APK test has strictly higher power~(right), where we consider all parameters~$\btheta$ such that~$\theta_\D>\theta_\C$ and both~$\theta_\C,\theta_\D$ are multiples of~0.01.   The average power comparison shows the difference in average power (\%) for the APK test over the column test, and in brackets the absolute difference in average power (\%) for parameters where the column test has higher power (left) and for parameters where the APK test has higher power (right), where ``$-$'' indicates there are no such parameter values. 
The one-sided significance level equals~$2.5\%$.}
    \centering 
    \footnotesize\begin{tabular}{lllll}
\toprule
 &  &  & UX Z$_{\text{U}}$ & MPK \\
$n$ & $(n_{\text{C}}, n_{\text{D}})$ & Power comparison &  &  \\
\midrule
\multirow[t]{6}{*}{20} & \multirow[t]{2}{*}{(10, 10)} & \% Pointwise higher (column/APK) & ${ 0.00/ 0.00}$ & ${ 0.00/ 0.00}$ \\
 &  & Avg. diff. total (APK worse/better) & \textcolor[RGB]{65,105,225}{${0.00~(-/-)}$} & \textcolor[RGB]{65,105,225}{${0.00~(-/-)}$} \\
\cline{2-5}
 & \multirow[t]{2}{*}{(7, 13)} & \% Pointwise higher (column/APK) & ${ 0.00/ 97.53}$ & ${ 18.21/ 79.32}$ \\
 &  & Avg. diff. total (APK worse/better) & \textcolor[RGB]{128,116,150}{${5.57~(-/5.71)}$} & \textcolor[RGB]{103,112,180}{${2.69~(2.32/3.93)}$} \\
\cline{2-5}
 & \multirow[t]{2}{*}{(4, 16)} & \% Pointwise higher (column/APK) & ${ 0.00/ 0.00}$ & ${ 0.00/ 0.00}$ \\
 &  & Avg. diff. total (APK worse/better) & \textcolor[RGB]{65,105,225}{${0.00~(-/-)}$} & \textcolor[RGB]{65,105,225}{${0.00~(-/-)}$} \\
\cline{1-5} \cline{2-5}
\multirow[t]{6}{*}{50} & \multirow[t]{2}{*}{(25, 25)} & \% Pointwise higher (column/APK) & ${ 0.00/ 0.00}$ & ${ 54.26/ 41.27}$ \\
 &  & Avg. diff. total (APK worse/better) & \textcolor[RGB]{65,105,225}{${0.00~(-/-)}$} & \textcolor[RGB]{65,105,225}{${0.01~(0.70/0.95)}$} \\
\cline{2-5}
 & \multirow[t]{2}{*}{(17, 33)} & \% Pointwise higher (column/APK) & ${ 0.00/ 97.78}$ & ${ 44.74/ 54.95}$ \\
 &  & Avg. diff. total (APK worse/better) & \textcolor[RGB]{115,114,165}{${3.65~(-/3.73)}$} & \textcolor[RGB]{65,105,225}{${0.49~(0.90/1.63)}$} \\
\cline{2-5}
 & \multirow[t]{2}{*}{(10, 40)} & \% Pointwise higher (column/APK) & ${ 14.21/ 85.57}$ & ${ 0.00/ 0.00}$ \\
 &  & Avg. diff. total (APK worse/better) & \textcolor[RGB]{242,137,14}{${13.72~(2.75/16.49)}$} & \textcolor[RGB]{65,105,225}{${0.00~(-/-)}$} \\
\cline{1-5} \cline{2-5}
\multirow[t]{6}{*}{100} & \multirow[t]{2}{*}{(50, 50)} & \% Pointwise higher (column/APK) & ${ 16.51/ 75.96}$ & ${ 19.47/ 74.65}$ \\
 &  & Avg. diff. total (APK worse/better) & \textcolor[RGB]{77,107,209}{${0.56~(0.08/0.76)}$} & \textcolor[RGB]{90,109,195}{${2.33~(0.05/3.13)}$} \\
\cline{2-5}
 & \multirow[t]{2}{*}{(35, 65)} & \% Pointwise higher (column/APK) & ${ 0.00/ 95.72}$ & ${ 26.25/ 68.77}$ \\
 &  & Avg. diff. total (APK worse/better) & \textcolor[RGB]{115,114,165}{${3.95~(-/4.13)}$} & \textcolor[RGB]{90,109,195}{${1.83~(0.15/2.72)}$} \\
\cline{2-5}
 & \multirow[t]{2}{*}{(20, 80)} & \% Pointwise higher (column/APK) & ${ 0.00/ 97.81}$ & ${ 39.23/ 55.92}$ \\
 &  & Avg. diff. total (APK worse/better) & \textcolor[RGB]{255,140,0}{${15.27~(-/15.61)}$} & \textcolor[RGB]{77,107,209}{${1.32~(0.13/2.45)}$} \\
\cline{1-5} \cline{2-5}
\multirow[t]{6}{*}{300} & \multirow[t]{2}{*}{(150, 150)} & \% Pointwise higher (column/APK) & ${ 10.02/ 62.14}$ & ${ 7.03/ 68.02}$ \\
 &  & Avg. diff. total (APK worse/better) & \textcolor[RGB]{65,105,225}{${0.11~(0.00/0.17)}$} & \textcolor[RGB]{90,109,195}{${1.90~(0.04/2.80)}$} \\
\cline{2-5}
 & \multirow[t]{2}{*}{(100, 200)} & \% Pointwise higher (column/APK) & ${ 0.00/ 77.90}$ & ${ 7.88/ 69.39}$ \\
 &  & Avg. diff. total (APK worse/better) & \textcolor[RGB]{90,109,195}{${2.42~(-/3.11)}$} & \textcolor[RGB]{90,109,195}{${1.69~(0.02/2.44)}$} \\
\cline{2-5}
 & \multirow[t]{2}{*}{(60, 240)} & \% Pointwise higher (column/APK) & ${ 0.77/ 88.26}$ & ${ 9.96/ 74.08}$ \\
 &  & Avg. diff. total (APK worse/better) & \textcolor[RGB]{217,133,45}{${12.42~(1.67/14.08)}$} & \textcolor[RGB]{77,107,209}{${1.09~(0.04/1.47)}$} \\
\cline{1-5} \cline{2-5}

\end{tabular}
     \label{tab:comparison_table_marginsup}\\
{\footnotesize  UX Z$_\text{U}$: unconditional exact  Z-unpooled test, MPK: maximin power knapsack, APK: average power knapsack}
\end{table*}
\FloatBarrier

\section{Comparison to non-exact tests}\label{sect:comparison_non_exact}
In this section, we compare the unconditional exact tests from the main paper with non-exact tests known from the literature. This comparison was not included in the main paper to keep the main comparison fair, as non-exact tests often show power gains over exact tests because they do not provide the same strong type I error rate control guarantees. Making a fair and objective comparison of non-exact versus exact tests is difficult because an acceptable degree of type I error rate inflation is subjective.

The included non-exact tests are the Z-unpooled test, which rejects when the test statistic in~\eqref{stat_unpooled}~(for~$\delta_n=0$) is lower than the $\alpha$-quantile~$z_\alpha$ of the standard normal distribution. This test equals the Wald test for the two-sample binomial testing setting. The second non-exact test is the pooled-Z test~(Z$_\text{P}$), which rejects when the Z-pooled statistic as defined in, e.g.,~\citet{mehrotra2003cautionary} is lower than~$z_\alpha$. This test equals the score test in the two-sample binomial testing setting. The third non-exact test is the signed root likelihood ratio~(SRLR) test, defined as in~\citet{diciccio2001simple} and based on the likelihood ratio from the two-sided two-sample binomial testing problem. 
The fourth non-exact test is Fisher's mid-p value~(FMP) test which is defined as in Section~3.4 but where we now reject if this mid-p value is less than~$\alpha.$ Finally, the fifth and sixth non-exact tests we consider are versions of the SRLR and Z$_\text{P}$ tests where we base the critical value on the estimated common success rate under the null hypothesis~(the E-Z$_\text{P}$ and E-SRLR tests) as described in~\citet{ripamonti_quatto_2017}. We have also calculated such versions for the~Z$_\text{U}$ test, FET and FMP test, however it was observed that the latter two tests had a very high correlation with the E-SRLR test, while the E-Z$_\text{U}$ test gave a substantially worse performance in comparison to the E-Z$_\text{P}$ and E-SRLR tests. Hence, we chose to consider only the E-Z$_\text{P}$ and E-SRLR tests.

\begin{figure}

    \centering
    \hspace*{-13mm}\includegraphics[width=1.1\linewidth]{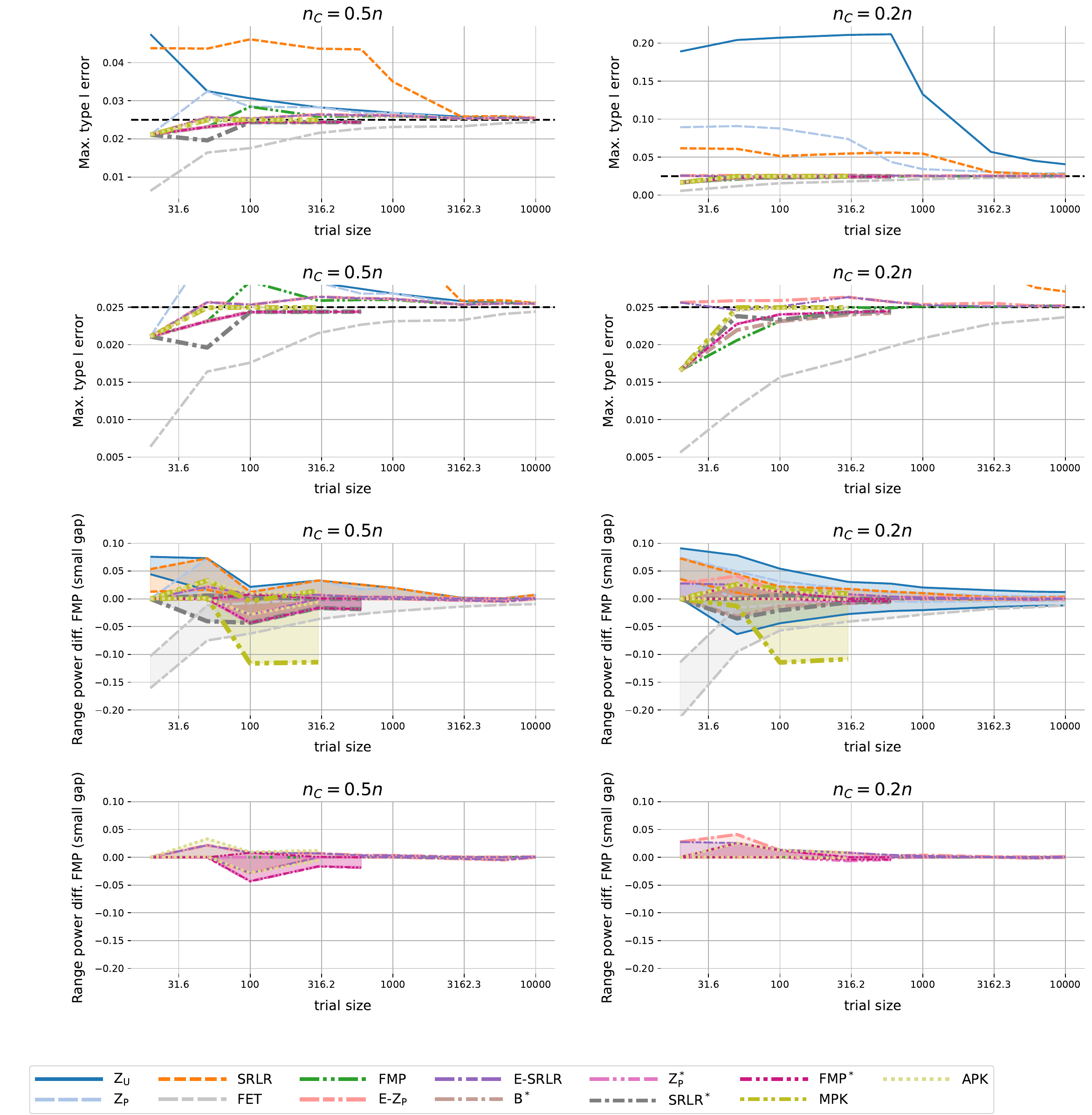}
    \caption{Top two rows: Maximum type I error rates~(for common success rates that are multiples of~1\%) for several non-exact and exact tests versus the trial size (log-scale) for the balanced case~($n_\C=0.5n$) and an unbalanced case~($n_\C= 0.2n$). The second row is a zoomed-in version of the top row. Bottom two rows: ranges of power differences with Fisher's mid-p-value test versus the trial size~$n$ (log scale) for different non-exact and exact tests. Power differences are calculated over parameters~$(\theta_{\C}, \theta_{\C} + \Delta_{\bn})$ where~$\theta_{\C}\in\{0.01, 0.02,\dots, 1-\Delta_{\bn}\}$ and~$\Delta_{\bn} = \frac{1}{2}(z_{97.5\%} + z_{80\%})\sqrt{1/n_{\D} + 1/n_{\C}}.$}
    \label{fig:non_exact_comparison}
\end{figure}

The top two rows in \autoref{fig:non_exact_comparison} show the maximum type I error rates over common success rates ranging from~0\% to 100\% with steps of 1\% versus the trial size~$n$ for the different tests. 
The bottom two rows display the range in power differences with the FMP test versus the trial size. The power differences are calculated over parameters~$(\theta_{\C}, \theta_{\C} + \Delta_{\bn})$ such that~$\theta_{\C}\in\{0.01, 0.02,\dots, 1-\Delta_{\bn}\}$ and~$\Delta_{\bn} = \frac{1}{2}(z_{97.5\%} + z_{80\%})\sqrt{1/n_{\D} + 1/n_{\C}}$ leading to approximately~$80\%$ power through a normal approximation.  
The left column of plots shows the results for balanced treatment groups, while the plots on the right show results for unbalanced treatment groups, where the control group contains 20\% of the trial participants (a 1:4 ratio).

From the top row in \autoref{fig:non_exact_comparison} we observe that the Z$_\text{U}$, Z$_\text{P}$
 and SRLR tests show substantial type I error inflation, where in the unbalanced treatment groups setting the Z$_\text{U}$ test still shows a type I error around~5\% for~$n=10,000$, hence these tests still yield non-negligible type I error inflation for very high trial sizes and are not recommended.  
Using our cutoff of about 2 hours of computation time, we recommend  the APK test for trial sizes up to 300 as it is an exact test and shows power values comparable to the (``near-exact'') E-SRLR, E-Z$_\text{P}$ and FMP tests (bottom row of \autoref{fig:non_exact_comparison}), showing that at the cost of longer computation times we can get similar power values as non-exact tests, but with the guarantee of type I error rate control. 
 For larger trial sizes, the best test depends on the tolerance for type I error rate inflation. If type I error rate control is a hard constraint, then the FMP$^*$ and~Z$_{\text{P}}^*$ p-values can likely be calculated up to relatively high trial sizes for single datasets (see \autoref{fig:comptime}); the exact evaluation of rejection rates for these tests may require a large computation time as it requires maximization over the nuisance parameter for each considered 2x2 table, but power evaluation may be tractable through simulation. If a slight type I error rate inflation is allowed, then we would recommend the FMP test, as it shows the lowest degree of type I error inflation out of all considered tests in this setting (for trial sizes up to 10,000), while it shows similar or higher power values comparable with the estimated p-value tests E-SRLR and E-Z$_P$.  
\FloatBarrier

\end{document}